\documentclass[floats,floatfix,showpacs,amssymb,prd,twocolumn,superscriptaddress,nofootinbib,nolongbibliography,reprint]{revtex4-2}

\usepackage{amssymb,amsmath, verbatim,mathtools,needspace,enumitem,etoolbox,graphicx, physics,microtype,afterpage,bigints,gensymb,tabularx,xspace, bm}

\usepackage[dvipsnames, usenames]{xcolor}
\definecolor{linkcolor}{rgb}{0.0,0.3,0.5}
\definecolor{dodgerblue}{HTML}{1E90FF}
\usepackage[unicode, colorlinks=true, linkcolor=linkcolor, citecolor=linkcolor, filecolor=linkcolor,urlcolor=linkcolor, pdfusetitle]{hyperref}
\usepackage[all]{hypcap}
\usepackage[T1]{fontenc}
\usepackage[utf8]{inputenc}
\usepackage{orcidlink}

\newcommand{\new}[1]{#1}

\renewcommand{\citet}[1]{\citeauthor{#1}~\cite{#1}}

\newcommand{\event}{GW200208\_222617\xspace}

\begin{document}
%%%%%%%%%%%%%%%%%%%%%%%%%%%%%%%%%%%%%%%%%%%%%%%%%%%%%%%%%%%%%%%%%%%%

% \title{On the eccentric black-hole merger \event}
\title{\event as an eccentric black-hole binary merger: \\ properties and astrophysical implications}

\author{Isobel Romero-Shaw$\,$\orcidlink{0000-0002-4181-8090}}
\email{isobel.romeroshaw@gmail.com}
\affiliation{Department of Applied Mathematics and Theoretical Physics, Cambridge CB3 0WA, United Kingdom}
\affiliation{Kavli Institute for Cosmology Cambridge, Madingley Road, Cambridge CB3 0HA, United Kingdom}
\affiliation{H.H. Wills Physics Laboratory, Tyndall Avenue, Bristol BS8 1TL, United Kingdom}

\author{Jakob Stegmann$\,$\orcidlink{0000-0003-2340-8140}}
\affiliation{Max Planck Institute for Astrophysics, 
Karl-Schwarzschild-Str.~1, 85748 Garching, Germany}

\author{Hiromichi Tagawa}
\affiliation{Shanghai Astronomical Observatory, Shanghai, 200030, People's Republic of China}

\author{Davide Gerosa$\,$\orcidlink{0000-0002-0933-3579}}
\affiliation{Dipartimento di Fisica ``G. Occhialini'', Universit\`a degli Studi di Milano-Bicocca, Piazza della Scienza 3, 20126 Milano, Italy}
\affiliation{INFN, Sezione di Milano-Bicocca, Piazza della Scienza 3, 20126 Milano, Italy}

\author{Johan Samsing}
\affiliation{Niels Bohr International Academy, The Niels Bohr Institute, Blegdamsvej 17, DK-2100, Copenhagen, Denmark}

\author{Nihar Gupte$\,$\orcidlink{0000-0002-7287-5151}}
\affiliation{Max Planck Institute for Gravitational Physics (Albert Einstein Institute), Am M\"uhlenberg 1, Potsdam 14476, Germany}
\affiliation{Department of Physics, University of Maryland, College Park, MD 20742, USA}

\author{Stephen R. Green}
\affiliation{School of Mathematical Sciences, University of Nottingham, University Park, Nottingham NG7 2RD, United Kingdom}

\begin{abstract}

Detecting orbital eccentricity in a stellar-mass black-hole merger would point to a non-isolated formation channel. Eccentric binaries can form in dense
stellar environments such as globular clusters or active galactic nuclei, or from triple stellar systems in the Galactic field. However, confidently
measuring eccentricity is challenging—short signals from high-mass eccentric mergers can mimic spin-induced precession, making the two effects hard to disentangle. This
degeneracy weakens considerably for longer-duration signals. Here, \event\ provides a rare opportunity. Originating from a relatively low-mass binary with
source-frame chirp mass $\sim20$~M$_\odot$, its gravitational-wave signal spanned $\sim14$ orbital cycles in band, with no indication of data quality issues. Previous analyses for quasi-circular binaries found no evidence for spin precession, and multiple subsequent studies found the data to favour an eccentric merger despite notable technical differences. All in all, we believe \event is the black-hole merger event from GWTC-3 with the least ambiguous detection of eccentricity. We present a critical discussion of properties and astrophysical interpretation of \event as an eccentric black-hole merger using models of field triples, globular clusters, and active galactic nuclei. We find that if \event was indeed eccentric, its origin is consistent with a field triple or globular cluster. Formation in the inner regions of an active galactic nucleus is disfavoured. The outer regions of such a disk remain a viable origin for \event; we demonstrate how future detections of eccentric mergers formed in such environments could be powerful tools for constraining the disk geometry.

\end{abstract}

\maketitle

%\dg{broad comment, should we use the full telephone number instead of cutting it short at \event?} \irs{Yes, we probably should. Its full name is GW200208\_222617.}

%\dg{For all figures, can I recommend setting "usetex=True" in Python so fonts match the rest of the draft? }

\section{Introduction}

%\irs{Outline of introduction:}
Measurable orbital eccentricity in a compact binary observed via its gravitational-wave (GW) emission in current ground-based detectors is considered smoking-gun evidence that the binary was externally driven to merge, either through interactions in dynamical environments or in field multiples \citep[e.g.,][]{Lower2018, Zevin2021}. This is wholly inconsistent with isolated compact binary evolution, which only yields circularized mergers at $10$~Hz GW frequency, where sources enter the sensitivity range of current detectors~\citep{Peters1964, 2024PhRvD.110f3012F}. %at $10$~Hz 
Evidence for orbital eccentricity in the current population of detected LIGO-Virgo-KAGRA (LVK) %~\citep{Aasi13}  \dg{wrong citation? This is a review on future upgrades}
binary black holes (BBHs) has been claimed by multiple groups \citep[e.g.,][]{2020:Romero-Shaw:GW190521, 2020:Gayathri:GW190521, 2023:Gamba:GW190521, 2024:Iglesias:Eccentric-re-analysis, 2024:Gupte:GWTC-3-ecc, Lluc:2025:eccentricity}. However, the robustness of these claims is debated due to a lack of complete and efficient inspiral-merger-ringdown gravitational waveform models with the effects of eccentricity, spin-induced precession, and higher-order modes included. 

The unusually massive BBH progenitor of the GW signal GW190521 was the first to be touted as potentially having measurable orbital eccentricity \citep{2020:Romero-Shaw:GW190521, 2020:Gayathri:GW190521, 2023:Gamba:GW190521}, although these results are in tension with other studies~\citep{2024:Iglesias:Eccentric-re-analysis, 2024:Gupte:GWTC-3-ecc}. Unfortunately, the short duration of GW190521 makes it difficult to determine its eccentricity using existing GW approximants for merger-dominated signals, which model either eccentricity or spin-induced precession---two effects which are somewhat degenerate---but not both \citep{2023:Romero-Shaw:Ecc-or-precc, 2023:XuHamilton:Ecc-or-precc}. 
Without complete waveform approximants containing both eccentricity and spin-precession, the best we can do is to compare the evidence for eccentricity to the evidence for precession~\citep{2020:Romero-Shaw:GW190521, 2023:Romero-Shaw:Ecc-or-precc} or spot-check the data against numerical relativity simulations over a reduced parameter space \citep{2020:Gayathri:GW190521}. When comparing evidences, short signals like GW190521 are ambiguous: eccentricity modulates the amplitude and phase of the signal over shorter timescales than in-plane spins, but if the waveform is too short to observe a full eccentricity cycle, the effects cannot be distinguished from one another. Long signals from low-mass events with more visible cycles are more promising.

Both \citet{2022:Romero-Shaw:GWTC-3-ecc} and \citet{2024:Gupte:GWTC-3-ecc}, using different analysis methods and waveform models, found non-negligible evidence for non-zero eccentricity in a BBH with chirp mass $\mathcal{M}_c\simeq 20$~M$_\odot$ and $\sim 14$ orbital cycles in band: \event. Crucially, unlike several other events found to have possible evidence for eccentricity by either study, this event had no prior evidence for significant spin-precession nor strongly negative spins from the initial LVK analysis \citep{GWTC-3}. 

In this paper, we investigate the properties and astrophysical implications of \event as a putative detection of an eccentric BBH. 
%We compare the results of \citet{2022:Romero-Shaw:GWTC-3-ecc} and \citet{2024:Gupte:GWTC-3-ecc} for this event and highlight the similarities and differences in their findings. We discuss why results for other events differed between the two studies.  
We argue that, due to the lack of evidence for strong misaligned spins, its long inspiral, the consistent properties as inferred in both Refs.~\cite{2022:Romero-Shaw:GWTC-3-ecc} and \cite{2024:Gupte:GWTC-3-ecc}, and cleanliness from data quality issues, \event represents the BBH candidate from GWTC-3 with the least ambiguous detection of eccentricity. 
We discuss the inferred properties of \event in section \ref{sec:properties}, review possible astrophysical formation scenarios in section \ref{sec:astro}, and demonstrate the importance of scattering environment geometry in section \ref{sec:geometry}. We discuss the astrophysical implications of \event in section \ref{sec:conclusion}.  

\section{Properties of \event}
\label{sec:properties}

\begin{figure}
    \centering
    \includegraphics[width=0.99\columnwidth]{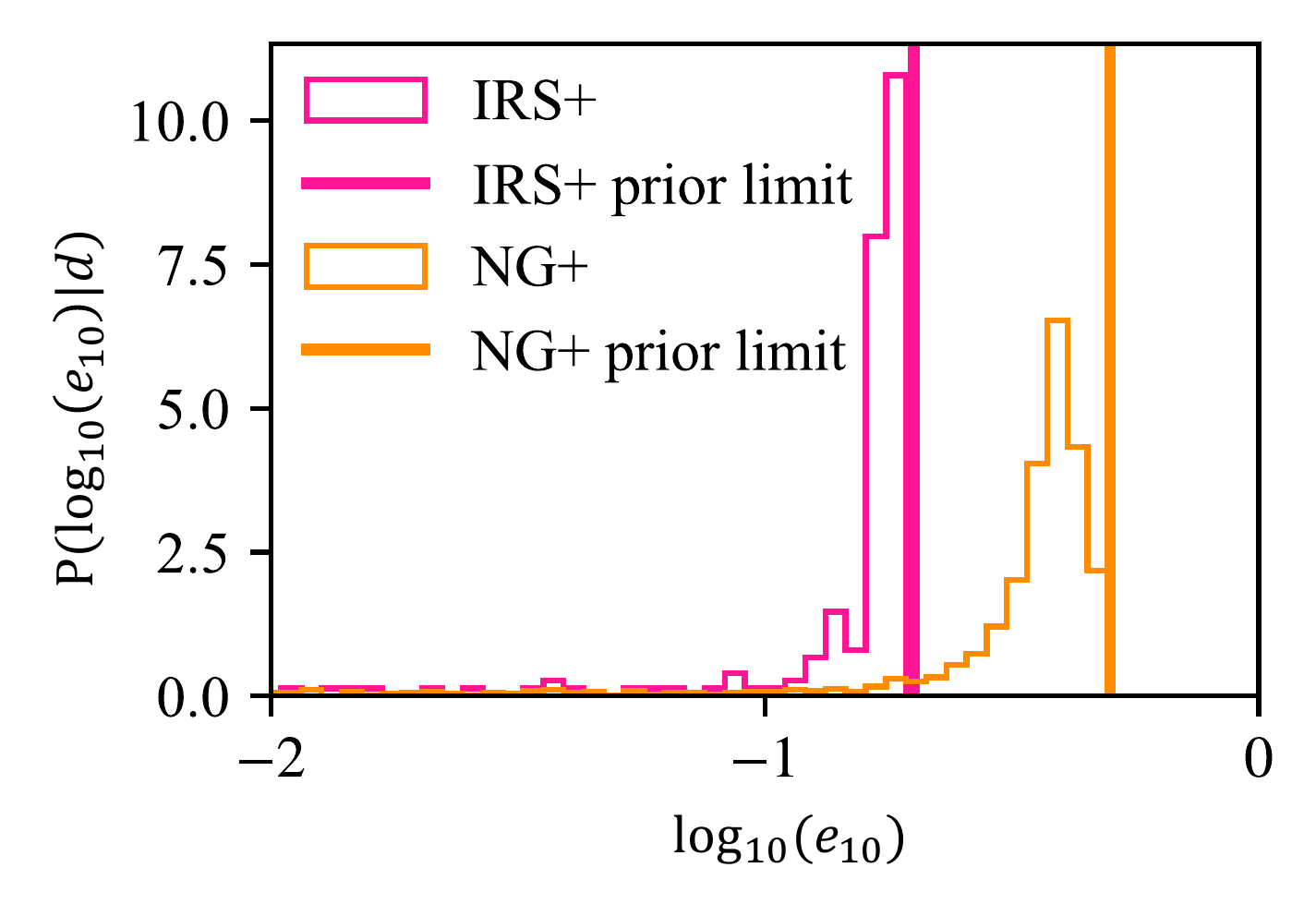}
    \caption{Comparison of eccentricity measurements at $10$~Hz obtained with (i) waveform approximant SEOBNRE by \citet{2022:Romero-Shaw:GWTC-3-ecc} (IRS+; pink histogram) and a log-uniform prior in the range $10^{-4} \leq e_{10} \leq 0.2$, and (ii) waveform approximant SEOBNRv4EHM by \citet{2024:Gupte:GWTC-3-ecc} (NG+; orange histogram) and a log-uniform prior in the range $10^{-4} \leq e_{10} \leq 0.5$.
    We plot these posteriors in log scale as they were obtained with log-uniform eccentricity priors, with upper limits indicated by thick vertical lines.
    }
    \label{fig:ecc-posts}
\end{figure}
\renewcommand{\arraystretch}{1.3} % 1.0 = default
\begin{table}[]
    \centering
\begin{tabular}{l|c|c|c}
    & LVK & IRS+ & NG+ \\
     \hline \hline
    Total mass $M_\mathrm{T}$ [M$_\odot$] & $63^{+100}_{-26}$ & $42^{+10}_{-15}$ & $46^{+17}_{-8}$ \\
    Mass ratio $q$ & $0.21^{+0.67}_{-0.16}$ & $0.65^{+0.30}_{-0.40}$ & $0.55^{+0.40}_{-0.35}$ \\
    Effective spin $\chi_\mathrm{eff}$ & $0.45^{+0.42}_{-0.46}$ & $0.05^{+0.23}_{-0.13}$ & $0.13^{+0.29}_{-0.23}$ \\
    % Effective precession spin $\chi_\mathrm{p}$ &  & N/A & N/A \\
    Eccentricity $e_{10}$ &  --- & $0.18^{+0.02}_{-0.18}$ & $0.39^{+0.09}_{-0.39}$ \\
    % Relativistic anomaly $l$ & N/A & N/A  & \\
    Luminosity dist. $d_\mathrm{L}$ [Gpc] & $4.1^{+4.4}_{-2.0}$ & $2.6^{+1.6}_{-1.1}$ & $2.4^{+1.9}_{-1.3}$ \\
    % Right ascension $\alpha$ [rad] &  & & \\
    % Declination $\delta$ [rad] &  & & \\
    % Inclination $\iota$ [rad] &  & & \\
    % Reference phase $\phi$ [rad] &  & & \\
    % Polarisation $\psi$ [rad] &  & & \\
\end{tabular}
    \caption{\new{Medians and $90\%$ credible intervals for parameters of interest inferred through the analyses of the LVK, IRS+ and NG+. Total mass is given in the source frame.}}
   % \dg{There are too many digits in those numbers. I would restrict to significant digits, so for instance $0.18^{+0.02}_{-0.18}$ instead of $0.179^{+0.019}_{-0.178}$}
    \label{tab:parameters}
\end{table}

There have been several recent studies that systematically analyze current GW events looking for signs of orbital eccentricity.
In this paper, we compare the following two investigations (but see also, e.g., Refs.~\cite{2024:Iglesias:Eccentric-re-analysis, Lluc:2025:eccentricity}):

\begin{itemize}
\item

\citet{2022:Romero-Shaw:GWTC-3-ecc}, henceforth IRS+, presented measurements of orbital eccentricity in events from GWTC-3 using likelihood reweighting \citep{2019:Romero-Shaw:GWTC-1-ecc,2019:Payne:Likelihood-reweighting} in conjunction with the eccentric waveform model SEOBNRE \citep{2017:CaoHan:SEOBNRE}. SEOBNRE is an accurate but computationally inefficient model not suitable for conventional Bayesian inference, thus requiring approximate strategies. 
In particular, IRS+ flagged four specific events as candidates for non-zero eccentricity: 
GW190521, GW190620, GW191109, and \event.\footnote{All events referred to in this work were the only events on the day they occurred, except GW190521 and \event. GW190521 was a high-profile exceptional event \citep{GW190521-disco} and we thereefore refer to it by its shortened name as in Ref.~\cite{GWTC-2}.} The eccentric aligned-spin versus quasi-circular spin-precessing Bayes factors for these events were in the range $0.1 \lesssim \mathcal{B}_\mathrm{E/P} \lesssim 10$ using a log-uniform prior on $e_{10}$ in the range $10^{-4} \leq e_{10} \leq 0.2$. 

\item
\citet{2024:Gupte:GWTC-3-ecc}, henceforth NG+, performed a full re-analysis of the same events using SEOBNRv4EHM \cite{Ramos-Buades:2021adz}---a more sophisticated model that includes higher-order modes and enables variation of both the eccentricity and the relativistic anomaly. While still not adequately efficient for traditional stochastic samplers, analyzing data with such models is now accessible via deep-learning strategies, most notably DINGO~\citep{Dax:Dingo:2021}. 
They found three events with evidence for non-zero eccentricity that had compelling Bayes factors compared with the spin-precessing, quasi-circular hypothesis: GW190701, GW200129, and \event. The eccentric aligned-spin versus quasi-circular spin-precessing Bayes factors for these events were in the range $3 \lesssim \mathcal{B}_\mathrm{E/P} \lesssim 10^4$ 
using a log-uniform prior on $e_{10}$  in the range $10^{-4} \leq e_{10} \leq 0.5$.
\end{itemize}

The keen-eyed reader will note that only one event was confidently identified in both studies: \event. This was found with $\mathcal{B}_\mathrm{E/P} = 2.6$ by IRS+ and $\mathcal{B}_\mathrm{E/P} = 3.0$ by NG+ with log-uniform priors on $e_{10}$, where $e_{10}$ is the eccentricity provided to the waveform model at a reference frequency of $10$~Hz. These moderate Bayes factors do not overwhelmingly favour the eccentric hypothesis. However, with a uniform eccentricity prior over the same range, NG+ found $\mathcal{B}_\mathrm{E/P} = 17$. A critical discussion of the differences between these two studies is presented in Appendix~\ref{differences}; variations in results for other eccentric BBH candidates are discussed in Appendix~\ref{other}. 

\event was first reported in GWTC-3 \citep{GWTC-3} as a low signal-to-noise ratio (SNR) ($\rho \simeq 7.5$) signal from a BBH with median detector-frame (source-frame) chirp mass of $\mathcal{M}_c\simeq 45$~M$_\odot$ ($\simeq 20$~M$_\odot$), mass ratio $q \simeq 0.21$, and  effective spin parameter $\chi_{\rm eff}\simeq 0.45$.  GWTC-3 identified significant multimodality in the posterior distribution of \event, with two modes in the masses and spins marginals: one favouring more unequal mass ratios and higher spins, and one favouring more equal mass ratios and lower spins. 

\event had a false alarm rate (FAR) of $160$, $420$ and $4.8$ per year and $p_{\text{astro}}$ of $<0.01$, $<0.01$ and $0.7$ when reported when using the \texttt{gstLAL}, \texttt{MBTA} and \texttt{pycbc-BBH} search pipelines respectively 
\cite{GWTC-3, Messick:2016aqy, Adams:2015ulm, Usman:2015kfa}. Both the FAR and $p_{\text{astro}}$ are computed using a quasi-circular template bank, and estimating these quantities using an eccentric template bank requires further investigation. We note, however, that the SNR for this event increases when an eccentric model with higher-order modes is used. The SEOBNRv4EHM analysis of NG+ increases the median two-detector (LIGO Hanford and LIGO Livingston) SNR from 6.87 to 8.30, which is higher than the three-detector (as above plus Virgo) median network SNR of 7.4 reported in Ref.~\cite{GWTC-3}. The quasi-circular IMRPhenomD~\citep{IMRPhenomD} analysis and the SEOBNRE results, reweighted from IMRPhenomD, of IRS+ find slightly lower median three-detector SNRs of $\sim6.9$, which may be attributed to the lack of higher-order modes in these waveforms.

There are four key reasons that we consider measurement of eccentricity in \event to be of particular interest:
\begin{itemize}
    \item \event has a relatively long duration. Eccentricity is stronger the further away from merger it is measured \citep{Peters1964}. For a fixed SNR, evidence for eccentricity also grows with the number of orbital cycles visible in-band, so long inspirals represent a less ambiguous opportunity to measure eccentricity~\citep{2021:Romero-Shaw:GWTC-2-ecc, 2023:Romero-Shaw:Ecc-or-precc}. 
    \item \event has shown no evidence for spin-induced precession in analyses that include this effect \citep[e.g.,][]{GWTC-3}. This makes it unlikely that a spurious measurement of eccentricity would be made as a result of neglecting spin-precession in eccentric analyses \citep[e.g.,][]{2020:Romero-Shaw:GW190521, 2023:Romero-Shaw:Ecc-or-precc}.
    \item \event has been reported by several analyses using different inference methods and waveform models to have evidence for eccentricity. In addition to the two studies we compare in this work, see Ref.~\citep{Lluc:2025:eccentricity}.
    \item \event does not contain data quality issues, unlike GW200129 and GW190701, which were reported as eccentric candidates with higher Bayes factors in NG+.
\end{itemize}
The analyses of IRS+ and NG+ both favor the less extreme mass (higher chirp mass, more equal mass ratio) and spin ($\chi_\mathrm{eff}$ more consistent with 0) modes reported in Ref.~\citep{GWTC-3}. The two eccentric analyses find \new{consistent} parameters for \event: the BBH has detector-frame (source-frame) chirp mass of $\mathcal{M}_c\simeq25$~M$_\odot$ ($\simeq17$~M$_\odot$), mass ratio $q\simeq 0.5$, and spins consistent with $\chi_\mathrm{eff} \simeq 0$ with a slight skew to positive values. Both favour luminosity distances $d_{\rm L}\simeq 2.5$~Gpc, lower than the distance of $\simeq4.1$~Gpc found in the GWTC-3 result \citep{GWTC-3}. Both also find distinctly non-zero posteriors for eccentricity as defined within the waveform model at $10$~Hz, $e_{10}$. \new{We compare median and $90\%$ credible intervals recovered parameters of interest between these two analyses and the analysis of the LVK in Table \ref{tab:parameters}}.
% Specifically, the analysis of IRS+ find $\mathcal{M}_c=25^{+3}_{-9}$~M$_\odot$, $q=0.65^{+0.30}_{-0.40}$, $\chi_\mathrm{eff}=0.05^{+0.23}_{-0.13}$, $d_\mathrm{L} = 2.6^{+1.6}_{-1.1}$~Gpc, and $e_{10} = 0.179^{+0.019}_{-0.178}$, while NG+ find $\mathcal{M}_c=26^{+5}_{-3}$~M$_\odot$, $q=0.55^{+0.40}_{-0.35}$, $\chi_\mathrm{eff}=0.13^{+0.29}_{-0.23}$, $d_\mathrm{L} = 2.4^{+1.9}_{-1.3}$~Gpc, and $e_{10} = 0.386^{+0.091}_{-0.385}$ (we quote medians and 90\% credible intervals).

The posteriors on $e_{10}$ from the two analyses we consider are quantitatively distinct, as demonstrated in Figure \ref{fig:ecc-posts}. While different waveform models have different eccentricity definitions \citep[e.g.,][]{2022ApJ...936..172K}, the difference we see here can be almost entirely attributed to different prior limits: the IRS+ posterior strongly rails at the upper bound, suggesting that the true peak lies beyond $e_{10}=0.2$, whereas the priors of NG+ appear to capture the peak of the posterior at $e_{10}\approx0.4$ while still railing at the upper limit of $0.5$. Qualitatively, both posteriors demonstrate the same astrophysical conclusion: considerable preference for non-zero eccentricity, indicating a non-isolated formation scenario. 

\section{Astrophysical implications of \event}
\label{sec:astro}

If \event is indeed an eccentric binary, it is virtually impossible that it formed in a fully isolated binary scenario \citep{Peters1964, 2024PhRvD.110f3012F}. There are, however, several formation pathways that may produce BBH mergers with measurable eccentricity in the LVK band and properties like those measured for \event. 

Eccentricities predicted in simulations are commonly defined at a GW peak frequency of $10$~Hz. The peak frequency is approximated as \citep{Wen2003}
\begin{align}
\label{eq:peak}
f_p & = \frac{(1+e)^{1.1954}}{\pi (1-e^2)^{3/2}} \sqrt{GM/a^3} \nonumber\\
        & \approx \pi^{-1} \sqrt{GM/r^3},
\end{align}
where $e$ is the orbital eccentricity, $a$ the semi-major axis, $M$ the total binary mass, and $r=a(1-e)$ the pericenter distance. The last expression illustrates that the peak frequency is predominantly determined by the pericenter distance; we will use this in section \ref{sec:geometry}. Both eccentricity and peak frequency are ill-defined in the high-eccentricity limit in general \citep[e.g.,][]{Vijaykumar2024, 2025PhRvD.112b4012F, 2025PhRvD.111b4008B}, but resolving this is far beyond the scope of this work, where we focus on the astrophysical implications of measurable eccentricity for \event. Since the ``measurability'' threshold of $e_{10} \approx 0.05$ for BBHs in current ground-based detectors \citep{Lower2018} is below the threshold at which eccentricity definitions start to deviate substantially \citep[e.g.,][]{Vijaykumar2024}, the current definitions are adequate for our purposes.

Below, we review formation scenarios that could have formed \event if it was an eccentric merger, and discuss the astrophysical implications in each case. We find that \event could plausibly have originated in a hierarchical field triple or in a globular cluster, while its formation in an active galactic nucleus depends on its location in the disk and the disk geometry. 

\subsection{Hierarchical stellar triples in the field}
The majority of massive stars that are progenitors to black holes are found in close inner binaries which are orbited by distant tertiary companion \citep{Moe2017}. In these hierarchical triples, the gravitational perturbation from the companion can cause long-term, large-amplitude von Zeipel-Kozai-Lidov (ZKL) oscillations of the inner binary eccentricity and inclination \citep{Zeipel1910,Kozai1962,Lidov1962}. Once the inner binary forms a BBH, tertiary-driven ZKL eccentricity oscillations can promote a merger of the inner BBH by increasing the energy loss due to GWs at close pericenter passage. While most BBH mergers formed in this way would largely circularize due to GW emission upon entering the frequency band of ground-based interferometers, triple population synthesis studies consistently predict that a considerable fraction ($\sim1$~--~$30\,\%$) of such events retain a residual eccentricity $e_{p,10}\gtrsim 0.1$, where $e_{p,10}$ is the eccentricity at a GW peak frequency of $10$~Hz 
~\citep{Antonini2014,Antognini2014,Antonini2016,Antonini2017,Liu2019,Martinez2020,Fragione2020}.

In hierarchical triples, it is generally expected that at BBH formation the black-hole spins start out nearly aligned to each other and to the inner binary orbital angular momentum vector. This is a result of tidal interactions between the progenitor stars (just like in the classical isolated binary channel \cite{1981A&A....99..126H,2018PhRvD..98h4036G}). Small misalignments  may only be expected due to natal kicks at black-hole formation \cite{2000ApJ...541..319K} and are typically less than a few degrees \citep{Antonini2018}. The subsequent spin evolution from BBH formation to merger is determined by the evolution of the binary orbital angular momentum vector, whose direction oscillates due to the ZKL effect, and relativistic spin-orbit and spin-spin couplings \citep{Rodriguez2018}.

BBHs that merge with residual eccentricity undergo exceptionally strong ZKL oscillations, while their spin evolution is highly non-adiabatic \citep{Stegmann:2025shr}. That is, the orbital angular momentum vector oscillates several orders of magnitude faster than the spins can follow. \new{These BBHs eventually merge due to highly efficient GW emission at a near-radial eccentricity. While their spins remain strongly correlated and closely aligned with each other, the angle between the spin vectors and the orbital angular momentum is randomized. Therefore, one expects a uniform distribution for $\mathrm{cos}~\theta_i$}
%the spin-orbit angles are effectively randomized but the spins remain strongly correlated and close to each other 
\citep{Antonini2018,Rodriguez2018,Liu2018,Liu2019}. In other words, the distribution of $\cos\theta_i = \mathbf{\hat{L}}\cdot\mathbf{\hat{S}}_i$ ($i=1,2$) for BBHs with $e_{p,10}>10^{-3}$ 
and $q=1$ is expected to be roughly flat between the limits of the spin magnitude (so between $-1$ and $1$ if the spin magnitudes are maximal)
, while the difference $|\cos\theta_1 - \cos\theta_2|$ is expected to peak at zero~\citep{Rodriguez2018,Liu2019}.

In general, BBHs with $\cos\theta_1 \sim \cos\theta_2$ result in a broader $\chi_{\rm eff}$ distribution. 
The $\chi_{\rm eff}$ distribution from such sources is less strongly peaked around zero compared to cases where the spins are randomly oriented with respect to one another (as expected in spherically-symmetric star clusters). More details on this argument are provided in a dedicated paper~\citep{Stegmann:2025shr}. 

The mass ratio distribution of specifically eccentric tertiary-driven BBH mergers remains largely unexplored.  
However, field triples are generally expected to produce lower ($q = m_2 / m_1 < 1$) mass ratio BBH mergers like \event more efficiently than isolated BBH mergers can. For instance, Ref.~\cite{Martinez2022} finds about an order of magnitude more detectable mergers at $q\sim0.5$ from ZKL-driven BBH mergers in triples than in non-ZKL-driven BBH mergers. 
\new{From, e.g., Refs.~\citep{Martinez2022, Dorozsmai2025}, more than $90\%$ of tertiary-driven BBH mergers have $q > 0.3$, and from Ref. \citep{Stegmann:2025shr}, more than $90\%$ of detectably-eccentric mergers from triples have $|\chi_\mathrm{eff}| < 0.5$. The posterior distributions on \event from NG+ and IRS+ have $83\%$ and $94\%$ support in this region, respectively.}

Ref.~\cite{Liu2019} studied the formation of highly-eccentric compact object mergers by focusing on a fiducial BBH merger with $m_1=30\,\rm M_\odot$ and $m_2=20\,\rm M_\odot$ (so $q\simeq 0.66$, \new{consistent with} that inferred for \event, but with higher component masses) and a fiducial neutron-star black-hole merger with the same $m_1$ but $m_2=1.4\,\rm M_\odot$ (so $q\simeq 0.05$), and found that more than twice as many neutron-star black-hole mergers form with $e_{p,10}$ compared to BBHs. We therefore speculate that there could be a preference for eccentric mergers with a more unequal mass ratios from field triples. The recent claim of eccentricity detected in neutron star-black hole merger GW200105 \citep{2025arXiv250315393M,2025arXiv250601760D} may indeed imply a significant contribution to the compact-object merger population from field triples \citep{StegmannKlencki2025}, of which \event could be another example.

\subsection{Star Clusters}

\begin{figure}
    \centering
    \includegraphics[width=0.5\textwidth]{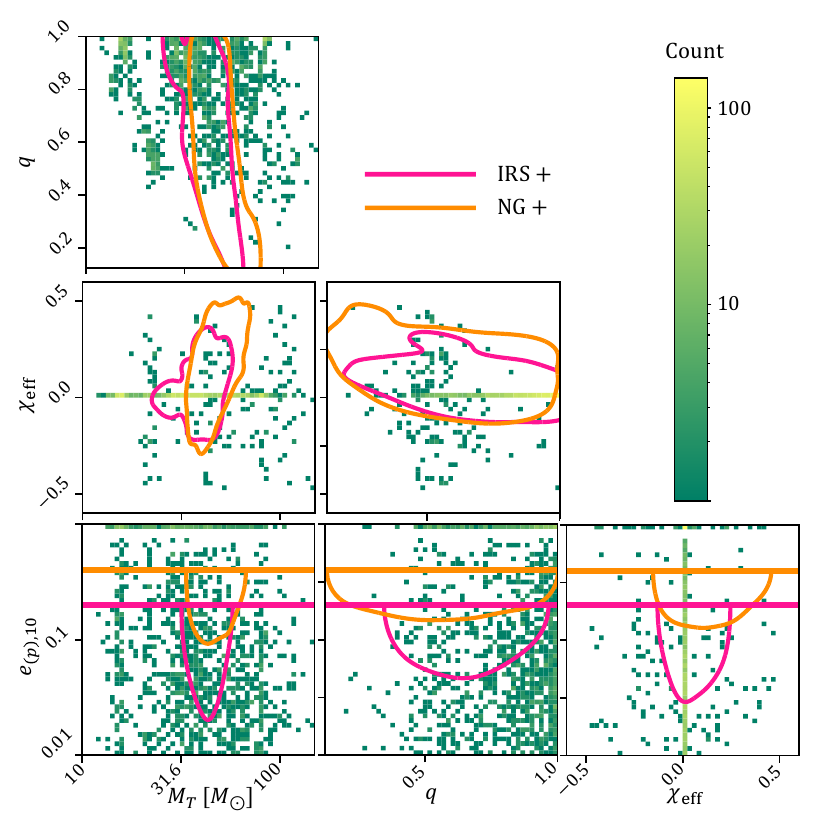}
    \caption{\label{fig:cmc} Distribution of few-body, in-cluster mergers and single-single GW-capture mergers from the \texttt{Cluster Monte Carlo Catalog} \citep{Kremer:CMC:2020} (scatter points) which models globular clusters consistent with those observed in the Milky Way. 
    %Mergers occurring after the binary is ejected from the cluster are not included. 
    \new{Only mergers with $e_{p,10} > 0.01$ are included in this plot.}
    The high-count ``spikes'' at  $\chi_\mathrm{eff}=0$ are first-generation mergers with BH natal spins assumed to be zero, while higher-generation mergers present a broader range of $\chi_\mathrm{eff}$. The colour scale is logarithmic; there are orders of magnitude more first-generation than hierarchical mergers. Since \event has properties consistent with both first- and higher-generation mergers, the relative abundance of first-generation mergers makes this the more probable of the cluster formation scenarios. 
    The 90\% credible intervals of \event are overlain in pink and orange using posterior samples from IRS+ and NG+, respectively.
    \new{In the lowest row, thick horizontal bars sit at the upper prior limit on $e_{10}$ for each analysis.}
   The definition of the eccentricities plotted here differ between simulation and posterior: eccentricities $e_{p,10}$ are extracted from the cluster simulations at the GW peak frequency of $10$~Hz, while we plot the values of eccentricity input to the waveform models $e_{10}$ for the IRS+ and NG+ results. The differences between these definitions is expected to be small for the vast majority of the parameter space shown here \citep{Vijaykumar2024}.
   }
\end{figure}

Star clusters---including globular, nuclear, young and open clusters---are potential formation environments for merging BBHs \cite{
Samsing14, RodriguezChatterjee2016, Askar2017, 2018PhRvD..97j3014S, JSDJ18, Banerjee2018, Banerjee2021, Mapelli:2021MNRAS, 2024A&A...683A.186D, 2023:Chattopadhyay:NSCs}. Globular clusters  and nuclear clusters are of particular interest for the formation of systems with measurable eccentricity. In these environments, BBHs may enter current ground-based detectors with detectable eccentricity if they merge inside the cluster due to few-body interactions or single-single capture \cite{2018PhRvD..97j3014S, 2020PhRvD.101l3010S, Zevin2021, 2024A&A...683A.186D, 2023:Chattopadhyay:NSCs}. Such mergers take place between BHs that have not interacted during their stellar evolution: as such, their spin-spin and spin-orbit angles should be isotropically distributed. 

Mass segregation is expected to preferentially pair up mergers with $q \sim 1$, and we can anticipate more massive BHs in mergers in both nuclear and globular clusters compared to both isolated binaries and triples, since merger remnants that are retained in the cluster can go on to merge again, building up larger binary components \cite{2019PhRvD.100d3027R, 2021ApJ...923..126S, 2024A&A...683A.186D,2021ApJ...915...56G}. The maximum mass that can form through hierarchical mergers depends on the escape velocity of the cluster: a higher escape velocity facilitates more generations of mergers before the final remnant is ejected \cite{2019PhRvD.100d1301G,2020ApJ...893...35D,2021ApJ...915...56G,2021ApJ...915L..35K,Mapelli:2021MNRAS,2025arXiv250321278B}. The escape velocity also influences the eccentricity of BBHs, as in-cluster mergers have higher eccentricities than those that are ejected before they merge~\cite{2018PhRvD..97j3014S, Zevin2021}. 

The natal spins of BHs influence both the spins and masses we would expect to detect from clusters. If natal spins are small, the distribution of $\chi_\mathrm{eff}$ for first-generation mergers will, of course, be peaked at $0$. In contrast, mergers containing merger remnants will have a broader spin distribution and higher values of the effective precession parameter $\chi_p$ \citep{2024ApJ...966L..16P}, since the dimensionless spin magnitude distribution of remnants is centred at $0.7$ \cite{2017PhRvD..95l4046G,2017ApJ...840L..24F}; \event shows no signs of having strongly spinning components, disfavouring it as a higher-generation merger if it formed in a cluster. 
It can be shown analytically that formation in clusters gives rise to a near-uniform symmetric $\chi_\mathrm{eff}$ distribution for mergers with one second-generation (2g) and one first-generation (1g) component, with maxima and minima at $\chi_\mathrm{eff} \simeq 0.45$ \citep{2021PhRvD.104h4002B,Antonini:cluster-pop:2025}. If natal spins are high, the relative rate of higher-generation mergers decreases, as high spins increase the merger kick such that more merger products are kicked out of the cluster \cite{2019PhRvD.100d1301G,2025arXiv250321278B,Antonini:cluster-pop:2025}. 

\new{The small spins inferred for \event, if it is a first-generation globular cluster merger, would imply a relatively high rate of higher-generation mergers: small natal spins are shown to lead to more remnants being retained in-cluster and hence an increased rate of hierarchical mergers \cite{2019PhRvD.100d1301G}, although the relative rate is also dependent on cluster escape velocity. Reference~\cite{Antonini:cluster-pop:2025} suggests that the population spin distribution is consistent with $\sim1\%$ of the population comprising hierarchical mergers in star clusters, implying that $\simeq20\%$ of the observed population is of star cluster origin. Since $\approx 5\%$ of star cluster BBH mergers are detectably eccentric \citep{Zevin2021}, we would therefore expect $\mathcal{O}(5\% \times 20\% = 1\%)$ of the observed BBHs to be detectably eccentric.}
% The small spins inferred for \event, if it is a first-generation cluster merger, would imply a relatively high rate of higher-generation mergers, implying that other binaries in the population could contain merger remnants, as suggested by Ref.~\cite{Antonini:cluster-pop:2025}. 

Figure \ref{fig:cmc} compares the properties of \event as an eccentric BBH as reported by IRS+ and NG+ against \new{BBH merger events with $e_{p, 10} > 0.01$ in} globular cluster simulations from the \texttt{Cluster Monte Carlo Catalog} \citep{Kremer:CMC:2020}. Broadly speaking, the properties of this event are more consistent with those of first-generation (1g+1g) BBHs in globulars, as opposed to those containing the products of a previous merger. This is largely because 1g+1g mergers dominate the merger rate of BBHs in clusters, though properties like those seen in \event may arise in both first- and higher-generation mergers. 
\new{More than $90\%$ of the \texttt{Cluster Monte Carlo Catalog} samples with $e_{p, 10} > 0.01$ have $0.45 < q$, $|\chi_\mathrm{eff}| < 0.12$; the posterior samples from the NG+ and IRS+ analyses of \event have $32\%$ and $64\%$ support in these regions, respectively.}
We stress that, while Fig.~\ref{fig:cmc} is indicative, a more sophisticated analysis should be performed to better quantify the statements above (see e.g. Refs.~\cite{2019MNRAS.486.1086M,2023MNRAS.525.3986M} for the necessary statistical framework to compare single GW events against large simulated populations).

\subsection{Active Galactic Nuclei}

In active galactic nuclei (AGN), stellar-mass black holes can both be captured by an accretion disk due to gas drag~\citep{Syer1999} and form through in-situ star formation \citep{Levin2003,Stone2017,EpsteinMartin2025}. 
Within the AGN disk, black holes can form binaries and be brought to merger by gaseous torques~\citep{Goldreich2002,Tagawa2020}. 
Potential locations for efficient mergers include migration traps~\citep{Bellovary2016,2024MNRAS.530.3689G,Vaccaro2024,Grishin2024} or gaps in the disk \citep{Tagawa2020}, where black holes tend to accumulate. 
The properties of these mergers are influenced by the location of accumulation points in the disk; binary-single interactions and gaseous torques become 
more efficient in the outer and inner regions, respectively. 
The existence of migration traps is also strongly dependent multiple variables, e.g.,the structure of the disk and the mass of the central supermassive black hole \citep{Pan2021_EMRI}. 

%In the inner regions, BBH mergers are rapidly driven by gas dynamical friction and GW emission \citep{Bartos2017,McKernan2018}. 
\new{In the inner regions, BBH evolution may be strongly influenced by gas dynamical friction and torques from circum-single and -binary disks~\citep{Bartos2017,McKernan2018}.}
Due to frequent hierarchical mergers, 
$|\chi_{\rm eff}|$ for mergers in the inner regions is \new{generally} $\gtrsim 0.2$ \citep{Yang2019_spin,2023PhRvD.108h3033S, Vaccaro2024},
and the mass ratio distribution is predicted to peak away from unity, $q\lesssim 0.4$ \citep{Yang2019, Vaccaro2024}; \new{the posteriors on \event from NG+ have $17\%$ of their support in this region, while the posteriors from IRS+ have $4\%$ of their support in this region}. 
These predictions assume that 
the accretion onto black holes is limited by the Eddington rate due to radiation feedback and wind mass loss \citep{Blandford1999,Pan2021}, 
and spin-up is less efficient 
compared to scenarios where all captured gas accretes onto and spins the black hole up (see Ref.~\cite{Tagawa2022} for issues in the latter case). 
\new{When gas hardening is allowed, the $\chi_\mathrm{eff}$ distribution is drastically shifted from symmetrical about $\chi_\mathrm{eff}=0$ to $\chi_\mathrm{eff} \gtrsim 0.1$, and the mass ratio distribution skews heavily towards $q<0.1$~\cite{Vaccaro2024}; the posteriors on \event from NG+ and IRS+  have $0\%$ support in this region.}

\new{The eccentricity at binary formation can be as large as $e\gtrsim 0.9$ \citep{DeLaurentiis2023,LiJiaru2023,Whitehead2024}, 
and gas torques may maintain high eccentricities even as the separation shrinks, particularly if the orbit is retrograde at formation~\cite{Roedig2011, Zrake2021,DOrazio2021,Siwek:CBDs:2023, Franchini2024, Calcino2024, Dittmann2025} as may be common in AGN disks \cite{LiJiaru2023, Whitehead2024}.}
%at a non-zero attractor value that closely depends on the mass ratio of the binary even as the separation shrinks ($e\sim 0.45$ for $q=1$)~\citep{Zrake2021,DOrazio2021,Siwek:CBDs:2023}.
Once \new{gravitational radiation begins to dominate the binary evolution}, the eccentricity decreases via GW emission, but some retrograde binaries could still merge on highly-eccentric orbits during orbital flipping~\citep{Rowan2023}.
%these predictions are not compatible with properties of \event. 

Meanwhile, more than $97\%$ of mergers in AGN are predicted to occur in the outer regions of the disk \citep{Tagawa2020}.
For binaries in these outer regions, binary-single interactions are predicted to occur frequently before GW emissions drive mergers \citep{Tagawa2020,Xue2025}. 
\new{If} the angular momentum of binaries during binary-single interactions \new{is randomized}, the distribution of $\chi_{\rm eff}$ directly after an interaction is expected to be symmetric around $\sim 0$ \citep{Tagawa2020_spin,2017Natur.548..426F}. However, gas torques can re-align the spins, so higher-eccentricity sources may be more likely to have spin tilts drawn from a random distribution than non-eccentric sources in AGN, because these have more recently been through a randomising interaction \citep{Samsing2022}, similarly to the case for field triples. As in dense clusters, close to equal mass ratios are favoured in the outer regions of AGN disks, though $q\simeq 0.5$ is possible through hierarchical mergers \citep{Tagawa2021_gap}.

It is worth noting that many input parameters in the AGN disk channel have not been well constrained or explored, resulting in less robust predictions compared to cluster and triple models. 
\new{It is therefore not so meaningful to compare inferred parameters for \event to the distributions predicted from different AGN outer-disk models, since the variation is so high (compare, for example, the predictions of different models demonstrated in Ref.~\cite{Gayathri2021}).}
\new{The main distinguishing factor for mergers in AGN disks is likely to be an overabundance of highly-eccentric BBH mergers.}

The fraction of mergers with an almost-parabolic orbit (i.e., with very high eccentricity) is significant for BBH mergers in the outer regions of the AGN disk. 
For four representative models (M1, M2, M4, and M12  from Ref.~\cite{Tagawa2021_ecc}), among mergers with `detectable' $e_{p,10}\geq 0.03$ \citep{Lower2018}, 
the fractions of mergers with $e_{p,10}\geq 0.1$, $0.4$, and $0.9$ are $\sim 50$--$90\%$, $\sim 20$--$60\%$, and $\sim 20$--$40\%$, respectively. 
The uncertainties mostly originate from the geometry of binary-single interactions \citep{Tagawa2021_ecc,Samsing2022}, which we explore in section \ref{sec:geometry}. 
These eccentricity distributions are significantly different from those of other models, e.g. Refs.~\cite{Antonini2014,Rodriguez2018}: in models with binary-single interactions restricted to the plane, eccentric mergers are the norm rather than the exception. 
If this is the case---i.e., most mergers produced by AGN have detectable eccentricity---and \event originated from an AGN, the overall contribution from AGN to the merger rate would be extremely low, as we have only detected a sparse handful of events that even show tentative evidence for eccentricity \citep[e.g.,][]{2022:Romero-Shaw:GWTC-3-ecc,2024:Gupte:GWTC-3-ecc,Lluc:2025:eccentricity}. 
% In most models of Ref.~\cite{Tagawa2021_ecc}, though, highly-eccentric mergers constitute a relatively small fraction of the overall rate. 
For mergers with detectable eccentricity in the models of Ref.~\cite{Tagawa2021_ecc}, there are more than $\sim 0.7$--$4$ times as many mergers with $e_{p,10}$ in the range $0.6$--$1.0$ than with $e_{p,10}\sim 0.03$--$0.6$, implying that for every \event-like \new{moderate-eccentricity} merger there should be a similar number of more highly-eccentric events, \new{which may be more easily missed by traditional searches with quasi-circular waveform templates \citep[e.g.,][]{Zevin2021, 2025PhRvD.111l3032B}}.

% Comparing predictions from models M1, M2, M4 and M12 in ~\cite{Tagawa2021_ecc} to \new{roughly the $50\%$ credible} inferred parameters of \event \new{from both IRS+ and NG+}, the fraction of mergers with $-0.2 \lesssim \chi_{\rm eff} \lesssim 0.4$, $0.3 \lesssim q \lesssim 0.7$ and $0.3 \lesssim e_{\rm p, 10} \lesssim 0.5$ is $\sim 0.2$--$2~\%$ for an AGN lifetime of $3$--$10~{\rm Myr}$ (see Ref.~\cite{Gayathri2021} for the mass, mass ratio, and spin distributions). This small fraction could be enhanced by increasing the lifetime of the AGN disk. 

\section{Importance of Scattering Environment Geometry}
\label{sec:geometry}

The estimated eccentricity of \event\ can be considered in relation to general properties of the
underlying formation environment. To illustrate this, we demonstrate the difference between
the outcomes of scattering interactions for objects interacting in a 2D planar geometry
(the extremal case for an AGN disk-like environment) versus a 3D
spherical geometry (a globular cluster-like environment). \new{In reality, the interaction geometry for binaries confined within an AGN disk is unlikely to be completely planar \cite[e.g.,][]{2020MNRAS.499.2608F, 2023MNRAS.522.5393N, 2024MNRAS.528.4958W}, and gas turbulence may further misalign binary orbits \cite{Whitehead2024}. We focus here on the extremal cases of 2D and 3D geometries, but note that realistic AGN properties likely lead to something in between the two.}

The probability that a population of isotropically incoming BHs
encounter another BH with a pericenter distance $r'$ smaller than $r$ is given by
\begin{equation}
    P(r'<r) \propto \sqrt{r},
\end{equation}
\begin{equation}
    P(r'<r) \propto r,
\end{equation}
for the 2D and 3D cases, respectively \cite{Samsing2022}. Mergers occurring due to scattering interactions in 2D geometries, therefore, tend to occur with smaller initial pericenter compared to those in 3D geometries. Hence, as we now describe, 2D geometries tend to lead to higher-eccentricity outcomes. 

For BHs that lose enough energy through GWs after their first pericenter passage with another BH to form a merging binary, one can relate the distance at which that initial pericenter passage occurred, $r$, to the corresponding eccentricity, $e_f$, the binary will have after evolving to GW frequency $f$ \citep{Wen2003, 2018ApJ...855..124S}. In the case where we specify the binary properties at the GW peak frequency $f_p \approx \pi^{-1} \sqrt{GM/r^3}$ (see equation \ref{eq:peak}), 
one finds the following relation between the initial pericenter distance, $r_p$ (the $p$ subscript here denotes that the quantity is computed using the peak frequency), $e_p$, 
and $f_p$:
\begin{equation}
r_p \approx C \frac{(1+e_p)}{e_p^{12/19}} \left[ \frac{425}{304} \left(1 + \frac{121}{304}e_p^2 \right)^{-1} \right]^\frac{870}{2299} f_p^{-\frac{2}{3}} ,
\label{eq:r_EM}
\end{equation}
where $C = \frac{1}{2}(GM/{\pi}^2)^{1/3}$.

If we instead wish to use as a reference the 22-mode GW frequency defined as $f_{22} = 2/T = \pi^{-1} \sqrt{GM/a^3}$,
 using $e_{22}$, the eccentricity extracted at $f_{22}$, the expression for the corresponding initial pericenter distance $r_{22}$ is
\begin{equation}
r_{22} \approx C \frac{(1-e_{22}^2)}{e_{22}^{12/19}} \left[ \frac{425}{304} \left(1 + \frac{121}{304}e_{22}^2 \right)^{-1} \right]^\frac{870}{2299} f_{22}^{-\frac{2}{3}}.
\label{eq:r_EM2}
\end{equation}

We stress that both $r_p$ and $r_{22}$ are the initial periastron distance, computed using different reference frequencies.
 With these relations, one can now produce a distribution of values for the initial pericenter distance $r$ to a distribution of values for $e_f$ at a given GW frequency $f$ by either definition.

\begin{figure}
    \centering
    \includegraphics[width=0.5\textwidth]{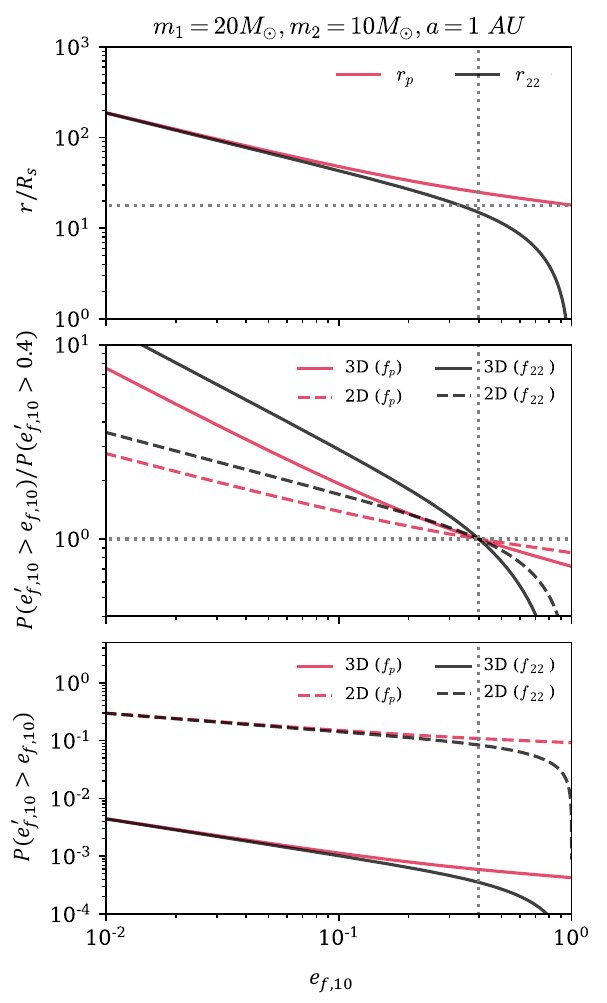}
    \caption{\label{fig:scattering} \textit{Top:} The results of Eqs. \ref{eq:r_EM} (red) and \ref{eq:r_EM2} (grey) for $r$ as a function of $e_f$, with $r$ in units of the Schwarzschild radius $R_s$ of a BH with mass $M = m_1 + m_2$. The horizontal dotted line shows the value of $r_p$ for which the peak frequency $f_p=10$~Hz. \textit{Middle:} The probability of detecting an eccentricity larger than $e_{f,10}$, normalised such that $P(e_{f,10}' > 0.4) = 1$, due to scattering interactions in 2D (dashed) and 3D (solid) geometries. The steeper decline in the 3D case demonstrates the preference for lower-eccentricity outcomes relative to the 2D case. \textit{Bottom:} The absolute probability of detecting an eccentricity larger than $e_{f,10}$. In all panels the vertical dotted line shows $e_{f,10}=0.4$, the approximate value found by NG+ for \event.}
\end{figure}

Figure \ref{fig:scattering} shows results for 3D and 2D scattering environments
with $m_1 = 20$~M$_\odot$, $m_2 = 10$~M$_\odot$ (so $q=0.5$ and $M=30$~M$_\odot$, \new{consistent with parameters inferred for} \event) at $f=10$~Hz.
The top plot shows $r_p$ and $r_{22}$, the middle plot shows $P(e_f'>e_f)$ normalized to $P(e_f'>0.4)$, and the bottom plot shows the absolute $P(e_f'>e_f)$ for the specific binary-single interaction setup. As seen in the top plot, $r_p$ and $r_{22}$ asymptote to the same values for low $e_f$, and deviate as expected when $e_f$ approaches $1$. 
Interestingly, for our considered \event-like binary, $r_{22}$ must be smaller than the pericenter distance for which $f_p=10$~Hz (horizontal dotted line) for $e_f \gtrsim 0.3$, which implies such a source would produce a burst-like signal with $f_p > 10$~Hz before reaching $f_{22} = 10$~Hz (since $f_{22} < f_p$).
Support for the eccentricity of systems like \event with $e_{22, 10} \gtrsim 0.3$ could be increased if evidence for GW bursts was uncovered in the data preceding the chirp signal, although these are likely to be below the SNR threshold for detection for \event.

We see in the middle plot of Fig.~\ref{fig:scattering} that there is a slight difference in the scaling of probability with eccentricity between the 2D and 3D cases for $e_f \lesssim 0.2$:  $\propto e_f^{-6/19}$ and $\propto e_f^{-12/19}$, respectively. 
In both geometries, for every event similar to \event, one expects an additional $\sim$few events with detectable $e_f>0.05$ \citep{Lower2018}.
In addition, for every $\sim 2$ events similar to \event\, one may expect $\sim1$ to appear initially as a burst-like source with $f_p>10$ Hz, shown by the fact that $P(e_p' > 1)$ is non-zero and close to the amplitude of $P(e_p' > 0.4)$. 
Most systems that become bound with $f_p>10$~Hz have $f_{22}<10$~Hz, and evolve to have lower eccentricities as measured from the waveform at $10$~Hz \citep{2023:Shaikh:ecc, 2025PhRvD.111b4008B, Vijaykumar2024}. 

The bottom plot shows the absolute probability for $e_f$ as the outcome of a chaotic triple scattering, modelled as a binary with semi-major axis equal to $1$ AU being reshuffled $N=20$ times according to the eccentricity distributions
$e/\sqrt{1-e^2}$ and $2e$, in the 2D and 3D cases, respectively. This %clearly
 illustrates that mergers occurring due to scattering interactions in 2D geometries should be much more likely to be eccentric, compared to the mergers from the 3D case \citep{Samsing2022}. 

\section{Discussion \& Conclusion}
\label{sec:conclusion}

It remains to be seen if an eccentric \textit{and} precessing analysis would produce a measurement of detectable eccentricity for \event. Nonetheless, analyses with spin-precessing waveform models \citep[e.g.,][]{GWTC-3} found negligible precession, while studies with aligned-spin eccentric waveform models \citep[e.g.,][]{2022:Romero-Shaw:GWTC-3-ecc, 2024:Gupte:GWTC-3-ecc, Lluc:2025:eccentricity} have found evidence for eccentricity. 

In this paper, we have reviewed the channels that we consider most promising in the production of eccentric BBH mergers detected by LVK. We conclude, based on these discussions, that a triple or dense cluster origin is consistent with the measured properties of \event under the eccentric hypothesis\new{; there is $\gtrsim 80\%$ and $\gtrsim 30\%$ support in both sets of posteriors for regions consistent with eccentric mergers from triples and clusters, respectively}. Meanwhile, simulations of AGN imply that \event is \new{less likely} to have formed in the inner regions of an AGN disk\new{, with $0\%$-$17\%$ support in the posteriors for consistent parameters, depending on gas hardening}. \event could have formed in the outer regions of the disk\new{, although the consistency of the posterior distributions is unclear given the variation in model predictions. If \event did originate in an AGN, this} would suggest that our searches may have missed BBH mergers with larger eccentricities, or that detections of such highly-eccentric BBH mergers will occur soon. 

\new{A rigorous assessment of the preferred formation environment for \event should include a comparison of the environment-specific merger rates for \event-like BBHs. Such a calculation would require self-consistent simulations of the many alternate formation channels, a worthwhile but computationally arduous endeavour that goes beyond the scope of this paper; we postpone this for future work.}

We have shown that the ratio of highly-eccentric to moderately-eccentric mergers is sensitively dependent on disk geometry \citep[see also, e.g.,][]{Tagawa2021_ecc, Samsing2022}. If a subset of BBH mergers can be confidently attributed to an AGN origin, then their eccentricity distribution could constrain such astrophysical uncertainties. 

If \event formed via any of the aforementioned channels, this implies that a non-negligible fraction of the rest of the population also formed in similar environments.
For example, if \event formed in a globular cluster and assuming this is the only detectably-eccentric of 83 observed BBHs \citep{GWTC-3}, then using findings from Ref.~\cite{Zevin2021}, the percentage of detected events that formed in globulars is constrained to $7\%\lesssim\beta_\mathrm{GC}\lesssim100\%$.
If \event formed in a ZKL triple, we may expect that the percentage of BBHs from this channel with detectable eccentricity is three times higher \citep{Wen2003, Antognini2014, Antonini2014}, so estimate using the same procedure that $2\%\lesssim\beta_\mathrm{triple}\lesssim50\%$ (although we acknowledge that the eccentricity detectability estimates in Ref.~\cite{Zevin2021} are tied to other properties predicted for mergers in dense clusters and, as such, our assumption that the recovered fraction of the population as a function of eccentricity is the same for triples is not necessarily valid).

\new{Comparing predictions for the masses and mass ratios of BBHs produced in field triples to those produced in globular clusters in Ref.~\citep{Dorozsmai2025}, it seems that an \event-like system is consistent with either a field triple or a globular cluster origin, especially given the breadth of the posteriors. 
The merger rate of BBHs consistent with this binary's properties could be of paramount importance for identifying the most probable origin environment; however, consulting the merger rate comparison plots of Ref.~\citep{Dorozsmai2025}, rates of local-Universe mergers with $e_{10} > 0.05$ are similar in both channels}.

Meanwhile, AGN with disks that constrain scattering interactions to two dimensions are expected to produce larger numbers of highly-eccentric sources than moderately-eccentric sources like \event. If this event originated in the outer regions of such an AGN disk, signals from higher-eccentricity mergers may be lurking undetected in the data.

\new{ {\it Note added.}
While this paper was under review, we became aware of the preprint \cite{2025arXiv250722862M}, in which the authors independently identify \event as an eccentric BH merger.}

\section*{Acknowledgements}
\new{We thank our anonymous referees, whose comments improved the manuscript.} We thank the Sexten Center for Astrophysics, where this work was kickstarted.
We thank  %alph order
Matteo Boschini, 
Tristan Bruel,
Alessandra Buonanno, 
Giulia Fumagalli, 
Nicholas Loutrel,  
Arif Shaikh,
and 
Fabio Antonini
for discussions.
I.M.R.-S. acknowledges the support of the Herchel Smith fund, and the Science and Technology Facilities Council grant numbers ST/Y001990/1 and UKRI2423. 
D.G. is supported by
ERC Starting Grant No.~945155--GWmining, 
Cariplo Foundation Grant No.~2021-0555, 
MUR PRIN Grant No.~2022-Z9X4XS, 
Italian-French University (UIF/UFI) Grant No.~2025-C3-386,
MUR Grant ``Progetto Dipartimenti di Eccellenza 2023-2027'' (BiCoQ),
MSCA Fellowship No. 101064542–StochRewind, MSCA Fellowship No. 101149270–ProtoBH, MUR Young Researchers Grant No. SOE2024-0000125, and the ICSC National Research Centre funded by NextGenerationEU. 
S.R.G. is supported by a UKRI Future Leaders Fellowship (grant no. MR/Y018060/1). 
J.Samsing acknowledges support from  ERC Starting Grant No. 121817–BlackHoleMergs and  Villum Fonden grant No. 29466.
Computational work was performed at CINECA with allocations 
through INFN and Bicocca, and at NVIDIA with allocations through the Academic Grant program.

\appendix

\section{Technical differences between eccentric analyses}
\label{differences}

\subsection{Waveform models}

SEOBNRE \citep{2017:CaoHan:SEOBNRE} and SEOBNRv4EHM \citep{Ramos-Buades:2021adz}
are both waveform models of the SEOB family \citep{1999PhRvD..59h4006B}.
\new{The SEOB models use an effective one-body approach to describe the motion and gravitational radiation of a coalescing compact binary; see Ref.~\cite{DamourNagar2009} for a detailed review of this formalism. SEOBNRv4EHM is a newer waveform model than SEOBNRE and is therefore more sophisticated, featuring several improvements:} 

\new{
\begin{itemize}
    \item SEOBNRv4EHM is built on v4 of the SEOBNR models \citep{Cotesta:2018fcv} whereas SEOBNRE is built on the v1 of the SEOBNR models \cite{Taracchini:2012ig} - there are many differences between these baseline models, which can be studied in the references provided; 
    \item SEOBNRv4EHM models higher-order $(\ell, m)$ modes beyond the $(\ell=2, m=2)$ mode, while SEOBNRE does not include higher-order modes;
    \item SEOBNRv4EHM includes the effect of a variable relativistic anomaly, while SEOBNRE fixes this parameter, initialising all binaries at periapsis;
    \item SEOBNRv4EHM applies eccentric corrections to the factorized waveform modes, whereas SEOBNRE adds an eccentric perturbation to the quasi-circular modes;
    \item SEOBNRV4EHM models spin-orbit and spin-spin eccentric corrections;
    \item SEOBNRv4EHM does not modify radiation reaction force with eccentric corrections, to preserve v4 baseline quasi-circular calibration.
\end{itemize}
}

Results obtained with SEOBNRv4EHM should therefore be considered
more reliable, as the risk of biases due to missing higher-order modes or
anomaly variations are removed. \new{Caution must be taken when comparing the eccentric distributions of SEOBNRE and SEOBNRv4EHM; they use different parametrizations of the eccentricity and measure the eccentricity using different definitions of the reference frequency.} Nonetheless, the similarities between the
results of both analyses for \event lead us to believe that these potential
systematics do not have a large impact for astrophysical interpretations of this event. 

\subsection{Inference strategies}

Reference~\cite{2019:Romero-Shaw:GWTC-1-ecc} and subsequent papers from the same group used likelihood reweighting, and in one case parallel Bilby \citep{2020:Romero-Shaw:GW190521}, to circumvent the inefficiencies of the SEOBNRE model. Results obtained using likelihood reweighting have an inherent drawback: in order to achieve high enough reweighting efficiency and sample the eccentric posterior appropriately, this posterior must either (i) overlap substantially with the quasi-circular posterior, or (ii) have orders of magnitude more initial samples than one intends to end up with in the final posterior. In the case of GW190521, which had poor sampling efficiency, a large enough number of samples were obtained to achieve a well-sampled eccentric posterior; this was confirmed with a very computationally costly parallel Bilby run, which returned a compatible posterior distribution~\citep{2022:Romero-Shaw:GWTC-3-ecc}. \event had a reweighting efficiency of $\sim0.1\%$, indicating that its eccentric posterior overlapped slightly with the quasi-circular posterior, and a substantial number of initial samples were needed to obtain a reliable eccentric posterior.  

While still not sufficiently efficient for standard Bayesian inference on a large number of events, SEOBNRv4EHM is nonetheless cheaper to run than SEOBNRE. Indeed, SEOBNRv4EHM has been successfully used for inference on both real and simulated data using both parallel Bilby and DINGO \citep{Ramos-Buades:2023yhy,2024:Gupte:GWTC-3-ecc}. With DINGO, one trains a neural network to directly learn the mapping between the GW strain and the posterior 
distribution, enabling fast inference for observed data. To ensure accuracy and agreement with traditional sampling, the DINGO posterior is importance-sampled to the standard likelihood times prior in post-processing; this procedure also gives the Bayesian evidence \citep{Dax:2022pxd}. 
%\sg{Updated to give more details and mention evidence.}

Priors and other inference settings differ between the two studies  considered here \citep{2022:Romero-Shaw:GWTC-3-ecc,2024:Gupte:GWTC-3-ecc}; we refer readers to those papers for full details. The results compared in this paper have both been obtained with log-uniform priors on $e_{10}$, although NG+ had a higher upper limit of $0.5$ in comparison to the upper limit of $0.2$ used by IRS+. The consistent results obtained by these two studies, despite their different inference methodologies and settings, further supports the hypothesis that \event is a signal from an eccentric merger.

\section{Other eccentric BBH candidates}
\label{other}

%\dg{I moved this subsection to an appendix. This discussion on the other events is necessary but kind of lengthy; I think it's distracting in the main paper. }  
We briefly discussed the other events that were flagged by either IRS+ or NG+ as eccentric BBH candidates. In the following we refer to three papers by the IRS+ authors: Ref.~\citep{2020:Romero-Shaw:GW190521} for an eccentric analysis of GW190521 only; Ref.~\citep{2021:Romero-Shaw:GWTC-2-ecc} for an eccentric analysis of events in GWTC-2; and Ref.~\citep{2022:Romero-Shaw:GWTC-3-ecc} for events in GWTC-3, including \event, which we have previously referred to, and continue to refer to here, as `IRS+'.

\begin{itemize}
\item {GW190521} originated from a high-mass binary, with source-frame total mass 
$M\simeq 150$~M$_\odot$~\citep{GW190521-disco}, and its signal is dominated by its merger and ringdown. The merger frequency is $\simeq60$~Hz, and there is only $\sim1$ orbital cycle ($\sim2$ GW cycles) in-band prior to the merger. This makes it challenging to analyze for eccentricity: like other waveform models, SEOBNRE and SEOBNRv4EHM assume the system has circularized by merger, and so their merger and ringdown portions are identical to those of a non-eccentric system. The eccentricity posteriors recovered for GW190521 are very different between the two approaches: that of Ref.~\cite{2020:Romero-Shaw:GW190521} has a strong peak above $e_{10}\gtrsim0.1$, while that of NG+ predominantly returns the prior. The non-eccentric spin-precessing analysis of the LVK Collaboration~\cite{GW190521-disco} found GW190521 to show moderate evidence for spin precession; as the orientation of the orbit was close to face-on, distinguishing spin precession from eccentricity becomes even less likely~\citep{2023:Romero-Shaw:Ecc-or-precc}. The differences in the recovered posteriors for GW190521 between the two studies are unlikely to arise solely from the neglect of higher-order modes in Ref.~\citep{2020:Romero-Shaw:GW190521}, as NG+ and Ref.~\cite{Ramos-Buades:2023yhy} both analyzed GW190521 without higher modes (using SEOBNRv4E) and did not find evidence for eccentricity. More likely, this difference might be due to the explicit modeling of the mean anomaly in SEOBNRv4EHM. In such a short signal, fixing this parameter at periastron, as is done in SEOBNRE, may have enough of an effect to yield spurious measurements of eccentricity. There are hints of this effect in Ref.~\cite{Ramos-Buades:2023yhy}, which shows some correlation between mean anomaly and eccentricity measurements.

%\irs{can / should we test this further?}  

\item {GW190620} was first reported with a non-eccentric spin-precessing analysis to originate from a relatively high-mass binary, $M \simeq 90$~M$_\odot$, with a positive $\chi_\mathrm{eff} = 0.33^{+0.22}_{-0.25}$ constrained away from $0$ \cite{GWTC-2}. 
The eccentric analysis of Ref.~\cite{2021:Romero-Shaw:GWTC-2-ecc} prefers lower spins closer to $\chi_\mathrm{eff}=0$. The shapes of the posteriors obtained in Ref.~\cite{2021:Romero-Shaw:GWTC-2-ecc} and NG+ are similar for GW190620; however, the Bayes factor for the eccentric hypothesis is considered too low in the latter study for this event to be considered a significant candidate.

\item {GW190701} was found by NG+ to have support for eccentricity, while Ref.~\cite{2021:Romero-Shaw:GWTC-2-ecc} did not. We speculate, based on the eccentricity posterior in Fig.~7 of NG+, that this may be due to the sharp increase in posterior support for eccentricities above $e_{\mathrm{GW, 10}} \gtrsim 0.4$, which lies well outside the waveform-enforced prior upper bound of $e_{10}=0.2$ used \cite{2021:Romero-Shaw:GWTC-2-ecc}. It is likely that the analysis of Ref.~\cite{2021:Romero-Shaw:GWTC-2-ecc} simply missed this peak due to the restricted prior, and saw no tell-tale railing of the posterior at the upper prior bound due to the relative flatness of the eccentricity posterior within the prior-supported range ($e_{10} \lesssim 0.2$).

\item {GW191109} is another high-mass event, $M \simeq 112$~M$_\odot$, and is also dominated by its merger and ringdown. GW191109 has a tentative measurement of negative effective spin  $\chi_\mathrm{eff} =-0.29^{+0.42}_{-0.31}$ with a non-eccentric spin-precessing model~\cite{GWTC-3}. Once more, the eccentric analysis of IRS+ prefers effective spins more consistent with $0$. Like GW190620, the eccentricity posterior recovered by NG+ for this event is qualitatively very similar to that recovered by IRS+, but with a Bayes factor below their significance threshold.

\item {GW200129} was found by NG+ to have $e_\mathrm{10} \approx 0.22$--$0.35$ under several different analyses with different eccentricity priors and glitch mitigation strategies, $M = 69.4^{+4.2}_{-3.1}$~M$_\odot$, and a small effective spin $\chi_\mathrm{eff} = 0.02^{+0.12}_{-0.11}$, assuming no spin precession. The non-eccentric spin-precessing analysis of the LVK Collaboration found $M = 65.0^{+12.6}_{-8.2}$ \citep{GWTC-3}. Some follow-up work \citep{2022:Hannam:GW200129,2022PhRvL.128s1102V} found this event to show significant evidence for spin precession; the signal is overlaid by a glitch, whose mitigation can significantly reduce the evidence for spin precession depending on the adopted technique \citep{2022:Payne:glitch, 2024:Macas:glitch}. However, in the case of eccentricity, irrespective of the glitch mitigation strategy, the eccentric aligned-spin hypothesis is preferred over the quasi-circular precessing hypothesis \cite{2024:Gupte:GWTC-3-ecc,Lluc:2025:eccentricity}. Fig.~12 by IRS+ shows that the samples for this event are skewed towards $e_{10} \gtrsim 0.1$, with the large weights causing this posterior to be undersampled occurring at these high eccentricities. This points toward some evidence for eccentricity in the data, but also a lack of overlap between the eccentric and non-eccentric posteriors that needs to be further explored.
\end{itemize}

%%%%%%%%%%%%%%%%%%%%%%%%%%%%%%%%%%%%%%%%%%%%%%%%%%%%%%%%%%%%%%%%%%%%
%%%%%%%%%%%%%%%%%%%%%%%%%%%%%%%%%%%%%%%%%%%%%%%%%%%%%%%%%%%%%%%%%%%%
\bibliography{gw20020822}

\begin{thebibliography}{134}%
\makeatletter
\providecommand \@ifxundefined [1]{%
 \@ifx{#1\undefined}
}%
\providecommand \@ifnum [1]{%
 \ifnum #1\expandafter \@firstoftwo
 \else \expandafter \@secondoftwo
 \fi
}%
\providecommand \@ifx [1]{%
 \ifx #1\expandafter \@firstoftwo
 \else \expandafter \@secondoftwo
 \fi
}%
\providecommand \natexlab [1]{#1}%
\providecommand \enquote  [1]{``#1''}%
\providecommand \bibnamefont  [1]{#1}%
\providecommand \bibfnamefont [1]{#1}%
\providecommand \citenamefont [1]{#1}%
\providecommand \href@noop [0]{\@secondoftwo}%
\providecommand \href [0]{\begingroup \@sanitize@url \@href}%
\providecommand \@href[1]{\@@startlink{#1}\@@href}%
\providecommand \@@href[1]{\endgroup#1\@@endlink}%
\providecommand \@sanitize@url [0]{\catcode `\\12\catcode `\$12\catcode `\&12\catcode `\#12\catcode `\^12\catcode `\_12\catcode `\%12\relax}%
\providecommand \@@startlink[1]{}%
\providecommand \@@endlink[0]{}%
\providecommand \url  [0]{\begingroup\@sanitize@url \@url }%
\providecommand \@url [1]{\endgroup\@href {#1}{\urlprefix }}%
\providecommand \urlprefix  [0]{URL }%
\providecommand \Eprint [0]{\href }%
\providecommand \doibase [0]{https://doi.org/}%
\providecommand \selectlanguage [0]{\@gobble}%
\providecommand \bibinfo  [0]{\@secondoftwo}%
\providecommand \bibfield  [0]{\@secondoftwo}%
\providecommand \translation [1]{[#1]}%
\providecommand \BibitemOpen [0]{}%
\providecommand \bibitemStop [0]{}%
\providecommand \bibitemNoStop [0]{.\EOS\space}%
\providecommand \EOS [0]{\spacefactor3000\relax}%
\providecommand \BibitemShut  [1]{\csname bibitem#1\endcsname}%
\let\auto@bib@innerbib\@empty
%</preamble>
\bibitem [{\citenamefont {{Lower}}\ \emph {et~al.}(2018)\citenamefont {{Lower}}, \citenamefont {{Thrane}}, \citenamefont {{Lasky}},\ and\ \citenamefont {{Smith}}}]{Lower2018}%
  \BibitemOpen
  \bibfield  {author} {\bibinfo {author} {\bibfnamefont {M.~E.}\ \bibnamefont {{Lower}}}, \bibinfo {author} {\bibfnamefont {E.}~\bibnamefont {{Thrane}}}, \bibinfo {author} {\bibfnamefont {P.~D.}\ \bibnamefont {{Lasky}}},\ and\ \bibinfo {author} {\bibfnamefont {R.}~\bibnamefont {{Smith}}},\ }\href {https://doi.org/10.1103/PhysRevD.98.083028} {\bibfield  {journal} {\bibinfo  {journal} {Phys. Rev. D}\ }\textbf {\bibinfo {volume} {98}},\ \bibinfo {eid} {083028} (\bibinfo {year} {2018})},\ \Eprint {https://arxiv.org/abs/1806.05350} {arXiv:1806.05350 [astro-ph.HE]} \BibitemShut {NoStop}%
\bibitem [{\citenamefont {{Zevin}}\ \emph {et~al.}(2021)\citenamefont {{Zevin}}, \citenamefont {{Romero-Shaw}}, \citenamefont {{Kremer}}, \citenamefont {{Thrane}},\ and\ \citenamefont {{Lasky}}}]{Zevin2021}%
  \BibitemOpen
  \bibfield  {author} {\bibinfo {author} {\bibfnamefont {M.}~\bibnamefont {{Zevin}}}, \bibinfo {author} {\bibfnamefont {I.~M.}\ \bibnamefont {{Romero-Shaw}}}, \bibinfo {author} {\bibfnamefont {K.}~\bibnamefont {{Kremer}}}, \bibinfo {author} {\bibfnamefont {E.}~\bibnamefont {{Thrane}}},\ and\ \bibinfo {author} {\bibfnamefont {P.~D.}\ \bibnamefont {{Lasky}}},\ }\href {https://doi.org/10.3847/2041-8213/ac32dc} {\bibfield  {journal} {\bibinfo  {journal} {Astrophys. J. Lett.}\ }\textbf {\bibinfo {volume} {921}},\ \bibinfo {eid} {L43} (\bibinfo {year} {2021})},\ \Eprint {https://arxiv.org/abs/2106.09042} {arXiv:2106.09042 [astro-ph.HE]} \BibitemShut {NoStop}%
\bibitem [{\citenamefont {Peters}(1964)}]{Peters1964}%
  \BibitemOpen
  \bibfield  {author} {\bibinfo {author} {\bibfnamefont {P.~C.}\ \bibnamefont {Peters}},\ }\href {https://doi.org/10.1103/PhysRev.136.B1224} {\bibfield  {journal} {\bibinfo  {journal} {Phys. Rev.}\ }\textbf {\bibinfo {volume} {136}},\ \bibinfo {pages} {1224} (\bibinfo {year} {1964})}\BibitemShut {NoStop}%
\bibitem [{\citenamefont {{Fumagalli}}\ \emph {et~al.}(2024)\citenamefont {{Fumagalli}}, \citenamefont {{Romero-Shaw}}, \citenamefont {{Gerosa}}, \citenamefont {{De Renzis}}, \citenamefont {{Kritos}},\ and\ \citenamefont {{Olejak}}}]{2024PhRvD.110f3012F}%
  \BibitemOpen
  \bibfield  {author} {\bibinfo {author} {\bibfnamefont {G.}~\bibnamefont {{Fumagalli}}}, \bibinfo {author} {\bibfnamefont {I.}~\bibnamefont {{Romero-Shaw}}}, \bibinfo {author} {\bibfnamefont {D.}~\bibnamefont {{Gerosa}}}, \bibinfo {author} {\bibfnamefont {V.}~\bibnamefont {{De Renzis}}}, \bibinfo {author} {\bibfnamefont {K.}~\bibnamefont {{Kritos}}},\ and\ \bibinfo {author} {\bibfnamefont {A.}~\bibnamefont {{Olejak}}},\ }\href {https://doi.org/10.1103/PhysRevD.110.063012} {\bibfield  {journal} {\bibinfo  {journal} {Phys. Rev. D}\ }\textbf {\bibinfo {volume} {110}},\ \bibinfo {eid} {063012} (\bibinfo {year} {2024})},\ \Eprint {https://arxiv.org/abs/2405.14945} {arXiv:2405.14945 [astro-ph.HE]} \BibitemShut {NoStop}%
\bibitem [{\citenamefont {{Romero-Shaw}}\ \emph {et~al.}(2020)\citenamefont {{Romero-Shaw}}, \citenamefont {{Lasky}}, \citenamefont {{Thrane}},\ and\ \citenamefont {{Calder{\'o}n Bustillo}}}]{2020:Romero-Shaw:GW190521}%
  \BibitemOpen
  \bibfield  {author} {\bibinfo {author} {\bibfnamefont {I.}~\bibnamefont {{Romero-Shaw}}}, \bibinfo {author} {\bibfnamefont {P.~D.}\ \bibnamefont {{Lasky}}}, \bibinfo {author} {\bibfnamefont {E.}~\bibnamefont {{Thrane}}},\ and\ \bibinfo {author} {\bibfnamefont {J.}~\bibnamefont {{Calder{\'o}n Bustillo}}},\ }\href {https://doi.org/10.3847/2041-8213/abbe26} {\bibfield  {journal} {\bibinfo  {journal} {Astrophys. J. Lett.}\ }\textbf {\bibinfo {volume} {903}},\ \bibinfo {eid} {L5} (\bibinfo {year} {2020})},\ \Eprint {https://arxiv.org/abs/2009.04771} {arXiv:2009.04771 [astro-ph.HE]} \BibitemShut {NoStop}%
\bibitem [{\citenamefont {{Gayathri}}\ \emph {et~al.}(2022)\citenamefont {{Gayathri}}, \citenamefont {{Healy}}, \citenamefont {{Lange}}, \citenamefont {{O'Brien}}, \citenamefont {{Szczepanczyk}}, \citenamefont {{Bartos}}, \citenamefont {{Campanelli}}, \citenamefont {{Klimenko}}, \citenamefont {{Lousto}},\ and\ \citenamefont {{O'Shaughnessy}}}]{2020:Gayathri:GW190521}%
  \BibitemOpen
  \bibfield  {author} {\bibinfo {author} {\bibfnamefont {V.}~\bibnamefont {{Gayathri}}}, \bibinfo {author} {\bibfnamefont {J.}~\bibnamefont {{Healy}}}, \bibinfo {author} {\bibfnamefont {J.}~\bibnamefont {{Lange}}}, \bibinfo {author} {\bibfnamefont {B.}~\bibnamefont {{O'Brien}}}, \bibinfo {author} {\bibfnamefont {M.}~\bibnamefont {{Szczepanczyk}}}, \bibinfo {author} {\bibfnamefont {I.}~\bibnamefont {{Bartos}}}, \bibinfo {author} {\bibfnamefont {M.}~\bibnamefont {{Campanelli}}}, \bibinfo {author} {\bibfnamefont {S.}~\bibnamefont {{Klimenko}}}, \bibinfo {author} {\bibfnamefont {C.}~\bibnamefont {{Lousto}}},\ and\ \bibinfo {author} {\bibfnamefont {R.}~\bibnamefont {{O'Shaughnessy}}},\ }\href {https://doi.org/10.1038/s41550-021-01568-w} {\bibfield  {journal} {\bibinfo  {journal} {Nat. Astron.}\ }\textbf {\bibinfo {volume} {6}},\ \bibinfo {pages} {pages344} (\bibinfo {year} {2022})},\ \Eprint {https://arxiv.org/abs/2009.05461} {arXiv:2009.05461 [astro-ph.HE]} \BibitemShut {NoStop}%
\bibitem [{\citenamefont {{Gamba}}\ \emph {et~al.}(2023)\citenamefont {{Gamba}}, \citenamefont {{Breschi}}, \citenamefont {{Carullo}}, \citenamefont {{Albanesi}}, \citenamefont {{Rettegno}}, \citenamefont {{Bernuzzi}},\ and\ \citenamefont {{Nagar}}}]{2023:Gamba:GW190521}%
  \BibitemOpen
  \bibfield  {author} {\bibinfo {author} {\bibfnamefont {R.}~\bibnamefont {{Gamba}}}, \bibinfo {author} {\bibfnamefont {M.}~\bibnamefont {{Breschi}}}, \bibinfo {author} {\bibfnamefont {G.}~\bibnamefont {{Carullo}}}, \bibinfo {author} {\bibfnamefont {S.}~\bibnamefont {{Albanesi}}}, \bibinfo {author} {\bibfnamefont {P.}~\bibnamefont {{Rettegno}}}, \bibinfo {author} {\bibfnamefont {S.}~\bibnamefont {{Bernuzzi}}},\ and\ \bibinfo {author} {\bibfnamefont {A.}~\bibnamefont {{Nagar}}},\ }\href {https://doi.org/10.1038/s41550-022-01813-w} {\bibfield  {journal} {\bibinfo  {journal} {Nat. Astron.}\ }\textbf {\bibinfo {volume} {7}},\ \bibinfo {pages} {11} (\bibinfo {year} {2023})},\ \Eprint {https://arxiv.org/abs/2106.05575} {arXiv:2106.05575 [gr-qc]} \BibitemShut {NoStop}%
\bibitem [{\citenamefont {{Iglesias}}\ \emph {et~al.}(2024)\citenamefont {{Iglesias}}, \citenamefont {{Lange}}, \citenamefont {{Bartos}}, \citenamefont {{Bhaumik}}, \citenamefont {{Gamba}}, \citenamefont {{Gayathri}}, \citenamefont {{Jan}}, \citenamefont {{Nowicki}}, \citenamefont {{O'Shaughnessy}}, \citenamefont {{Shoemaker}}, \citenamefont {{Venkataramanan}},\ and\ \citenamefont {{Wagner}}}]{2024:Iglesias:Eccentric-re-analysis}%
  \BibitemOpen
  \bibfield  {author} {\bibinfo {author} {\bibfnamefont {H.~L.}\ \bibnamefont {{Iglesias}}}, \bibinfo {author} {\bibfnamefont {J.}~\bibnamefont {{Lange}}}, \bibinfo {author} {\bibfnamefont {I.}~\bibnamefont {{Bartos}}}, \bibinfo {author} {\bibfnamefont {S.}~\bibnamefont {{Bhaumik}}}, \bibinfo {author} {\bibfnamefont {R.}~\bibnamefont {{Gamba}}}, \bibinfo {author} {\bibfnamefont {V.}~\bibnamefont {{Gayathri}}}, \bibinfo {author} {\bibfnamefont {A.}~\bibnamefont {{Jan}}}, \bibinfo {author} {\bibfnamefont {R.}~\bibnamefont {{Nowicki}}}, \bibinfo {author} {\bibfnamefont {R.}~\bibnamefont {{O'Shaughnessy}}}, \bibinfo {author} {\bibfnamefont {D.~M.}\ \bibnamefont {{Shoemaker}}}, \bibinfo {author} {\bibfnamefont {R.}~\bibnamefont {{Venkataramanan}}},\ and\ \bibinfo {author} {\bibfnamefont {K.}~\bibnamefont {{Wagner}}},\ }\href {https://doi.org/10.3847/1538-4357/ad5ff6} {\bibfield  {journal} {\bibinfo  {journal} {Astrophys. J.}\ }\textbf {\bibinfo {volume} {972}},\ \bibinfo {eid} {65} (\bibinfo {year} {2024})},\
  \Eprint {https://arxiv.org/abs/2208.01766} {arXiv:2208.01766 [gr-qc]} \BibitemShut {NoStop}%
\bibitem [{\citenamefont {{Gupte}}\ \emph {et~al.}(2024)\citenamefont {{Gupte}}, \citenamefont {{Ramos-Buades}}, \citenamefont {{Buonanno}}, \citenamefont {{Gair}}, \citenamefont {{Miller}}, \citenamefont {{Dax}}, \citenamefont {{Green}}, \citenamefont {{P{\"u}rrer}}, \citenamefont {{Wildberger}}, \citenamefont {{Macke}}, \citenamefont {{Romero-Shaw}},\ and\ \citenamefont {{Sch{\"o}lkopf}}}]{2024:Gupte:GWTC-3-ecc}%
  \BibitemOpen
  \bibfield  {author} {\bibinfo {author} {\bibfnamefont {N.}~\bibnamefont {{Gupte}}}, \bibinfo {author} {\bibfnamefont {A.}~\bibnamefont {{Ramos-Buades}}}, \bibinfo {author} {\bibfnamefont {A.}~\bibnamefont {{Buonanno}}}, \bibinfo {author} {\bibfnamefont {J.}~\bibnamefont {{Gair}}}, \bibinfo {author} {\bibfnamefont {M.~C.}\ \bibnamefont {{Miller}}}, \bibinfo {author} {\bibfnamefont {M.}~\bibnamefont {{Dax}}}, \bibinfo {author} {\bibfnamefont {S.~R.}\ \bibnamefont {{Green}}}, \bibinfo {author} {\bibfnamefont {M.}~\bibnamefont {{P{\"u}rrer}}}, \bibinfo {author} {\bibfnamefont {J.}~\bibnamefont {{Wildberger}}}, \bibinfo {author} {\bibfnamefont {J.}~\bibnamefont {{Macke}}}, \bibinfo {author} {\bibfnamefont {I.~M.}\ \bibnamefont {{Romero-Shaw}}},\ and\ \bibinfo {author} {\bibfnamefont {B.}~\bibnamefont {{Sch{\"o}lkopf}}},\ }\href@noop {} {\bibfield  {journal} {\bibinfo  {journal} {{}}\ } (\bibinfo {year} {2024})},\ \Eprint {https://arxiv.org/abs/2404.14286} {arXiv:2404.14286 [gr-qc]} \BibitemShut {NoStop}%
\bibitem [{\citenamefont {{de Lluc Planas}}\ \emph {et~al.}(2025{\natexlab{a}})\citenamefont {{de Lluc Planas}}, \citenamefont {{Ramos-Buades}}, \citenamefont {{Garc{\'\i}a-Quir{\'o}s}}, \citenamefont {{Estell{\'e}s}}, \citenamefont {{Husa}},\ and\ \citenamefont {{Haney}}}]{Lluc:2025:eccentricity}%
  \BibitemOpen
  \bibfield  {author} {\bibinfo {author} {\bibfnamefont {M.}~\bibnamefont {{de Lluc Planas}}}, \bibinfo {author} {\bibfnamefont {A.}~\bibnamefont {{Ramos-Buades}}}, \bibinfo {author} {\bibfnamefont {C.}~\bibnamefont {{Garc{\'\i}a-Quir{\'o}s}}}, \bibinfo {author} {\bibfnamefont {H.}~\bibnamefont {{Estell{\'e}s}}}, \bibinfo {author} {\bibfnamefont {S.}~\bibnamefont {{Husa}}},\ and\ \bibinfo {author} {\bibfnamefont {M.}~\bibnamefont {{Haney}}},\ }\href@noop {} {\bibfield  {journal} {\bibinfo  {journal} {{}}\ } (\bibinfo {year} {2025}{\natexlab{a}})},\ \Eprint {https://arxiv.org/abs/2504.15833} {arXiv:2504.15833 [gr-qc]} \BibitemShut {NoStop}%
\bibitem [{\citenamefont {{Romero-Shaw}}\ \emph {et~al.}(2023)\citenamefont {{Romero-Shaw}}, \citenamefont {{Gerosa}},\ and\ \citenamefont {{Loutrel}}}]{2023:Romero-Shaw:Ecc-or-precc}%
  \BibitemOpen
  \bibfield  {author} {\bibinfo {author} {\bibfnamefont {I.~M.}\ \bibnamefont {{Romero-Shaw}}}, \bibinfo {author} {\bibfnamefont {D.}~\bibnamefont {{Gerosa}}},\ and\ \bibinfo {author} {\bibfnamefont {N.}~\bibnamefont {{Loutrel}}},\ }\href {https://doi.org/10.1093/mnras/stad031} {\bibfield  {journal} {\bibinfo  {journal} {Mon. Not. R. Astron. Soc.}\ }\textbf {\bibinfo {volume} {519}},\ \bibinfo {pages} {5352} (\bibinfo {year} {2023})},\ \Eprint {https://arxiv.org/abs/2211.07528} {arXiv:2211.07528 [astro-ph.HE]} \BibitemShut {NoStop}%
\bibitem [{\citenamefont {{Xu}}\ and\ \citenamefont {{Hamilton}}(2023)}]{2023:XuHamilton:Ecc-or-precc}%
  \BibitemOpen
  \bibfield  {author} {\bibinfo {author} {\bibfnamefont {Y.}~\bibnamefont {{Xu}}}\ and\ \bibinfo {author} {\bibfnamefont {E.}~\bibnamefont {{Hamilton}}},\ }\href {https://doi.org/10.1103/PhysRevD.107.103049} {\bibfield  {journal} {\bibinfo  {journal} {Phys. Rev. D}\ }\textbf {\bibinfo {volume} {107}},\ \bibinfo {eid} {103049} (\bibinfo {year} {2023})},\ \Eprint {https://arxiv.org/abs/2211.09561} {arXiv:2211.09561 [gr-qc]} \BibitemShut {NoStop}%
\bibitem [{\citenamefont {{Romero-Shaw}}\ \emph {et~al.}(2022)\citenamefont {{Romero-Shaw}}, \citenamefont {{Lasky}},\ and\ \citenamefont {{Thrane}}}]{2022:Romero-Shaw:GWTC-3-ecc}%
  \BibitemOpen
  \bibfield  {author} {\bibinfo {author} {\bibfnamefont {I.}~\bibnamefont {{Romero-Shaw}}}, \bibinfo {author} {\bibfnamefont {P.~D.}\ \bibnamefont {{Lasky}}},\ and\ \bibinfo {author} {\bibfnamefont {E.}~\bibnamefont {{Thrane}}},\ }\href {https://doi.org/10.3847/1538-4357/ac9798} {\bibfield  {journal} {\bibinfo  {journal} {Astrophys. J.}\ }\textbf {\bibinfo {volume} {940}},\ \bibinfo {eid} {171} (\bibinfo {year} {2022})},\ \Eprint {https://arxiv.org/abs/2206.14695} {arXiv:2206.14695 [astro-ph.HE]} \BibitemShut {NoStop}%
\bibitem [{\citenamefont {{Abbott}}\ \emph {et~al.}(2023)\citenamefont {{Abbott}} \emph {et~al.}}]{GWTC-3}%
  \BibitemOpen
  \bibfield  {author} {\bibinfo {author} {\bibfnamefont {R.}~\bibnamefont {{Abbott}}} \emph {et~al.},\ }\href {https://doi.org/10.1103/PhysRevX.13.041039} {\bibfield  {journal} {\bibinfo  {journal} {Phys. Rev. X}\ }\textbf {\bibinfo {volume} {13}},\ \bibinfo {eid} {041039} (\bibinfo {year} {2023})},\ \Eprint {https://arxiv.org/abs/2111.03606} {arXiv:2111.03606 [gr-qc]} \BibitemShut {NoStop}%
\bibitem [{\citenamefont {{Romero-Shaw}}\ \emph {et~al.}(2019)\citenamefont {{Romero-Shaw}}, \citenamefont {{Lasky}},\ and\ \citenamefont {{Thrane}}}]{2019:Romero-Shaw:GWTC-1-ecc}%
  \BibitemOpen
  \bibfield  {author} {\bibinfo {author} {\bibfnamefont {I.~M.}\ \bibnamefont {{Romero-Shaw}}}, \bibinfo {author} {\bibfnamefont {P.~D.}\ \bibnamefont {{Lasky}}},\ and\ \bibinfo {author} {\bibfnamefont {E.}~\bibnamefont {{Thrane}}},\ }\href {https://doi.org/10.1093/mnras/stz2996} {\bibfield  {journal} {\bibinfo  {journal} {Mon. Not. R. Astron. Soc.}\ }\textbf {\bibinfo {volume} {490}},\ \bibinfo {pages} {5210} (\bibinfo {year} {2019})},\ \Eprint {https://arxiv.org/abs/1909.05466} {arXiv:1909.05466 [astro-ph.HE]} \BibitemShut {NoStop}%
\bibitem [{\citenamefont {{Payne}}\ \emph {et~al.}(2019)\citenamefont {{Payne}}, \citenamefont {{Talbot}},\ and\ \citenamefont {{Thrane}}}]{2019:Payne:Likelihood-reweighting}%
  \BibitemOpen
  \bibfield  {author} {\bibinfo {author} {\bibfnamefont {E.}~\bibnamefont {{Payne}}}, \bibinfo {author} {\bibfnamefont {C.}~\bibnamefont {{Talbot}}},\ and\ \bibinfo {author} {\bibfnamefont {E.}~\bibnamefont {{Thrane}}},\ }\href {https://doi.org/10.1103/PhysRevD.100.123017} {\bibfield  {journal} {\bibinfo  {journal} {Phys. Rev. D}\ }\textbf {\bibinfo {volume} {100}},\ \bibinfo {eid} {123017} (\bibinfo {year} {2019})},\ \Eprint {https://arxiv.org/abs/1905.05477} {arXiv:1905.05477 [astro-ph.IM]} \BibitemShut {NoStop}%
\bibitem [{\citenamefont {{Cao}}\ and\ \citenamefont {{Han}}(2017)}]{2017:CaoHan:SEOBNRE}%
  \BibitemOpen
  \bibfield  {author} {\bibinfo {author} {\bibfnamefont {Z.}~\bibnamefont {{Cao}}}\ and\ \bibinfo {author} {\bibfnamefont {W.-B.}\ \bibnamefont {{Han}}},\ }\href {https://doi.org/10.1103/PhysRevD.96.044028} {\bibfield  {journal} {\bibinfo  {journal} {Phys. Rev. D}\ }\textbf {\bibinfo {volume} {96}},\ \bibinfo {eid} {044028} (\bibinfo {year} {2017})},\ \Eprint {https://arxiv.org/abs/1708.00166} {arXiv:1708.00166 [gr-qc]} \BibitemShut {NoStop}%
\bibitem [{\citenamefont {{Abbott}}\ \emph {et~al.}(2020)\citenamefont {{Abbott}} \emph {et~al.}}]{GW190521-disco}%
  \BibitemOpen
  \bibfield  {author} {\bibinfo {author} {\bibfnamefont {R.}~\bibnamefont {{Abbott}}} \emph {et~al.},\ }\href {https://doi.org/10.1103/PhysRevLett.125.101102} {\bibfield  {journal} {\bibinfo  {journal} {Phys. Rev. Lett.}\ }\textbf {\bibinfo {volume} {125}},\ \bibinfo {eid} {101102} (\bibinfo {year} {2020})},\ \Eprint {https://arxiv.org/abs/2009.01075} {arXiv:2009.01075 [gr-qc]} \BibitemShut {NoStop}%
\bibitem [{\citenamefont {{Abbott}}\ \emph {et~al.}(2021)\citenamefont {{Abbott}} \emph {et~al.}}]{GWTC-2}%
  \BibitemOpen
  \bibfield  {author} {\bibinfo {author} {\bibfnamefont {R.}~\bibnamefont {{Abbott}}} \emph {et~al.},\ }\href {https://doi.org/10.1103/PhysRevX.11.021053} {\bibfield  {journal} {\bibinfo  {journal} {Phys. Rev. X}\ }\textbf {\bibinfo {volume} {11}},\ \bibinfo {eid} {021053} (\bibinfo {year} {2021})},\ \Eprint {https://arxiv.org/abs/2010.14527} {arXiv:2010.14527 [gr-qc]} \BibitemShut {NoStop}%
\bibitem [{\citenamefont {Ramos-Buades}\ \emph {et~al.}(2022)\citenamefont {Ramos-Buades}, \citenamefont {Buonanno}, \citenamefont {Khalil},\ and\ \citenamefont {Ossokine}}]{Ramos-Buades:2021adz}%
  \BibitemOpen
  \bibfield  {author} {\bibinfo {author} {\bibfnamefont {A.}~\bibnamefont {Ramos-Buades}}, \bibinfo {author} {\bibfnamefont {A.}~\bibnamefont {Buonanno}}, \bibinfo {author} {\bibfnamefont {M.}~\bibnamefont {Khalil}},\ and\ \bibinfo {author} {\bibfnamefont {S.}~\bibnamefont {Ossokine}},\ }\href {https://doi.org/10.1103/PhysRevD.105.044035} {\bibfield  {journal} {\bibinfo  {journal} {Phys. Rev. D}\ }\textbf {\bibinfo {volume} {105}},\ \bibinfo {pages} {044035} (\bibinfo {year} {2022})},\ \Eprint {https://arxiv.org/abs/2112.06952} {arXiv:2112.06952 [gr-qc]} \BibitemShut {NoStop}%
\bibitem [{\citenamefont {{Dax}}\ \emph {et~al.}(2021)\citenamefont {{Dax}}, \citenamefont {{Green}}, \citenamefont {{Gair}}, \citenamefont {{Macke}}, \citenamefont {{Buonanno}},\ and\ \citenamefont {{Sch{\"o}lkopf}}}]{Dax:Dingo:2021}%
  \BibitemOpen
  \bibfield  {author} {\bibinfo {author} {\bibfnamefont {M.}~\bibnamefont {{Dax}}}, \bibinfo {author} {\bibfnamefont {S.~R.}\ \bibnamefont {{Green}}}, \bibinfo {author} {\bibfnamefont {J.}~\bibnamefont {{Gair}}}, \bibinfo {author} {\bibfnamefont {J.~H.}\ \bibnamefont {{Macke}}}, \bibinfo {author} {\bibfnamefont {A.}~\bibnamefont {{Buonanno}}},\ and\ \bibinfo {author} {\bibfnamefont {B.}~\bibnamefont {{Sch{\"o}lkopf}}},\ }\href {https://doi.org/10.1103/PhysRevLett.127.241103} {\bibfield  {journal} {\bibinfo  {journal} {Phys. Rev. Lett.}\ }\textbf {\bibinfo {volume} {127}},\ \bibinfo {eid} {241103} (\bibinfo {year} {2021})},\ \Eprint {https://arxiv.org/abs/2106.12594} {arXiv:2106.12594 [gr-qc]} \BibitemShut {NoStop}%
\bibitem [{\citenamefont {Messick}\ \emph {et~al.}(2017)\citenamefont {Messick} \emph {et~al.}}]{Messick:2016aqy}%
  \BibitemOpen
  \bibfield  {author} {\bibinfo {author} {\bibfnamefont {C.}~\bibnamefont {Messick}} \emph {et~al.},\ }\href {https://doi.org/10.1103/PhysRevD.95.042001} {\bibfield  {journal} {\bibinfo  {journal} {Phys. Rev. D}\ }\textbf {\bibinfo {volume} {95}},\ \bibinfo {pages} {042001} (\bibinfo {year} {2017})},\ \Eprint {https://arxiv.org/abs/1604.04324} {arXiv:1604.04324 [astro-ph.IM]} \BibitemShut {NoStop}%
\bibitem [{\citenamefont {Adams}\ \emph {et~al.}(2016)\citenamefont {Adams}, \citenamefont {Buskulic}, \citenamefont {Germain}, \citenamefont {Guidi}, \citenamefont {Marion}, \citenamefont {Montani}, \citenamefont {Mours}, \citenamefont {Piergiovanni},\ and\ \citenamefont {Wang}}]{Adams:2015ulm}%
  \BibitemOpen
  \bibfield  {author} {\bibinfo {author} {\bibfnamefont {T.}~\bibnamefont {Adams}}, \bibinfo {author} {\bibfnamefont {D.}~\bibnamefont {Buskulic}}, \bibinfo {author} {\bibfnamefont {V.}~\bibnamefont {Germain}}, \bibinfo {author} {\bibfnamefont {G.~M.}\ \bibnamefont {Guidi}}, \bibinfo {author} {\bibfnamefont {F.}~\bibnamefont {Marion}}, \bibinfo {author} {\bibfnamefont {M.}~\bibnamefont {Montani}}, \bibinfo {author} {\bibfnamefont {B.}~\bibnamefont {Mours}}, \bibinfo {author} {\bibfnamefont {F.}~\bibnamefont {Piergiovanni}},\ and\ \bibinfo {author} {\bibfnamefont {G.}~\bibnamefont {Wang}},\ }\href {https://doi.org/10.1088/0264-9381/33/17/175012} {\bibfield  {journal} {\bibinfo  {journal} {Class. Quantum Grav.}\ }\textbf {\bibinfo {volume} {33}},\ \bibinfo {pages} {175012} (\bibinfo {year} {2016})},\ \Eprint {https://arxiv.org/abs/1512.02864} {arXiv:1512.02864 [gr-qc]} \BibitemShut {NoStop}%
\bibitem [{\citenamefont {Usman}\ \emph {et~al.}(2016)\citenamefont {Usman} \emph {et~al.}}]{Usman:2015kfa}%
  \BibitemOpen
  \bibfield  {author} {\bibinfo {author} {\bibfnamefont {S.~A.}\ \bibnamefont {Usman}} \emph {et~al.},\ }\href {https://doi.org/10.1088/0264-9381/33/21/215004} {\bibfield  {journal} {\bibinfo  {journal} {Class. Quantum Grav.}\ }\textbf {\bibinfo {volume} {33}},\ \bibinfo {pages} {215004} (\bibinfo {year} {2016})},\ \Eprint {https://arxiv.org/abs/1508.02357} {arXiv:1508.02357 [gr-qc]} \BibitemShut {NoStop}%
\bibitem [{\citenamefont {{Khan}}\ \emph {et~al.}(2016)\citenamefont {{Khan}}, \citenamefont {{Husa}}, \citenamefont {{Hannam}}, \citenamefont {{Ohme}}, \citenamefont {{P{\"u}rrer}}, \citenamefont {{Forteza}},\ and\ \citenamefont {{Boh{\'e}}}}]{IMRPhenomD}%
  \BibitemOpen
  \bibfield  {author} {\bibinfo {author} {\bibfnamefont {S.}~\bibnamefont {{Khan}}}, \bibinfo {author} {\bibfnamefont {S.}~\bibnamefont {{Husa}}}, \bibinfo {author} {\bibfnamefont {M.}~\bibnamefont {{Hannam}}}, \bibinfo {author} {\bibfnamefont {F.}~\bibnamefont {{Ohme}}}, \bibinfo {author} {\bibfnamefont {M.}~\bibnamefont {{P{\"u}rrer}}}, \bibinfo {author} {\bibfnamefont {X.~J.}\ \bibnamefont {{Forteza}}},\ and\ \bibinfo {author} {\bibfnamefont {A.}~\bibnamefont {{Boh{\'e}}}},\ }\href {https://doi.org/10.1103/PhysRevD.93.044007} {\bibfield  {journal} {\bibinfo  {journal} {Phys. Rev. D}\ }\textbf {\bibinfo {volume} {93}},\ \bibinfo {eid} {044007} (\bibinfo {year} {2016})},\ \Eprint {https://arxiv.org/abs/1508.07253} {arXiv:1508.07253 [gr-qc]} \BibitemShut {NoStop}%
\bibitem [{\citenamefont {{Romero-Shaw}}\ \emph {et~al.}(2021)\citenamefont {{Romero-Shaw}}, \citenamefont {{Lasky}},\ and\ \citenamefont {{Thrane}}}]{2021:Romero-Shaw:GWTC-2-ecc}%
  \BibitemOpen
  \bibfield  {author} {\bibinfo {author} {\bibfnamefont {I.}~\bibnamefont {{Romero-Shaw}}}, \bibinfo {author} {\bibfnamefont {P.~D.}\ \bibnamefont {{Lasky}}},\ and\ \bibinfo {author} {\bibfnamefont {E.}~\bibnamefont {{Thrane}}},\ }\href {https://doi.org/10.3847/2041-8213/ac3138} {\bibfield  {journal} {\bibinfo  {journal} {Astrophys. J. Lett.}\ }\textbf {\bibinfo {volume} {921}},\ \bibinfo {eid} {L31} (\bibinfo {year} {2021})},\ \Eprint {https://arxiv.org/abs/2108.01284} {arXiv:2108.01284 [astro-ph.HE]} \BibitemShut {NoStop}%
\bibitem [{\citenamefont {{Knee}}\ \emph {et~al.}(2022)\citenamefont {{Knee}}, \citenamefont {{Romero-Shaw}}, \citenamefont {{Lasky}}, \citenamefont {{McIver}},\ and\ \citenamefont {{Thrane}}}]{2022ApJ...936..172K}%
  \BibitemOpen
  \bibfield  {author} {\bibinfo {author} {\bibfnamefont {A.~M.}\ \bibnamefont {{Knee}}}, \bibinfo {author} {\bibfnamefont {I.~M.}\ \bibnamefont {{Romero-Shaw}}}, \bibinfo {author} {\bibfnamefont {P.~D.}\ \bibnamefont {{Lasky}}}, \bibinfo {author} {\bibfnamefont {J.}~\bibnamefont {{McIver}}},\ and\ \bibinfo {author} {\bibfnamefont {E.}~\bibnamefont {{Thrane}}},\ }\href {https://doi.org/10.3847/1538-4357/ac8b02} {\bibfield  {journal} {\bibinfo  {journal} {Astrophys. J.}\ }\textbf {\bibinfo {volume} {936}},\ \bibinfo {eid} {172} (\bibinfo {year} {2022})},\ \Eprint {https://arxiv.org/abs/2207.14346} {arXiv:2207.14346 [gr-qc]} \BibitemShut {NoStop}%
\bibitem [{\citenamefont {{Wen}}(2003)}]{Wen2003}%
  \BibitemOpen
  \bibfield  {author} {\bibinfo {author} {\bibfnamefont {L.}~\bibnamefont {{Wen}}},\ }\href {https://doi.org/10.1086/378794} {\bibfield  {journal} {\bibinfo  {journal} {Astrophys. J.}\ }\textbf {\bibinfo {volume} {598}},\ \bibinfo {pages} {419} (\bibinfo {year} {2003})},\ \Eprint {https://arxiv.org/abs/astro-ph/0211492} {arXiv:astro-ph/0211492 [astro-ph]} \BibitemShut {NoStop}%
\bibitem [{\citenamefont {{Vijaykumar}}\ \emph {et~al.}(2024)\citenamefont {{Vijaykumar}}, \citenamefont {{Hanselman}},\ and\ \citenamefont {{Zevin}}}]{Vijaykumar2024}%
  \BibitemOpen
  \bibfield  {author} {\bibinfo {author} {\bibfnamefont {A.}~\bibnamefont {{Vijaykumar}}}, \bibinfo {author} {\bibfnamefont {A.~G.}\ \bibnamefont {{Hanselman}}},\ and\ \bibinfo {author} {\bibfnamefont {M.}~\bibnamefont {{Zevin}}},\ }\href {https://doi.org/10.3847/1538-4357/ad4455} {\bibfield  {journal} {\bibinfo  {journal} {Astrophys. J.}\ }\textbf {\bibinfo {volume} {969}},\ \bibinfo {eid} {132} (\bibinfo {year} {2024})},\ \Eprint {https://arxiv.org/abs/2402.07892} {arXiv:2402.07892 [astro-ph.HE]} \BibitemShut {NoStop}%
\bibitem [{\citenamefont {{Fumagalli}}\ \emph {et~al.}(2025)\citenamefont {{Fumagalli}}, \citenamefont {{Loutrel}}, \citenamefont {{Gerosa}},\ and\ \citenamefont {{Boschini}}}]{2025PhRvD.112b4012F}%
  \BibitemOpen
  \bibfield  {author} {\bibinfo {author} {\bibfnamefont {G.}~\bibnamefont {{Fumagalli}}}, \bibinfo {author} {\bibfnamefont {N.}~\bibnamefont {{Loutrel}}}, \bibinfo {author} {\bibfnamefont {D.}~\bibnamefont {{Gerosa}}},\ and\ \bibinfo {author} {\bibfnamefont {M.}~\bibnamefont {{Boschini}}},\ }\href {https://doi.org/10.1103/znmj-6wvt} {\bibfield  {journal} {\bibinfo  {journal} {Phys. Rev. D}\ }\textbf {\bibinfo {volume} {112}},\ \bibinfo {eid} {024012} (\bibinfo {year} {2025})},\ \Eprint {https://arxiv.org/abs/2502.06952} {arXiv:2502.06952 [gr-qc]} \BibitemShut {NoStop}%
\bibitem [{\citenamefont {{Boschini}}\ \emph {et~al.}(2025)\citenamefont {{Boschini}}, \citenamefont {{Loutrel}}, \citenamefont {{Gerosa}},\ and\ \citenamefont {{Fumagalli}}}]{2025PhRvD.111b4008B}%
  \BibitemOpen
  \bibfield  {author} {\bibinfo {author} {\bibfnamefont {M.}~\bibnamefont {{Boschini}}}, \bibinfo {author} {\bibfnamefont {N.}~\bibnamefont {{Loutrel}}}, \bibinfo {author} {\bibfnamefont {D.}~\bibnamefont {{Gerosa}}},\ and\ \bibinfo {author} {\bibfnamefont {G.}~\bibnamefont {{Fumagalli}}},\ }\href {https://doi.org/10.1103/PhysRevD.111.024008} {\bibfield  {journal} {\bibinfo  {journal} {Phys. Rev. D}\ }\textbf {\bibinfo {volume} {111}},\ \bibinfo {eid} {024008} (\bibinfo {year} {2025})},\ \Eprint {https://arxiv.org/abs/2411.00098} {arXiv:2411.00098 [gr-qc]} \BibitemShut {NoStop}%
\bibitem [{\citenamefont {{Moe}}\ and\ \citenamefont {{Di Stefano}}(2017)}]{Moe2017}%
  \BibitemOpen
  \bibfield  {author} {\bibinfo {author} {\bibfnamefont {M.}~\bibnamefont {{Moe}}}\ and\ \bibinfo {author} {\bibfnamefont {R.}~\bibnamefont {{Di Stefano}}},\ }\href {https://doi.org/10.3847/1538-4365/aa6fb6} {\bibfield  {journal} {\bibinfo  {journal} {Astrophys. J. Supp. S.}\ }\textbf {\bibinfo {volume} {230}},\ \bibinfo {eid} {15} (\bibinfo {year} {2017})},\ \Eprint {https://arxiv.org/abs/1606.05347} {arXiv:1606.05347 [astro-ph.SR]} \BibitemShut {NoStop}%
\bibitem [{\citenamefont {Zeipel}(1909)}]{Zeipel1910}%
  \BibitemOpen
  \bibfield  {author} {\bibinfo {author} {\bibfnamefont {H.~V.}\ \bibnamefont {Zeipel}},\ }\href {https://doi.org/10.1002/asna.19091832202} {\bibfield  {journal} {\bibinfo  {journal} {Astronomische Nachrichten}\ }\textbf {\bibinfo {volume} {183}},\ \bibinfo {pages} {345} (\bibinfo {year} {1909})}\BibitemShut {NoStop}%
\bibitem [{\citenamefont {{Kozai}}(1962)}]{Kozai1962}%
  \BibitemOpen
  \bibfield  {author} {\bibinfo {author} {\bibfnamefont {Y.}~\bibnamefont {{Kozai}}},\ }\href {https://doi.org/10.1086/108790} {\bibfield  {journal} {\bibinfo  {journal} {Astron. J.}\ }\textbf {\bibinfo {volume} {67}},\ \bibinfo {pages} {591} (\bibinfo {year} {1962})}\BibitemShut {NoStop}%
\bibitem [{\citenamefont {{Lidov}}(1962)}]{Lidov1962}%
  \BibitemOpen
  \bibfield  {author} {\bibinfo {author} {\bibfnamefont {M.~L.}\ \bibnamefont {{Lidov}}},\ }\href {https://doi.org/10.1016/0032-0633(62)90129-0} {\bibfield  {journal} {\bibinfo  {journal} {Planet. Space Sci.}\ }\textbf {\bibinfo {volume} {9}},\ \bibinfo {pages} {719} (\bibinfo {year} {1962})}\BibitemShut {NoStop}%
\bibitem [{\citenamefont {{Antonini}}\ \emph {et~al.}(2014)\citenamefont {{Antonini}}, \citenamefont {{Murray}},\ and\ \citenamefont {{Mikkola}}}]{Antonini2014}%
  \BibitemOpen
  \bibfield  {author} {\bibinfo {author} {\bibfnamefont {F.}~\bibnamefont {{Antonini}}}, \bibinfo {author} {\bibfnamefont {N.}~\bibnamefont {{Murray}}},\ and\ \bibinfo {author} {\bibfnamefont {S.}~\bibnamefont {{Mikkola}}},\ }\href {https://doi.org/10.1088/0004-637X/781/1/45} {\bibfield  {journal} {\bibinfo  {journal} {Astrophys. J.}\ }\textbf {\bibinfo {volume} {781}},\ \bibinfo {eid} {45} (\bibinfo {year} {2014})},\ \Eprint {https://arxiv.org/abs/1308.3674} {arXiv:1308.3674 [astro-ph.HE]} \BibitemShut {NoStop}%
\bibitem [{\citenamefont {{Antognini}}\ \emph {et~al.}(2014)\citenamefont {{Antognini}}, \citenamefont {{Shappee}}, \citenamefont {{Thompson}},\ and\ \citenamefont {{Amaro-Seoane}}}]{Antognini2014}%
  \BibitemOpen
  \bibfield  {author} {\bibinfo {author} {\bibfnamefont {J.~M.}\ \bibnamefont {{Antognini}}}, \bibinfo {author} {\bibfnamefont {B.~J.}\ \bibnamefont {{Shappee}}}, \bibinfo {author} {\bibfnamefont {T.~A.}\ \bibnamefont {{Thompson}}},\ and\ \bibinfo {author} {\bibfnamefont {P.}~\bibnamefont {{Amaro-Seoane}}},\ }\href {https://doi.org/10.1093/mnras/stu039} {\bibfield  {journal} {\bibinfo  {journal} {Mon. Not. R. Astron. Soc.}\ }\textbf {\bibinfo {volume} {439}},\ \bibinfo {pages} {1079} (\bibinfo {year} {2014})},\ \Eprint {https://arxiv.org/abs/1308.5682} {arXiv:1308.5682 [astro-ph.HE]} \BibitemShut {NoStop}%
\bibitem [{\citenamefont {{Antonini}}\ \emph {et~al.}(2016)\citenamefont {{Antonini}}, \citenamefont {{Chatterjee}}, \citenamefont {{Rodriguez}}, \citenamefont {{Morscher}}, \citenamefont {{Pattabiraman}}, \citenamefont {{Kalogera}},\ and\ \citenamefont {{Rasio}}}]{Antonini2016}%
  \BibitemOpen
  \bibfield  {author} {\bibinfo {author} {\bibfnamefont {F.}~\bibnamefont {{Antonini}}}, \bibinfo {author} {\bibfnamefont {S.}~\bibnamefont {{Chatterjee}}}, \bibinfo {author} {\bibfnamefont {C.~L.}\ \bibnamefont {{Rodriguez}}}, \bibinfo {author} {\bibfnamefont {M.}~\bibnamefont {{Morscher}}}, \bibinfo {author} {\bibfnamefont {B.}~\bibnamefont {{Pattabiraman}}}, \bibinfo {author} {\bibfnamefont {V.}~\bibnamefont {{Kalogera}}},\ and\ \bibinfo {author} {\bibfnamefont {F.~A.}\ \bibnamefont {{Rasio}}},\ }\href {https://doi.org/10.3847/0004-637X/816/2/65} {\bibfield  {journal} {\bibinfo  {journal} {Astrophys. J.}\ }\textbf {\bibinfo {volume} {816}},\ \bibinfo {eid} {65} (\bibinfo {year} {2016})},\ \Eprint {https://arxiv.org/abs/1509.05080} {arXiv:1509.05080 [astro-ph.GA]} \BibitemShut {NoStop}%
\bibitem [{\citenamefont {{Antonini}}\ \emph {et~al.}(2017)\citenamefont {{Antonini}}, \citenamefont {{Toonen}},\ and\ \citenamefont {{Hamers}}}]{Antonini2017}%
  \BibitemOpen
  \bibfield  {author} {\bibinfo {author} {\bibfnamefont {F.}~\bibnamefont {{Antonini}}}, \bibinfo {author} {\bibfnamefont {S.}~\bibnamefont {{Toonen}}},\ and\ \bibinfo {author} {\bibfnamefont {A.~S.}\ \bibnamefont {{Hamers}}},\ }\href {https://doi.org/10.3847/1538-4357/aa6f5e} {\bibfield  {journal} {\bibinfo  {journal} {Astrophys. J.}\ }\textbf {\bibinfo {volume} {841}},\ \bibinfo {eid} {77} (\bibinfo {year} {2017})},\ \Eprint {https://arxiv.org/abs/1703.06614} {arXiv:1703.06614 [astro-ph.GA]} \BibitemShut {NoStop}%
\bibitem [{\citenamefont {{Liu}}\ \emph {et~al.}(2019)\citenamefont {{Liu}}, \citenamefont {{Lai}},\ and\ \citenamefont {{Wang}}}]{Liu2019}%
  \BibitemOpen
  \bibfield  {author} {\bibinfo {author} {\bibfnamefont {B.}~\bibnamefont {{Liu}}}, \bibinfo {author} {\bibfnamefont {D.}~\bibnamefont {{Lai}}},\ and\ \bibinfo {author} {\bibfnamefont {Y.-H.}\ \bibnamefont {{Wang}}},\ }\href {https://doi.org/10.3847/1538-4357/ab2dfb} {\bibfield  {journal} {\bibinfo  {journal} {Astrophys. J.}\ }\textbf {\bibinfo {volume} {881}},\ \bibinfo {eid} {41} (\bibinfo {year} {2019})},\ \Eprint {https://arxiv.org/abs/1905.00427} {arXiv:1905.00427 [astro-ph.HE]} \BibitemShut {NoStop}%
\bibitem [{\citenamefont {{Martinez}}\ \emph {et~al.}(2020)\citenamefont {{Martinez}}, \citenamefont {{Fragione}}, \citenamefont {{Kremer}}, \citenamefont {{Chatterjee}}, \citenamefont {{Rodriguez}}, \citenamefont {{Samsing}}, \citenamefont {{Ye}}, \citenamefont {{Weatherford}}, \citenamefont {{Zevin}}, \citenamefont {{Naoz}},\ and\ \citenamefont {{Rasio}}}]{Martinez2020}%
  \BibitemOpen
  \bibfield  {author} {\bibinfo {author} {\bibfnamefont {M.~A.~S.}\ \bibnamefont {{Martinez}}}, \bibinfo {author} {\bibfnamefont {G.}~\bibnamefont {{Fragione}}}, \bibinfo {author} {\bibfnamefont {K.}~\bibnamefont {{Kremer}}}, \bibinfo {author} {\bibfnamefont {S.}~\bibnamefont {{Chatterjee}}}, \bibinfo {author} {\bibfnamefont {C.~L.}\ \bibnamefont {{Rodriguez}}}, \bibinfo {author} {\bibfnamefont {J.}~\bibnamefont {{Samsing}}}, \bibinfo {author} {\bibfnamefont {C.~S.}\ \bibnamefont {{Ye}}}, \bibinfo {author} {\bibfnamefont {N.~C.}\ \bibnamefont {{Weatherford}}}, \bibinfo {author} {\bibfnamefont {M.}~\bibnamefont {{Zevin}}}, \bibinfo {author} {\bibfnamefont {S.}~\bibnamefont {{Naoz}}},\ and\ \bibinfo {author} {\bibfnamefont {F.~A.}\ \bibnamefont {{Rasio}}},\ }\href {https://doi.org/10.3847/1538-4357/abba25} {\bibfield  {journal} {\bibinfo  {journal} {Astrophys. J.}\ }\textbf {\bibinfo {volume} {903}},\ \bibinfo {eid} {67} (\bibinfo {year} {2020})},\ \Eprint {https://arxiv.org/abs/2009.08468} {arXiv:2009.08468
  [astro-ph.GA]} \BibitemShut {NoStop}%
\bibitem [{\citenamefont {{Fragione}}\ and\ \citenamefont {{Kocsis}}(2020)}]{Fragione2020}%
  \BibitemOpen
  \bibfield  {author} {\bibinfo {author} {\bibfnamefont {G.}~\bibnamefont {{Fragione}}}\ and\ \bibinfo {author} {\bibfnamefont {B.}~\bibnamefont {{Kocsis}}},\ }\href {https://doi.org/10.1093/mnras/staa443} {\bibfield  {journal} {\bibinfo  {journal} {Mon. Not. R. Astron. Soc.}\ }\textbf {\bibinfo {volume} {493}},\ \bibinfo {pages} {3920} (\bibinfo {year} {2020})},\ \Eprint {https://arxiv.org/abs/1910.00407} {arXiv:1910.00407 [astro-ph.GA]} \BibitemShut {NoStop}%
\bibitem [{\citenamefont {{Hut}}(1981)}]{1981A&A....99..126H}%
  \BibitemOpen
  \bibfield  {author} {\bibinfo {author} {\bibfnamefont {P.}~\bibnamefont {{Hut}}},\ }\href@noop {} {\bibfield  {journal} {\bibinfo  {journal} {Astron. Astrophys.}\ }\textbf {\bibinfo {volume} {99}},\ \bibinfo {pages} {126} (\bibinfo {year} {1981})}\BibitemShut {NoStop}%
\bibitem [{\citenamefont {{Gerosa}}\ \emph {et~al.}(2018)\citenamefont {{Gerosa}}, \citenamefont {{Berti}}, \citenamefont {{O'Shaughnessy}}, \citenamefont {{Belczynski}}, \citenamefont {{Kesden}}, \citenamefont {{Wysocki}},\ and\ \citenamefont {{Gladysz}}}]{2018PhRvD..98h4036G}%
  \BibitemOpen
  \bibfield  {author} {\bibinfo {author} {\bibfnamefont {D.}~\bibnamefont {{Gerosa}}}, \bibinfo {author} {\bibfnamefont {E.}~\bibnamefont {{Berti}}}, \bibinfo {author} {\bibfnamefont {R.}~\bibnamefont {{O'Shaughnessy}}}, \bibinfo {author} {\bibfnamefont {K.}~\bibnamefont {{Belczynski}}}, \bibinfo {author} {\bibfnamefont {M.}~\bibnamefont {{Kesden}}}, \bibinfo {author} {\bibfnamefont {D.}~\bibnamefont {{Wysocki}}},\ and\ \bibinfo {author} {\bibfnamefont {W.}~\bibnamefont {{Gladysz}}},\ }\href {https://doi.org/10.1103/PhysRevD.98.084036} {\bibfield  {journal} {\bibinfo  {journal} {Phys. Rev. D}\ }\textbf {\bibinfo {volume} {98}},\ \bibinfo {eid} {084036} (\bibinfo {year} {2018})},\ \Eprint {https://arxiv.org/abs/1808.02491} {arXiv:1808.02491 [astro-ph.HE]} \BibitemShut {NoStop}%
\bibitem [{\citenamefont {{Kalogera}}(2000)}]{2000ApJ...541..319K}%
  \BibitemOpen
  \bibfield  {author} {\bibinfo {author} {\bibfnamefont {V.}~\bibnamefont {{Kalogera}}},\ }\href {https://doi.org/10.1086/309400} {\bibfield  {journal} {\bibinfo  {journal} {Astrophys. J.}\ }\textbf {\bibinfo {volume} {541}},\ \bibinfo {pages} {319} (\bibinfo {year} {2000})},\ \Eprint {https://arxiv.org/abs/astro-ph/9911417} {arXiv:astro-ph/9911417 [astro-ph]} \BibitemShut {NoStop}%
\bibitem [{\citenamefont {{Antonini}}\ \emph {et~al.}(2018)\citenamefont {{Antonini}}, \citenamefont {{Rodriguez}}, \citenamefont {{Petrovich}},\ and\ \citenamefont {{Fischer}}}]{Antonini2018}%
  \BibitemOpen
  \bibfield  {author} {\bibinfo {author} {\bibfnamefont {F.}~\bibnamefont {{Antonini}}}, \bibinfo {author} {\bibfnamefont {C.~L.}\ \bibnamefont {{Rodriguez}}}, \bibinfo {author} {\bibfnamefont {C.}~\bibnamefont {{Petrovich}}},\ and\ \bibinfo {author} {\bibfnamefont {C.~L.}\ \bibnamefont {{Fischer}}},\ }\href {https://doi.org/10.1093/mnrasl/sly126} {\bibfield  {journal} {\bibinfo  {journal} {Mon. Not. R. Astron. Soc.}\ }\textbf {\bibinfo {volume} {480}},\ \bibinfo {pages} {L58} (\bibinfo {year} {2018})},\ \Eprint {https://arxiv.org/abs/1711.07142} {arXiv:1711.07142 [astro-ph.HE]} \BibitemShut {NoStop}%
\bibitem [{\citenamefont {{Rodriguez}}\ and\ \citenamefont {{Antonini}}(2018)}]{Rodriguez2018}%
  \BibitemOpen
  \bibfield  {author} {\bibinfo {author} {\bibfnamefont {C.~L.}\ \bibnamefont {{Rodriguez}}}\ and\ \bibinfo {author} {\bibfnamefont {F.}~\bibnamefont {{Antonini}}},\ }\href {https://doi.org/10.3847/1538-4357/aacea4} {\bibfield  {journal} {\bibinfo  {journal} {Astrophys. J.}\ }\textbf {\bibinfo {volume} {863}},\ \bibinfo {eid} {7} (\bibinfo {year} {2018})},\ \Eprint {https://arxiv.org/abs/1805.08212} {arXiv:1805.08212 [astro-ph.HE]} \BibitemShut {NoStop}%
\bibitem [{\citenamefont {Stegmann}\ \emph {et~al.}(2025)\citenamefont {Stegmann}, \citenamefont {Gerosa}, \citenamefont {Romero-Shaw}, \citenamefont {Fumagalli}, \citenamefont {Tagawa},\ and\ \citenamefont {Zwick}}]{Stegmann:2025shr}%
  \BibitemOpen
  \bibfield  {author} {\bibinfo {author} {\bibfnamefont {J.}~\bibnamefont {Stegmann}}, \bibinfo {author} {\bibfnamefont {D.}~\bibnamefont {Gerosa}}, \bibinfo {author} {\bibfnamefont {I.}~\bibnamefont {Romero-Shaw}}, \bibinfo {author} {\bibfnamefont {G.}~\bibnamefont {Fumagalli}}, \bibinfo {author} {\bibfnamefont {H.}~\bibnamefont {Tagawa}},\ and\ \bibinfo {author} {\bibfnamefont {L.}~\bibnamefont {Zwick}},\ }\href@noop {} {\bibfield  {journal} {\bibinfo  {journal} {{}}\ } (\bibinfo {year} {2025})},\ \Eprint {https://arxiv.org/abs/2505.13589} {arXiv:2505.13589 [astro-ph.HE]} \BibitemShut {NoStop}%
\bibitem [{\citenamefont {{Liu}}\ and\ \citenamefont {{Lai}}(2018)}]{Liu2018}%
  \BibitemOpen
  \bibfield  {author} {\bibinfo {author} {\bibfnamefont {B.}~\bibnamefont {{Liu}}}\ and\ \bibinfo {author} {\bibfnamefont {D.}~\bibnamefont {{Lai}}},\ }\href {https://doi.org/10.3847/1538-4357/aad09f} {\bibfield  {journal} {\bibinfo  {journal} {Astrophys. J.}\ }\textbf {\bibinfo {volume} {863}},\ \bibinfo {eid} {68} (\bibinfo {year} {2018})},\ \Eprint {https://arxiv.org/abs/1805.03202} {arXiv:1805.03202 [astro-ph.HE]} \BibitemShut {NoStop}%
\bibitem [{\citenamefont {{Martinez}}\ \emph {et~al.}(2022)\citenamefont {{Martinez}}, \citenamefont {{Rodriguez}},\ and\ \citenamefont {{Fragione}}}]{Martinez2022}%
  \BibitemOpen
  \bibfield  {author} {\bibinfo {author} {\bibfnamefont {M.~A.~S.}\ \bibnamefont {{Martinez}}}, \bibinfo {author} {\bibfnamefont {C.~L.}\ \bibnamefont {{Rodriguez}}},\ and\ \bibinfo {author} {\bibfnamefont {G.}~\bibnamefont {{Fragione}}},\ }\href {https://doi.org/10.3847/1538-4357/ac8d55} {\bibfield  {journal} {\bibinfo  {journal} {Astrophys. J.}\ }\textbf {\bibinfo {volume} {937}},\ \bibinfo {eid} {78} (\bibinfo {year} {2022})},\ \Eprint {https://arxiv.org/abs/2105.01671} {arXiv:2105.01671 [astro-ph.SR]} \BibitemShut {NoStop}%
\bibitem [{\citenamefont {{Dorozsmai}}\ \emph {et~al.}(2025)\citenamefont {{Dorozsmai}}, \citenamefont {{Romero-Shaw}}, \citenamefont {{Vijaykumar}}, \citenamefont {{Toonen}}, \citenamefont {{Antonini}}, \citenamefont {{Kremer}}, \citenamefont {{Zevin}},\ and\ \citenamefont {{Grishin}}}]{Dorozsmai2025}%
  \BibitemOpen
  \bibfield  {author} {\bibinfo {author} {\bibfnamefont {A.}~\bibnamefont {{Dorozsmai}}}, \bibinfo {author} {\bibfnamefont {I.~M.}\ \bibnamefont {{Romero-Shaw}}}, \bibinfo {author} {\bibfnamefont {A.}~\bibnamefont {{Vijaykumar}}}, \bibinfo {author} {\bibfnamefont {S.}~\bibnamefont {{Toonen}}}, \bibinfo {author} {\bibfnamefont {F.}~\bibnamefont {{Antonini}}}, \bibinfo {author} {\bibfnamefont {K.}~\bibnamefont {{Kremer}}}, \bibinfo {author} {\bibfnamefont {M.}~\bibnamefont {{Zevin}}},\ and\ \bibinfo {author} {\bibfnamefont {E.}~\bibnamefont {{Grishin}}},\ }\href {https://doi.org/10.48550/arXiv.2507.23212} {\bibfield  {journal} {\bibinfo  {journal} {arXiv e-prints}\ ,\ \bibinfo {eid} {arXiv:2507.23212}} (\bibinfo {year} {2025})},\ \Eprint {https://arxiv.org/abs/2507.23212} {arXiv:2507.23212 [astro-ph.GA]} \BibitemShut {NoStop}%
\bibitem [{\citenamefont {{Morras}}\ \emph {et~al.}(2025)\citenamefont {{Morras}}, \citenamefont {{Pratten}},\ and\ \citenamefont {{Schmidt}}}]{2025arXiv250315393M}%
  \BibitemOpen
  \bibfield  {author} {\bibinfo {author} {\bibfnamefont {G.}~\bibnamefont {{Morras}}}, \bibinfo {author} {\bibfnamefont {G.}~\bibnamefont {{Pratten}}},\ and\ \bibinfo {author} {\bibfnamefont {P.}~\bibnamefont {{Schmidt}}},\ }\href@noop {} {\bibfield  {journal} {\bibinfo  {journal} {{}}\ } (\bibinfo {year} {2025})},\ \Eprint {https://arxiv.org/abs/2503.15393} {arXiv:2503.15393 [astro-ph.HE]} \BibitemShut {NoStop}%
\bibitem [{\citenamefont {{de Lluc Planas}}\ \emph {et~al.}(2025{\natexlab{b}})\citenamefont {{de Lluc Planas}}, \citenamefont {{Husa}}, \citenamefont {{Ramos-Buades}},\ and\ \citenamefont {{Valencia}}}]{2025arXiv250601760D}%
  \BibitemOpen
  \bibfield  {author} {\bibinfo {author} {\bibfnamefont {M.}~\bibnamefont {{de Lluc Planas}}}, \bibinfo {author} {\bibfnamefont {S.}~\bibnamefont {{Husa}}}, \bibinfo {author} {\bibfnamefont {A.}~\bibnamefont {{Ramos-Buades}}},\ and\ \bibinfo {author} {\bibfnamefont {J.}~\bibnamefont {{Valencia}}},\ }\href@noop {} {\bibfield  {journal} {\bibinfo  {journal} {{}}\ } (\bibinfo {year} {2025}{\natexlab{b}})},\ \Eprint {https://arxiv.org/abs/2506.01760} {arXiv:2506.01760 [astro-ph.HE]} \BibitemShut {NoStop}%
\bibitem [{\citenamefont {{Stegmann}}\ and\ \citenamefont {{Klencki}}(2025)}]{StegmannKlencki2025}%
  \BibitemOpen
  \bibfield  {author} {\bibinfo {author} {\bibfnamefont {J.}~\bibnamefont {{Stegmann}}}\ and\ \bibinfo {author} {\bibfnamefont {J.}~\bibnamefont {{Klencki}}},\ }\href@noop {} {\bibfield  {journal} {\bibinfo  {journal} {{}}\ } (\bibinfo {year} {2025})},\ \Eprint {https://arxiv.org/abs/2506.09121} {arXiv:2506.09121 [astro-ph.HE]} \BibitemShut {NoStop}%
\bibitem [{\citenamefont {{Kremer}}\ \emph {et~al.}(2020)\citenamefont {{Kremer}}, \citenamefont {{Ye}}, \citenamefont {{Rui}}, \citenamefont {{Weatherford}}, \citenamefont {{Chatterjee}}, \citenamefont {{Fragione}}, \citenamefont {{Rodriguez}}, \citenamefont {{Spera}},\ and\ \citenamefont {{Rasio}}}]{Kremer:CMC:2020}%
  \BibitemOpen
  \bibfield  {author} {\bibinfo {author} {\bibfnamefont {K.}~\bibnamefont {{Kremer}}}, \bibinfo {author} {\bibfnamefont {C.~S.}\ \bibnamefont {{Ye}}}, \bibinfo {author} {\bibfnamefont {N.~Z.}\ \bibnamefont {{Rui}}}, \bibinfo {author} {\bibfnamefont {N.~C.}\ \bibnamefont {{Weatherford}}}, \bibinfo {author} {\bibfnamefont {S.}~\bibnamefont {{Chatterjee}}}, \bibinfo {author} {\bibfnamefont {G.}~\bibnamefont {{Fragione}}}, \bibinfo {author} {\bibfnamefont {C.~L.}\ \bibnamefont {{Rodriguez}}}, \bibinfo {author} {\bibfnamefont {M.}~\bibnamefont {{Spera}}},\ and\ \bibinfo {author} {\bibfnamefont {F.~A.}\ \bibnamefont {{Rasio}}},\ }\href {https://doi.org/10.3847/1538-4365/ab7919} {\bibfield  {journal} {\bibinfo  {journal} {Astrophys. J. Supp. S.}\ }\textbf {\bibinfo {volume} {247}},\ \bibinfo {eid} {48} (\bibinfo {year} {2020})},\ \Eprint {https://arxiv.org/abs/1911.00018} {arXiv:1911.00018 [astro-ph.HE]} \BibitemShut {NoStop}%
\bibitem [{\citenamefont {{Samsing}}\ \emph {et~al.}(2014)\citenamefont {{Samsing}}, \citenamefont {{MacLeod}},\ and\ \citenamefont {{Ramirez-Ruiz}}}]{Samsing14}%
  \BibitemOpen
  \bibfield  {author} {\bibinfo {author} {\bibfnamefont {J.}~\bibnamefont {{Samsing}}}, \bibinfo {author} {\bibfnamefont {M.}~\bibnamefont {{MacLeod}}},\ and\ \bibinfo {author} {\bibfnamefont {E.}~\bibnamefont {{Ramirez-Ruiz}}},\ }\href {https://doi.org/10.1088/0004-637X/784/1/71} {\bibfield  {journal} {\bibinfo  {journal} {Astrophys. J.}\ }\textbf {\bibinfo {volume} {784}},\ \bibinfo {eid} {71} (\bibinfo {year} {2014})},\ \Eprint {https://arxiv.org/abs/1308.2964} {arXiv:1308.2964 [astro-ph.HE]} \BibitemShut {NoStop}%
\bibitem [{\citenamefont {{Rodriguez}}\ \emph {et~al.}(2016)\citenamefont {{Rodriguez}}, \citenamefont {{Chatterjee}},\ and\ \citenamefont {{Rasio}}}]{RodriguezChatterjee2016}%
  \BibitemOpen
  \bibfield  {author} {\bibinfo {author} {\bibfnamefont {C.~L.}\ \bibnamefont {{Rodriguez}}}, \bibinfo {author} {\bibfnamefont {S.}~\bibnamefont {{Chatterjee}}},\ and\ \bibinfo {author} {\bibfnamefont {F.~A.}\ \bibnamefont {{Rasio}}},\ }\href {https://doi.org/10.1103/PhysRevD.93.084029} {\bibfield  {journal} {\bibinfo  {journal} {Phys. Rev. D}\ }\textbf {\bibinfo {volume} {93}},\ \bibinfo {eid} {084029} (\bibinfo {year} {2016})},\ \Eprint {https://arxiv.org/abs/1602.02444} {arXiv:1602.02444 [astro-ph.HE]} \BibitemShut {NoStop}%
\bibitem [{\citenamefont {{Askar}}\ \emph {et~al.}(2017)\citenamefont {{Askar}}, \citenamefont {{Szkudlarek}}, \citenamefont {{Gondek-Rosi{\'n}ska}}, \citenamefont {{Giersz}},\ and\ \citenamefont {{Bulik}}}]{Askar2017}%
  \BibitemOpen
  \bibfield  {author} {\bibinfo {author} {\bibfnamefont {A.}~\bibnamefont {{Askar}}}, \bibinfo {author} {\bibfnamefont {M.}~\bibnamefont {{Szkudlarek}}}, \bibinfo {author} {\bibfnamefont {D.}~\bibnamefont {{Gondek-Rosi{\'n}ska}}}, \bibinfo {author} {\bibfnamefont {M.}~\bibnamefont {{Giersz}}},\ and\ \bibinfo {author} {\bibfnamefont {T.}~\bibnamefont {{Bulik}}},\ }\href {https://doi.org/10.1093/mnrasl/slw177} {\bibfield  {journal} {\bibinfo  {journal} {Mon. Not. R. Astron. Soc.}\ }\textbf {\bibinfo {volume} {464}},\ \bibinfo {pages} {L36} (\bibinfo {year} {2017})},\ \Eprint {https://arxiv.org/abs/1608.02520} {arXiv:1608.02520 [astro-ph.HE]} \BibitemShut {NoStop}%
\bibitem [{\citenamefont {{Samsing}}(2018)}]{2018PhRvD..97j3014S}%
  \BibitemOpen
  \bibfield  {author} {\bibinfo {author} {\bibfnamefont {J.}~\bibnamefont {{Samsing}}},\ }\href {https://doi.org/10.1103/PhysRevD.97.103014} {\bibfield  {journal} {\bibinfo  {journal} {Phys. Rev. D}\ }\textbf {\bibinfo {volume} {97}},\ \bibinfo {eid} {103014} (\bibinfo {year} {2018})},\ \Eprint {https://arxiv.org/abs/1711.07452} {arXiv:1711.07452 [astro-ph.HE]} \BibitemShut {NoStop}%
\bibitem [{\citenamefont {{Samsing}}\ and\ \citenamefont {{D'Orazio}}(2018)}]{JSDJ18}%
  \BibitemOpen
  \bibfield  {author} {\bibinfo {author} {\bibfnamefont {J.}~\bibnamefont {{Samsing}}}\ and\ \bibinfo {author} {\bibfnamefont {D.~J.}\ \bibnamefont {{D'Orazio}}},\ }\href {https://doi.org/10.1093/mnras/sty2334} {\bibfield  {journal} {\bibinfo  {journal} {Mon. Not. R. Astron. Soc.}\ }\textbf {\bibinfo {volume} {481}},\ \bibinfo {pages} {5445} (\bibinfo {year} {2018})},\ \Eprint {https://arxiv.org/abs/1804.06519} {arXiv:1804.06519 [astro-ph.HE]} \BibitemShut {NoStop}%
\bibitem [{\citenamefont {{Banerjee}}(2018)}]{Banerjee2018}%
  \BibitemOpen
  \bibfield  {author} {\bibinfo {author} {\bibfnamefont {S.}~\bibnamefont {{Banerjee}}},\ }\href {https://doi.org/10.1093/mnras/stx2347} {\bibfield  {journal} {\bibinfo  {journal} {Mon. Not. R. Astron. Soc.}\ }\textbf {\bibinfo {volume} {473}},\ \bibinfo {pages} {909} (\bibinfo {year} {2018})},\ \Eprint {https://arxiv.org/abs/1707.00922} {arXiv:1707.00922 [astro-ph.HE]} \BibitemShut {NoStop}%
\bibitem [{\citenamefont {{Banerjee}}(2021)}]{Banerjee2021}%
  \BibitemOpen
  \bibfield  {author} {\bibinfo {author} {\bibfnamefont {S.}~\bibnamefont {{Banerjee}}},\ }\href {https://doi.org/10.1093/mnras/staa2392} {\bibfield  {journal} {\bibinfo  {journal} {Mon. Not. R. Astron. Soc.}\ }\textbf {\bibinfo {volume} {500}},\ \bibinfo {pages} {3002} (\bibinfo {year} {2021})},\ \Eprint {https://arxiv.org/abs/2004.07382} {arXiv:2004.07382 [astro-ph.HE]} \BibitemShut {NoStop}%
\bibitem [{\citenamefont {{Mapelli}}\ \emph {et~al.}(2021)\citenamefont {{Mapelli}}, \citenamefont {{Dall'Amico}}, \citenamefont {{Bouffanais}}, \citenamefont {{Giacobbo}}, \citenamefont {{Arca Sedda}}, \citenamefont {{Artale}}, \citenamefont {{Ballone}}, \citenamefont {{Di Carlo}}, \citenamefont {{Iorio}}, \citenamefont {{Santoliquido}},\ and\ \citenamefont {{Torniamenti}}}]{Mapelli:2021MNRAS}%
  \BibitemOpen
  \bibfield  {author} {\bibinfo {author} {\bibfnamefont {M.}~\bibnamefont {{Mapelli}}}, \bibinfo {author} {\bibfnamefont {M.}~\bibnamefont {{Dall'Amico}}}, \bibinfo {author} {\bibfnamefont {Y.}~\bibnamefont {{Bouffanais}}}, \bibinfo {author} {\bibfnamefont {N.}~\bibnamefont {{Giacobbo}}}, \bibinfo {author} {\bibfnamefont {M.}~\bibnamefont {{Arca Sedda}}}, \bibinfo {author} {\bibfnamefont {M.~C.}\ \bibnamefont {{Artale}}}, \bibinfo {author} {\bibfnamefont {A.}~\bibnamefont {{Ballone}}}, \bibinfo {author} {\bibfnamefont {U.~N.}\ \bibnamefont {{Di Carlo}}}, \bibinfo {author} {\bibfnamefont {G.}~\bibnamefont {{Iorio}}}, \bibinfo {author} {\bibfnamefont {F.}~\bibnamefont {{Santoliquido}}},\ and\ \bibinfo {author} {\bibfnamefont {S.}~\bibnamefont {{Torniamenti}}},\ }\href {https://doi.org/10.1093/mnras/stab1334} {\bibfield  {journal} {\bibinfo  {journal} {Mon. Not. R. Astron. Soc.}\ }\textbf {\bibinfo {volume} {505}},\ \bibinfo {pages} {339} (\bibinfo {year} {2021})},\ \Eprint {https://arxiv.org/abs/2103.05016}
  {arXiv:2103.05016 [astro-ph.HE]} \BibitemShut {NoStop}%
\bibitem [{\citenamefont {{Dall'Amico}}\ \emph {et~al.}(2024)\citenamefont {{Dall'Amico}}, \citenamefont {{Mapelli}}, \citenamefont {{Torniamenti}},\ and\ \citenamefont {{Arca Sedda}}}]{2024A&A...683A.186D}%
  \BibitemOpen
  \bibfield  {author} {\bibinfo {author} {\bibfnamefont {M.}~\bibnamefont {{Dall'Amico}}}, \bibinfo {author} {\bibfnamefont {M.}~\bibnamefont {{Mapelli}}}, \bibinfo {author} {\bibfnamefont {S.}~\bibnamefont {{Torniamenti}}},\ and\ \bibinfo {author} {\bibfnamefont {M.}~\bibnamefont {{Arca Sedda}}},\ }\href {https://doi.org/10.1051/0004-6361/202348745} {\bibfield  {journal} {\bibinfo  {journal} {Astron. Astrophys.}\ }\textbf {\bibinfo {volume} {683}},\ \bibinfo {eid} {A186} (\bibinfo {year} {2024})},\ \Eprint {https://arxiv.org/abs/2303.07421} {arXiv:2303.07421 [astro-ph.HE]} \BibitemShut {NoStop}%
\bibitem [{\citenamefont {{Chattopadhyay}}\ \emph {et~al.}(2023)\citenamefont {{Chattopadhyay}}, \citenamefont {{Stegmann}}, \citenamefont {{Antonini}}, \citenamefont {{Barber}},\ and\ \citenamefont {{Romero-Shaw}}}]{2023:Chattopadhyay:NSCs}%
  \BibitemOpen
  \bibfield  {author} {\bibinfo {author} {\bibfnamefont {D.}~\bibnamefont {{Chattopadhyay}}}, \bibinfo {author} {\bibfnamefont {J.}~\bibnamefont {{Stegmann}}}, \bibinfo {author} {\bibfnamefont {F.}~\bibnamefont {{Antonini}}}, \bibinfo {author} {\bibfnamefont {J.}~\bibnamefont {{Barber}}},\ and\ \bibinfo {author} {\bibfnamefont {I.~M.}\ \bibnamefont {{Romero-Shaw}}},\ }\href {https://doi.org/10.1093/mnras/stad3048} {\bibfield  {journal} {\bibinfo  {journal} {Mon. Not. R. Astron. Soc.}\ }\textbf {\bibinfo {volume} {526}},\ \bibinfo {pages} {4908} (\bibinfo {year} {2023})},\ \Eprint {https://arxiv.org/abs/2308.10884} {arXiv:2308.10884 [astro-ph.HE]} \BibitemShut {NoStop}%
\bibitem [{\citenamefont {{Samsing}}\ \emph {et~al.}(2020)\citenamefont {{Samsing}}, \citenamefont {{D'Orazio}}, \citenamefont {{Kremer}}, \citenamefont {{Rodriguez}},\ and\ \citenamefont {{Askar}}}]{2020PhRvD.101l3010S}%
  \BibitemOpen
  \bibfield  {author} {\bibinfo {author} {\bibfnamefont {J.}~\bibnamefont {{Samsing}}}, \bibinfo {author} {\bibfnamefont {D.~J.}\ \bibnamefont {{D'Orazio}}}, \bibinfo {author} {\bibfnamefont {K.}~\bibnamefont {{Kremer}}}, \bibinfo {author} {\bibfnamefont {C.~L.}\ \bibnamefont {{Rodriguez}}},\ and\ \bibinfo {author} {\bibfnamefont {A.}~\bibnamefont {{Askar}}},\ }\href {https://doi.org/10.1103/PhysRevD.101.123010} {\bibfield  {journal} {\bibinfo  {journal} {Phys. Rev. D}\ }\textbf {\bibinfo {volume} {101}},\ \bibinfo {eid} {123010} (\bibinfo {year} {2020})},\ \Eprint {https://arxiv.org/abs/1907.11231} {arXiv:1907.11231 [astro-ph.HE]} \BibitemShut {NoStop}%
\bibitem [{\citenamefont {{Rodriguez}}\ \emph {et~al.}(2019)\citenamefont {{Rodriguez}}, \citenamefont {{Zevin}}, \citenamefont {{Amaro-Seoane}}, \citenamefont {{Chatterjee}}, \citenamefont {{Kremer}}, \citenamefont {{Rasio}},\ and\ \citenamefont {{Ye}}}]{2019PhRvD.100d3027R}%
  \BibitemOpen
  \bibfield  {author} {\bibinfo {author} {\bibfnamefont {C.~L.}\ \bibnamefont {{Rodriguez}}}, \bibinfo {author} {\bibfnamefont {M.}~\bibnamefont {{Zevin}}}, \bibinfo {author} {\bibfnamefont {P.}~\bibnamefont {{Amaro-Seoane}}}, \bibinfo {author} {\bibfnamefont {S.}~\bibnamefont {{Chatterjee}}}, \bibinfo {author} {\bibfnamefont {K.}~\bibnamefont {{Kremer}}}, \bibinfo {author} {\bibfnamefont {F.~A.}\ \bibnamefont {{Rasio}}},\ and\ \bibinfo {author} {\bibfnamefont {C.~S.}\ \bibnamefont {{Ye}}},\ }\href {https://doi.org/10.1103/PhysRevD.100.043027} {\bibfield  {journal} {\bibinfo  {journal} {Phys. Rev. D}\ }\textbf {\bibinfo {volume} {100}},\ \bibinfo {eid} {043027} (\bibinfo {year} {2019})},\ \Eprint {https://arxiv.org/abs/1906.10260} {arXiv:1906.10260 [astro-ph.HE]} \BibitemShut {NoStop}%
\bibitem [{\citenamefont {{Samsing}}\ and\ \citenamefont {{Hotokezaka}}(2021)}]{2021ApJ...923..126S}%
  \BibitemOpen
  \bibfield  {author} {\bibinfo {author} {\bibfnamefont {J.}~\bibnamefont {{Samsing}}}\ and\ \bibinfo {author} {\bibfnamefont {K.}~\bibnamefont {{Hotokezaka}}},\ }\href {https://doi.org/10.3847/1538-4357/ac2b27} {\bibfield  {journal} {\bibinfo  {journal} {Astrophys. J.}\ }\textbf {\bibinfo {volume} {923}},\ \bibinfo {eid} {126} (\bibinfo {year} {2021})},\ \Eprint {https://arxiv.org/abs/2006.09744} {arXiv:2006.09744 [astro-ph.HE]} \BibitemShut {NoStop}%
\bibitem [{\citenamefont {{Gerosa}}\ \emph {et~al.}(2021)\citenamefont {{Gerosa}}, \citenamefont {{Giacobbo}},\ and\ \citenamefont {{Vecchio}}}]{2021ApJ...915...56G}%
  \BibitemOpen
  \bibfield  {author} {\bibinfo {author} {\bibfnamefont {D.}~\bibnamefont {{Gerosa}}}, \bibinfo {author} {\bibfnamefont {N.}~\bibnamefont {{Giacobbo}}},\ and\ \bibinfo {author} {\bibfnamefont {A.}~\bibnamefont {{Vecchio}}},\ }\href {https://doi.org/10.3847/1538-4357/ac00bb} {\bibfield  {journal} {\bibinfo  {journal} {Astrophys. J.}\ }\textbf {\bibinfo {volume} {915}},\ \bibinfo {eid} {56} (\bibinfo {year} {2021})},\ \Eprint {https://arxiv.org/abs/2104.11247} {arXiv:2104.11247 [astro-ph.HE]} \BibitemShut {NoStop}%
\bibitem [{\citenamefont {{Gerosa}}\ and\ \citenamefont {{Berti}}(2019)}]{2019PhRvD.100d1301G}%
  \BibitemOpen
  \bibfield  {author} {\bibinfo {author} {\bibfnamefont {D.}~\bibnamefont {{Gerosa}}}\ and\ \bibinfo {author} {\bibfnamefont {E.}~\bibnamefont {{Berti}}},\ }\href {https://doi.org/10.1103/PhysRevD.100.041301} {\bibfield  {journal} {\bibinfo  {journal} {Phys. Rev. D}\ }\textbf {\bibinfo {volume} {100}},\ \bibinfo {eid} {041301} (\bibinfo {year} {2019})},\ \Eprint {https://arxiv.org/abs/1906.05295} {arXiv:1906.05295 [astro-ph.HE]} \BibitemShut {NoStop}%
\bibitem [{\citenamefont {{Doctor}}\ \emph {et~al.}(2020)\citenamefont {{Doctor}}, \citenamefont {{Wysocki}}, \citenamefont {{O'Shaughnessy}}, \citenamefont {{Holz}},\ and\ \citenamefont {{Farr}}}]{2020ApJ...893...35D}%
  \BibitemOpen
  \bibfield  {author} {\bibinfo {author} {\bibfnamefont {Z.}~\bibnamefont {{Doctor}}}, \bibinfo {author} {\bibfnamefont {D.}~\bibnamefont {{Wysocki}}}, \bibinfo {author} {\bibfnamefont {R.}~\bibnamefont {{O'Shaughnessy}}}, \bibinfo {author} {\bibfnamefont {D.~E.}\ \bibnamefont {{Holz}}},\ and\ \bibinfo {author} {\bibfnamefont {B.}~\bibnamefont {{Farr}}},\ }\href {https://doi.org/10.3847/1538-4357/ab7fac} {\bibfield  {journal} {\bibinfo  {journal} {Astrophys. J.}\ }\textbf {\bibinfo {volume} {893}},\ \bibinfo {eid} {35} (\bibinfo {year} {2020})},\ \Eprint {https://arxiv.org/abs/1911.04424} {arXiv:1911.04424 [astro-ph.HE]} \BibitemShut {NoStop}%
\bibitem [{\citenamefont {{Kimball}}\ \emph {et~al.}(2021)\citenamefont {{Kimball}}, \citenamefont {{Talbot}}, \citenamefont {{Berry}}, \citenamefont {{Zevin}}, \citenamefont {{Thrane}}, \citenamefont {{Kalogera}}, \citenamefont {{Buscicchio}}, \citenamefont {{Carney}}, \citenamefont {{Dent}}, \citenamefont {{Middleton}}, \citenamefont {{Payne}}, \citenamefont {{Veitch}},\ and\ \citenamefont {{Williams}}}]{2021ApJ...915L..35K}%
  \BibitemOpen
  \bibfield  {author} {\bibinfo {author} {\bibfnamefont {C.}~\bibnamefont {{Kimball}}}, \bibinfo {author} {\bibfnamefont {C.}~\bibnamefont {{Talbot}}}, \bibinfo {author} {\bibfnamefont {C.~P.~L.}\ \bibnamefont {{Berry}}}, \bibinfo {author} {\bibfnamefont {M.}~\bibnamefont {{Zevin}}}, \bibinfo {author} {\bibfnamefont {E.}~\bibnamefont {{Thrane}}}, \bibinfo {author} {\bibfnamefont {V.}~\bibnamefont {{Kalogera}}}, \bibinfo {author} {\bibfnamefont {R.}~\bibnamefont {{Buscicchio}}}, \bibinfo {author} {\bibfnamefont {M.}~\bibnamefont {{Carney}}}, \bibinfo {author} {\bibfnamefont {T.}~\bibnamefont {{Dent}}}, \bibinfo {author} {\bibfnamefont {H.}~\bibnamefont {{Middleton}}}, \bibinfo {author} {\bibfnamefont {E.}~\bibnamefont {{Payne}}}, \bibinfo {author} {\bibfnamefont {J.}~\bibnamefont {{Veitch}}},\ and\ \bibinfo {author} {\bibfnamefont {D.}~\bibnamefont {{Williams}}},\ }\href {https://doi.org/10.3847/2041-8213/ac0aef} {\bibfield  {journal} {\bibinfo  {journal} {Astrophys. J. Lett.}\ }\textbf {\bibinfo {volume}
  {915}},\ \bibinfo {eid} {L35} (\bibinfo {year} {2021})},\ \Eprint {https://arxiv.org/abs/2011.05332} {arXiv:2011.05332 [astro-ph.HE]} \BibitemShut {NoStop}%
\bibitem [{\citenamefont {{Borchers}}\ \emph {et~al.}(2025)\citenamefont {{Borchers}}, \citenamefont {{Ye}},\ and\ \citenamefont {{Fishbach}}}]{2025arXiv250321278B}%
  \BibitemOpen
  \bibfield  {author} {\bibinfo {author} {\bibfnamefont {A.}~\bibnamefont {{Borchers}}}, \bibinfo {author} {\bibfnamefont {C.~S.}\ \bibnamefont {{Ye}}},\ and\ \bibinfo {author} {\bibfnamefont {M.}~\bibnamefont {{Fishbach}}},\ }\href@noop {} {\bibfield  {journal} {\bibinfo  {journal} {{}}\ } (\bibinfo {year} {2025})},\ \Eprint {https://arxiv.org/abs/2503.21278} {arXiv:2503.21278 [astro-ph.HE]} \BibitemShut {NoStop}%
\bibitem [{\citenamefont {{Payne}}\ \emph {et~al.}(2024)\citenamefont {{Payne}}, \citenamefont {{Kremer}},\ and\ \citenamefont {{Zevin}}}]{2024ApJ...966L..16P}%
  \BibitemOpen
  \bibfield  {author} {\bibinfo {author} {\bibfnamefont {E.}~\bibnamefont {{Payne}}}, \bibinfo {author} {\bibfnamefont {K.}~\bibnamefont {{Kremer}}},\ and\ \bibinfo {author} {\bibfnamefont {M.}~\bibnamefont {{Zevin}}},\ }\href {https://doi.org/10.3847/2041-8213/ad3e82} {\bibfield  {journal} {\bibinfo  {journal} {Astrophys. J. Lett.}\ }\textbf {\bibinfo {volume} {966}},\ \bibinfo {eid} {L16} (\bibinfo {year} {2024})},\ \Eprint {https://arxiv.org/abs/2402.15066} {arXiv:2402.15066 [gr-qc]} \BibitemShut {NoStop}%
\bibitem [{\citenamefont {{Gerosa}}\ and\ \citenamefont {{Berti}}(2017)}]{2017PhRvD..95l4046G}%
  \BibitemOpen
  \bibfield  {author} {\bibinfo {author} {\bibfnamefont {D.}~\bibnamefont {{Gerosa}}}\ and\ \bibinfo {author} {\bibfnamefont {E.}~\bibnamefont {{Berti}}},\ }\href {https://doi.org/10.1103/PhysRevD.95.124046} {\bibfield  {journal} {\bibinfo  {journal} {Phys. Rev. D}\ }\textbf {\bibinfo {volume} {95}},\ \bibinfo {eid} {124046} (\bibinfo {year} {2017})},\ \Eprint {https://arxiv.org/abs/1703.06223} {arXiv:1703.06223 [gr-qc]} \BibitemShut {NoStop}%
\bibitem [{\citenamefont {{Fishbach}}\ \emph {et~al.}(2017)\citenamefont {{Fishbach}}, \citenamefont {{Holz}},\ and\ \citenamefont {{Farr}}}]{2017ApJ...840L..24F}%
  \BibitemOpen
  \bibfield  {author} {\bibinfo {author} {\bibfnamefont {M.}~\bibnamefont {{Fishbach}}}, \bibinfo {author} {\bibfnamefont {D.~E.}\ \bibnamefont {{Holz}}},\ and\ \bibinfo {author} {\bibfnamefont {B.}~\bibnamefont {{Farr}}},\ }\href {https://doi.org/10.3847/2041-8213/aa7045} {\bibfield  {journal} {\bibinfo  {journal} {Astrophys. J. Lett.}\ }\textbf {\bibinfo {volume} {840}},\ \bibinfo {eid} {L24} (\bibinfo {year} {2017})},\ \Eprint {https://arxiv.org/abs/1703.06869} {arXiv:1703.06869 [astro-ph.HE]} \BibitemShut {NoStop}%
\bibitem [{\citenamefont {{Baibhav}}\ \emph {et~al.}(2021)\citenamefont {{Baibhav}}, \citenamefont {{Berti}}, \citenamefont {{Gerosa}}, \citenamefont {{Mould}},\ and\ \citenamefont {{Wong}}}]{2021PhRvD.104h4002B}%
  \BibitemOpen
  \bibfield  {author} {\bibinfo {author} {\bibfnamefont {V.}~\bibnamefont {{Baibhav}}}, \bibinfo {author} {\bibfnamefont {E.}~\bibnamefont {{Berti}}}, \bibinfo {author} {\bibfnamefont {D.}~\bibnamefont {{Gerosa}}}, \bibinfo {author} {\bibfnamefont {M.}~\bibnamefont {{Mould}}},\ and\ \bibinfo {author} {\bibfnamefont {K.~W.~K.}\ \bibnamefont {{Wong}}},\ }\href {https://doi.org/10.1103/PhysRevD.104.084002} {\bibfield  {journal} {\bibinfo  {journal} {Phys. Rev. D}\ }\textbf {\bibinfo {volume} {104}},\ \bibinfo {eid} {084002} (\bibinfo {year} {2021})},\ \Eprint {https://arxiv.org/abs/2105.12140} {arXiv:2105.12140 [gr-qc]} \BibitemShut {NoStop}%
\bibitem [{\citenamefont {{Antonini}}\ \emph {et~al.}(2025)\citenamefont {{Antonini}}, \citenamefont {{Romero-Shaw}},\ and\ \citenamefont {{Callister}}}]{Antonini:cluster-pop:2025}%
  \BibitemOpen
  \bibfield  {author} {\bibinfo {author} {\bibfnamefont {F.}~\bibnamefont {{Antonini}}}, \bibinfo {author} {\bibfnamefont {I.~M.}\ \bibnamefont {{Romero-Shaw}}},\ and\ \bibinfo {author} {\bibfnamefont {T.}~\bibnamefont {{Callister}}},\ }\href {https://doi.org/10.1103/PhysRevLett.134.011401} {\bibfield  {journal} {\bibinfo  {journal} {Phys. Rev. Lett.}\ }\textbf {\bibinfo {volume} {134}},\ \bibinfo {eid} {011401} (\bibinfo {year} {2025})},\ \Eprint {https://arxiv.org/abs/2406.19044} {arXiv:2406.19044 [astro-ph.HE]} \BibitemShut {NoStop}%
\bibitem [{\citenamefont {{Mandel}}\ \emph {et~al.}(2019)\citenamefont {{Mandel}}, \citenamefont {{Farr}},\ and\ \citenamefont {{Gair}}}]{2019MNRAS.486.1086M}%
  \BibitemOpen
  \bibfield  {author} {\bibinfo {author} {\bibfnamefont {I.}~\bibnamefont {{Mandel}}}, \bibinfo {author} {\bibfnamefont {W.~M.}\ \bibnamefont {{Farr}}},\ and\ \bibinfo {author} {\bibfnamefont {J.~R.}\ \bibnamefont {{Gair}}},\ }\href {https://doi.org/10.1093/mnras/stz896} {\bibfield  {journal} {\bibinfo  {journal} {Mon. Not. R. Astron. Soc.}\ }\textbf {\bibinfo {volume} {486}},\ \bibinfo {pages} {1086} (\bibinfo {year} {2019})},\ \Eprint {https://arxiv.org/abs/1809.02063} {arXiv:1809.02063 [physics.data-an]} \BibitemShut {NoStop}%
\bibitem [{\citenamefont {{Mould}}\ \emph {et~al.}(2023)\citenamefont {{Mould}}, \citenamefont {{Gerosa}}, \citenamefont {{Dall'Amico}},\ and\ \citenamefont {{Mapelli}}}]{2023MNRAS.525.3986M}%
  \BibitemOpen
  \bibfield  {author} {\bibinfo {author} {\bibfnamefont {M.}~\bibnamefont {{Mould}}}, \bibinfo {author} {\bibfnamefont {D.}~\bibnamefont {{Gerosa}}}, \bibinfo {author} {\bibfnamefont {M.}~\bibnamefont {{Dall'Amico}}},\ and\ \bibinfo {author} {\bibfnamefont {M.}~\bibnamefont {{Mapelli}}},\ }\href {https://doi.org/10.1093/mnras/stad2502} {\bibfield  {journal} {\bibinfo  {journal} {Mon. Not. R. Astron. Soc.}\ }\textbf {\bibinfo {volume} {525}},\ \bibinfo {pages} {3986} (\bibinfo {year} {2023})},\ \Eprint {https://arxiv.org/abs/2305.18539} {arXiv:2305.18539 [astro-ph.HE]} \BibitemShut {NoStop}%
\bibitem [{\citenamefont {{Syer}}\ \emph {et~al.}(1991)\citenamefont {{Syer}}, \citenamefont {{Clarke}},\ and\ \citenamefont {{Rees}}}]{Syer1999}%
  \BibitemOpen
  \bibfield  {author} {\bibinfo {author} {\bibfnamefont {D.}~\bibnamefont {{Syer}}}, \bibinfo {author} {\bibfnamefont {C.~J.}\ \bibnamefont {{Clarke}}},\ and\ \bibinfo {author} {\bibfnamefont {M.~J.}\ \bibnamefont {{Rees}}},\ }\href {https://doi.org/10.1093/mnras/250.3.505} {\bibfield  {journal} {\bibinfo  {journal} {Mon. Not. R. Astron. Soc.}\ }\textbf {\bibinfo {volume} {250}},\ \bibinfo {pages} {505} (\bibinfo {year} {1991})}\BibitemShut {NoStop}%
\bibitem [{\citenamefont {{Levin}}\ and\ \citenamefont {{Beloborodov}}(2003)}]{Levin2003}%
  \BibitemOpen
  \bibfield  {author} {\bibinfo {author} {\bibfnamefont {Y.}~\bibnamefont {{Levin}}}\ and\ \bibinfo {author} {\bibfnamefont {A.~M.}\ \bibnamefont {{Beloborodov}}},\ }\href {https://doi.org/10.1086/376675} {\bibfield  {journal} {\bibinfo  {journal} {Astrophys. J. Lett.}\ }\textbf {\bibinfo {volume} {590}},\ \bibinfo {pages} {L33} (\bibinfo {year} {2003})},\ \Eprint {https://arxiv.org/abs/astro-ph/0303436} {arXiv:astro-ph/0303436 [astro-ph]} \BibitemShut {NoStop}%
\bibitem [{\citenamefont {{Stone}}\ \emph {et~al.}(2017)\citenamefont {{Stone}}, \citenamefont {{Metzger}},\ and\ \citenamefont {{Haiman}}}]{Stone2017}%
  \BibitemOpen
  \bibfield  {author} {\bibinfo {author} {\bibfnamefont {N.~C.}\ \bibnamefont {{Stone}}}, \bibinfo {author} {\bibfnamefont {B.~D.}\ \bibnamefont {{Metzger}}},\ and\ \bibinfo {author} {\bibfnamefont {Z.}~\bibnamefont {{Haiman}}},\ }\href {https://doi.org/10.1093/mnras/stw2260} {\bibfield  {journal} {\bibinfo  {journal} {Mon. Not. R. Astron. Soc.}\ }\textbf {\bibinfo {volume} {464}},\ \bibinfo {pages} {946} (\bibinfo {year} {2017})},\ \Eprint {https://arxiv.org/abs/1602.04226} {arXiv:1602.04226 [astro-ph.GA]} \BibitemShut {NoStop}%
\bibitem [{\citenamefont {{Epstein-Martin}}\ \emph {et~al.}(2025)\citenamefont {{Epstein-Martin}}, \citenamefont {{Tagawa}}, \citenamefont {{Haiman}},\ and\ \citenamefont {{Perna}}}]{EpsteinMartin2025}%
  \BibitemOpen
  \bibfield  {author} {\bibinfo {author} {\bibfnamefont {M.}~\bibnamefont {{Epstein-Martin}}}, \bibinfo {author} {\bibfnamefont {H.}~\bibnamefont {{Tagawa}}}, \bibinfo {author} {\bibfnamefont {Z.}~\bibnamefont {{Haiman}}},\ and\ \bibinfo {author} {\bibfnamefont {R.}~\bibnamefont {{Perna}}},\ }\href {https://doi.org/10.1093/mnras/staf237} {\bibfield  {journal} {\bibinfo  {journal} {Mon. Not. R. Astron. Soc.}\ }\textbf {\bibinfo {volume} {537}},\ \bibinfo {pages} {3396} (\bibinfo {year} {2025})},\ \Eprint {https://arxiv.org/abs/2405.09380} {arXiv:2405.09380 [astro-ph.HE]} \BibitemShut {NoStop}%
\bibitem [{\citenamefont {{Goldreich}}\ \emph {et~al.}(2002)\citenamefont {{Goldreich}}, \citenamefont {{Lithwick}},\ and\ \citenamefont {{Sari}}}]{Goldreich2002}%
  \BibitemOpen
  \bibfield  {author} {\bibinfo {author} {\bibfnamefont {P.}~\bibnamefont {{Goldreich}}}, \bibinfo {author} {\bibfnamefont {Y.}~\bibnamefont {{Lithwick}}},\ and\ \bibinfo {author} {\bibfnamefont {R.}~\bibnamefont {{Sari}}},\ }\href {https://doi.org/10.1038/nature01227} {\bibfield  {journal} {\bibinfo  {journal} {Nature}\ }\textbf {\bibinfo {volume} {420}},\ \bibinfo {pages} {643} (\bibinfo {year} {2002})},\ \Eprint {https://arxiv.org/abs/astro-ph/0208490} {arXiv:astro-ph/0208490 [astro-ph]} \BibitemShut {NoStop}%
\bibitem [{\citenamefont {{Tagawa}}\ \emph {et~al.}(2020{\natexlab{a}})\citenamefont {{Tagawa}}, \citenamefont {{Haiman}},\ and\ \citenamefont {{Kocsis}}}]{Tagawa2020}%
  \BibitemOpen
  \bibfield  {author} {\bibinfo {author} {\bibfnamefont {H.}~\bibnamefont {{Tagawa}}}, \bibinfo {author} {\bibfnamefont {Z.}~\bibnamefont {{Haiman}}},\ and\ \bibinfo {author} {\bibfnamefont {B.}~\bibnamefont {{Kocsis}}},\ }\href {https://doi.org/10.3847/1538-4357/ab9b8c} {\bibfield  {journal} {\bibinfo  {journal} {Astrophys. J.}\ }\textbf {\bibinfo {volume} {898}},\ \bibinfo {eid} {25} (\bibinfo {year} {2020}{\natexlab{a}})},\ \Eprint {https://arxiv.org/abs/1912.08218} {arXiv:1912.08218 [astro-ph.GA]} \BibitemShut {NoStop}%
\bibitem [{\citenamefont {{Bellovary}}\ \emph {et~al.}(2016)\citenamefont {{Bellovary}}, \citenamefont {{Mac Low}}, \citenamefont {{McKernan}},\ and\ \citenamefont {{Ford}}}]{Bellovary2016}%
  \BibitemOpen
  \bibfield  {author} {\bibinfo {author} {\bibfnamefont {J.~M.}\ \bibnamefont {{Bellovary}}}, \bibinfo {author} {\bibfnamefont {M.-M.}\ \bibnamefont {{Mac Low}}}, \bibinfo {author} {\bibfnamefont {B.}~\bibnamefont {{McKernan}}},\ and\ \bibinfo {author} {\bibfnamefont {K.~E.~S.}\ \bibnamefont {{Ford}}},\ }\href {https://doi.org/10.3847/2041-8205/819/2/L17} {\bibfield  {journal} {\bibinfo  {journal} {Astrophys. J. Lett.}\ }\textbf {\bibinfo {volume} {819}},\ \bibinfo {eid} {L17} (\bibinfo {year} {2016})},\ \Eprint {https://arxiv.org/abs/1511.00005} {arXiv:1511.00005 [astro-ph.GA]} \BibitemShut {NoStop}%
\bibitem [{\citenamefont {{Gangardt}}\ \emph {et~al.}(2024)\citenamefont {{Gangardt}}, \citenamefont {{Trani}}, \citenamefont {{Bonnerot}},\ and\ \citenamefont {{Gerosa}}}]{2024MNRAS.530.3689G}%
  \BibitemOpen
  \bibfield  {author} {\bibinfo {author} {\bibfnamefont {D.}~\bibnamefont {{Gangardt}}}, \bibinfo {author} {\bibfnamefont {A.~A.}\ \bibnamefont {{Trani}}}, \bibinfo {author} {\bibfnamefont {C.}~\bibnamefont {{Bonnerot}}},\ and\ \bibinfo {author} {\bibfnamefont {D.}~\bibnamefont {{Gerosa}}},\ }\href {https://doi.org/10.1093/mnras/stae1117} {\bibfield  {journal} {\bibinfo  {journal} {Mon. Not. R. Astron. Soc.}\ }\textbf {\bibinfo {volume} {530}},\ \bibinfo {pages} {3689} (\bibinfo {year} {2024})},\ \Eprint {https://arxiv.org/abs/2403.00060} {arXiv:2403.00060 [astro-ph.HE]} \BibitemShut {NoStop}%
\bibitem [{\citenamefont {{Vaccaro}}\ \emph {et~al.}(2024)\citenamefont {{Vaccaro}}, \citenamefont {{Mapelli}}, \citenamefont {{P{\'e}rigois}}, \citenamefont {{Barone}}, \citenamefont {{Artale}}, \citenamefont {{Dall'Amico}}, \citenamefont {{Iorio}},\ and\ \citenamefont {{Torniamenti}}}]{Vaccaro2024}%
  \BibitemOpen
  \bibfield  {author} {\bibinfo {author} {\bibfnamefont {M.~P.}\ \bibnamefont {{Vaccaro}}}, \bibinfo {author} {\bibfnamefont {M.}~\bibnamefont {{Mapelli}}}, \bibinfo {author} {\bibfnamefont {C.}~\bibnamefont {{P{\'e}rigois}}}, \bibinfo {author} {\bibfnamefont {D.}~\bibnamefont {{Barone}}}, \bibinfo {author} {\bibfnamefont {M.~C.}\ \bibnamefont {{Artale}}}, \bibinfo {author} {\bibfnamefont {M.}~\bibnamefont {{Dall'Amico}}}, \bibinfo {author} {\bibfnamefont {G.}~\bibnamefont {{Iorio}}},\ and\ \bibinfo {author} {\bibfnamefont {S.}~\bibnamefont {{Torniamenti}}},\ }\href {https://doi.org/10.1051/0004-6361/202348509} {\bibfield  {journal} {\bibinfo  {journal} {Astron. Astrophys.}\ }\textbf {\bibinfo {volume} {685}},\ \bibinfo {eid} {A51} (\bibinfo {year} {2024})},\ \Eprint {https://arxiv.org/abs/2311.18548} {arXiv:2311.18548 [astro-ph.HE]} \BibitemShut {NoStop}%
\bibitem [{\citenamefont {{Grishin}}\ \emph {et~al.}(2024)\citenamefont {{Grishin}}, \citenamefont {{Gilbaum}},\ and\ \citenamefont {{Stone}}}]{Grishin2024}%
  \BibitemOpen
  \bibfield  {author} {\bibinfo {author} {\bibfnamefont {E.}~\bibnamefont {{Grishin}}}, \bibinfo {author} {\bibfnamefont {S.}~\bibnamefont {{Gilbaum}}},\ and\ \bibinfo {author} {\bibfnamefont {N.~C.}\ \bibnamefont {{Stone}}},\ }\href {https://doi.org/10.1093/mnras/stae828} {\bibfield  {journal} {\bibinfo  {journal} {Mon. Not. R. Astron. Soc.}\ }\textbf {\bibinfo {volume} {530}},\ \bibinfo {pages} {2114} (\bibinfo {year} {2024})},\ \Eprint {https://arxiv.org/abs/2307.07546} {arXiv:2307.07546 [astro-ph.HE]} \BibitemShut {NoStop}%
\bibitem [{\citenamefont {{Pan}}\ and\ \citenamefont {{Yang}}(2021{\natexlab{a}})}]{Pan2021_EMRI}%
  \BibitemOpen
  \bibfield  {author} {\bibinfo {author} {\bibfnamefont {Z.}~\bibnamefont {{Pan}}}\ and\ \bibinfo {author} {\bibfnamefont {H.}~\bibnamefont {{Yang}}},\ }\href {https://doi.org/10.1103/PhysRevD.103.103018} {\bibfield  {journal} {\bibinfo  {journal} {Phys. Rev. D}\ }\textbf {\bibinfo {volume} {103}},\ \bibinfo {eid} {103018} (\bibinfo {year} {2021}{\natexlab{a}})},\ \Eprint {https://arxiv.org/abs/2101.09146} {arXiv:2101.09146 [astro-ph.HE]} \BibitemShut {NoStop}%
\bibitem [{\citenamefont {{Bartos}}\ \emph {et~al.}(2017)\citenamefont {{Bartos}}, \citenamefont {{Kocsis}}, \citenamefont {{Haiman}},\ and\ \citenamefont {{M{\'a}rka}}}]{Bartos2017}%
  \BibitemOpen
  \bibfield  {author} {\bibinfo {author} {\bibfnamefont {I.}~\bibnamefont {{Bartos}}}, \bibinfo {author} {\bibfnamefont {B.}~\bibnamefont {{Kocsis}}}, \bibinfo {author} {\bibfnamefont {Z.}~\bibnamefont {{Haiman}}},\ and\ \bibinfo {author} {\bibfnamefont {S.}~\bibnamefont {{M{\'a}rka}}},\ }\href {https://doi.org/10.3847/1538-4357/835/2/165} {\bibfield  {journal} {\bibinfo  {journal} {Astrophys. J.}\ }\textbf {\bibinfo {volume} {835}},\ \bibinfo {eid} {165} (\bibinfo {year} {2017})},\ \Eprint {https://arxiv.org/abs/1602.03831} {arXiv:1602.03831 [astro-ph.HE]} \BibitemShut {NoStop}%
\bibitem [{\citenamefont {{McKernan}}\ \emph {et~al.}(2018)\citenamefont {{McKernan}}, \citenamefont {{Ford}}, \citenamefont {{Bellovary}}, \citenamefont {{Leigh}}, \citenamefont {{Haiman}}, \citenamefont {{Kocsis}}, \citenamefont {{Lyra}}, \citenamefont {{Mac Low}}, \citenamefont {{Metzger}}, \citenamefont {{O'Dowd}}, \citenamefont {{Endlich}},\ and\ \citenamefont {{Rosen}}}]{McKernan2018}%
  \BibitemOpen
  \bibfield  {author} {\bibinfo {author} {\bibfnamefont {B.}~\bibnamefont {{McKernan}}}, \bibinfo {author} {\bibfnamefont {K.~E.~S.}\ \bibnamefont {{Ford}}}, \bibinfo {author} {\bibfnamefont {J.}~\bibnamefont {{Bellovary}}}, \bibinfo {author} {\bibfnamefont {N.~W.~C.}\ \bibnamefont {{Leigh}}}, \bibinfo {author} {\bibfnamefont {Z.}~\bibnamefont {{Haiman}}}, \bibinfo {author} {\bibfnamefont {B.}~\bibnamefont {{Kocsis}}}, \bibinfo {author} {\bibfnamefont {W.}~\bibnamefont {{Lyra}}}, \bibinfo {author} {\bibfnamefont {M.~M.}\ \bibnamefont {{Mac Low}}}, \bibinfo {author} {\bibfnamefont {B.}~\bibnamefont {{Metzger}}}, \bibinfo {author} {\bibfnamefont {M.}~\bibnamefont {{O'Dowd}}}, \bibinfo {author} {\bibfnamefont {S.}~\bibnamefont {{Endlich}}},\ and\ \bibinfo {author} {\bibfnamefont {D.~J.}\ \bibnamefont {{Rosen}}},\ }\href {https://doi.org/10.3847/1538-4357/aadae5} {\bibfield  {journal} {\bibinfo  {journal} {Astrophys. J.}\ }\textbf {\bibinfo {volume} {866}},\ \bibinfo {eid} {66} (\bibinfo {year} {2018})},\ \Eprint
  {https://arxiv.org/abs/1702.07818} {arXiv:1702.07818 [astro-ph.HE]} \BibitemShut {NoStop}%
\bibitem [{\citenamefont {{Yang}}\ \emph {et~al.}(2019{\natexlab{a}})\citenamefont {{Yang}}, \citenamefont {{Bartos}}, \citenamefont {{Gayathri}}, \citenamefont {{Ford}}, \citenamefont {{Haiman}}, \citenamefont {{Klimenko}}, \citenamefont {{Kocsis}}, \citenamefont {{M{\'a}rka}}, \citenamefont {{M{\'a}rka}}, \citenamefont {{McKernan}},\ and\ \citenamefont {{O'Shaughnessy}}}]{Yang2019_spin}%
  \BibitemOpen
  \bibfield  {author} {\bibinfo {author} {\bibfnamefont {Y.}~\bibnamefont {{Yang}}}, \bibinfo {author} {\bibfnamefont {I.}~\bibnamefont {{Bartos}}}, \bibinfo {author} {\bibfnamefont {V.}~\bibnamefont {{Gayathri}}}, \bibinfo {author} {\bibfnamefont {K.~E.~S.}\ \bibnamefont {{Ford}}}, \bibinfo {author} {\bibfnamefont {Z.}~\bibnamefont {{Haiman}}}, \bibinfo {author} {\bibfnamefont {S.}~\bibnamefont {{Klimenko}}}, \bibinfo {author} {\bibfnamefont {B.}~\bibnamefont {{Kocsis}}}, \bibinfo {author} {\bibfnamefont {S.}~\bibnamefont {{M{\'a}rka}}}, \bibinfo {author} {\bibfnamefont {Z.}~\bibnamefont {{M{\'a}rka}}}, \bibinfo {author} {\bibfnamefont {B.}~\bibnamefont {{McKernan}}},\ and\ \bibinfo {author} {\bibfnamefont {R.}~\bibnamefont {{O'Shaughnessy}}},\ }\href {https://doi.org/10.1103/PhysRevLett.123.181101} {\bibfield  {journal} {\bibinfo  {journal} {Phys. Rev. Lett.}\ }\textbf {\bibinfo {volume} {123}},\ \bibinfo {eid} {181101} (\bibinfo {year} {2019}{\natexlab{a}})},\ \Eprint {https://arxiv.org/abs/1906.09281}
  {arXiv:1906.09281 [astro-ph.HE]} \BibitemShut {NoStop}%
\bibitem [{\citenamefont {{Santini}}\ \emph {et~al.}(2023)\citenamefont {{Santini}}, \citenamefont {{Gerosa}}, \citenamefont {{Cotesta}},\ and\ \citenamefont {{Berti}}}]{2023PhRvD.108h3033S}%
  \BibitemOpen
  \bibfield  {author} {\bibinfo {author} {\bibfnamefont {A.}~\bibnamefont {{Santini}}}, \bibinfo {author} {\bibfnamefont {D.}~\bibnamefont {{Gerosa}}}, \bibinfo {author} {\bibfnamefont {R.}~\bibnamefont {{Cotesta}}},\ and\ \bibinfo {author} {\bibfnamefont {E.}~\bibnamefont {{Berti}}},\ }\href {https://doi.org/10.1103/PhysRevD.108.083033} {\bibfield  {journal} {\bibinfo  {journal} {Phys. Rev. D}\ }\textbf {\bibinfo {volume} {108}},\ \bibinfo {eid} {083033} (\bibinfo {year} {2023})},\ \Eprint {https://arxiv.org/abs/2308.12998} {arXiv:2308.12998 [astro-ph.HE]} \BibitemShut {NoStop}%
\bibitem [{\citenamefont {{Yang}}\ \emph {et~al.}(2019{\natexlab{b}})\citenamefont {{Yang}}, \citenamefont {{Bartos}}, \citenamefont {{Haiman}}, \citenamefont {{Kocsis}}, \citenamefont {{M{\'a}rka}}, \citenamefont {{Stone}},\ and\ \citenamefont {{M{\'a}rka}}}]{Yang2019}%
  \BibitemOpen
  \bibfield  {author} {\bibinfo {author} {\bibfnamefont {Y.}~\bibnamefont {{Yang}}}, \bibinfo {author} {\bibfnamefont {I.}~\bibnamefont {{Bartos}}}, \bibinfo {author} {\bibfnamefont {Z.}~\bibnamefont {{Haiman}}}, \bibinfo {author} {\bibfnamefont {B.}~\bibnamefont {{Kocsis}}}, \bibinfo {author} {\bibfnamefont {Z.}~\bibnamefont {{M{\'a}rka}}}, \bibinfo {author} {\bibfnamefont {N.~C.}\ \bibnamefont {{Stone}}},\ and\ \bibinfo {author} {\bibfnamefont {S.}~\bibnamefont {{M{\'a}rka}}},\ }\href {https://doi.org/10.3847/1538-4357/ab16e3} {\bibfield  {journal} {\bibinfo  {journal} {Astrophys. J.}\ }\textbf {\bibinfo {volume} {876}},\ \bibinfo {eid} {122} (\bibinfo {year} {2019}{\natexlab{b}})},\ \Eprint {https://arxiv.org/abs/1903.01405} {arXiv:1903.01405 [astro-ph.HE]} \BibitemShut {NoStop}%
\bibitem [{\citenamefont {{Blandford}}\ and\ \citenamefont {{Begelman}}(1999)}]{Blandford1999}%
  \BibitemOpen
  \bibfield  {author} {\bibinfo {author} {\bibfnamefont {R.~D.}\ \bibnamefont {{Blandford}}}\ and\ \bibinfo {author} {\bibfnamefont {M.~C.}\ \bibnamefont {{Begelman}}},\ }\href {https://doi.org/10.1046/j.1365-8711.1999.02358.x} {\bibfield  {journal} {\bibinfo  {journal} {Mon. Not. R. Astron. Soc.}\ }\textbf {\bibinfo {volume} {303}},\ \bibinfo {pages} {L1} (\bibinfo {year} {1999})},\ \Eprint {https://arxiv.org/abs/astro-ph/9809083} {arXiv:astro-ph/9809083 [astro-ph]} \BibitemShut {NoStop}%
\bibitem [{\citenamefont {{Pan}}\ and\ \citenamefont {{Yang}}(2021{\natexlab{b}})}]{Pan2021}%
  \BibitemOpen
  \bibfield  {author} {\bibinfo {author} {\bibfnamefont {Z.}~\bibnamefont {{Pan}}}\ and\ \bibinfo {author} {\bibfnamefont {H.}~\bibnamefont {{Yang}}},\ }\href {https://doi.org/10.3847/1538-4357/ac249c} {\bibfield  {journal} {\bibinfo  {journal} {Astrophys. J.}\ }\textbf {\bibinfo {volume} {923}},\ \bibinfo {eid} {173} (\bibinfo {year} {2021}{\natexlab{b}})},\ \Eprint {https://arxiv.org/abs/2108.00267} {arXiv:2108.00267 [astro-ph.HE]} \BibitemShut {NoStop}%
\bibitem [{\citenamefont {{Tagawa}}\ \emph {et~al.}(2022)\citenamefont {{Tagawa}}, \citenamefont {{Kimura}}, \citenamefont {{Haiman}}, \citenamefont {{Perna}}, \citenamefont {{Tanaka}},\ and\ \citenamefont {{Bartos}}}]{Tagawa2022}%
  \BibitemOpen
  \bibfield  {author} {\bibinfo {author} {\bibfnamefont {H.}~\bibnamefont {{Tagawa}}}, \bibinfo {author} {\bibfnamefont {S.~S.}\ \bibnamefont {{Kimura}}}, \bibinfo {author} {\bibfnamefont {Z.}~\bibnamefont {{Haiman}}}, \bibinfo {author} {\bibfnamefont {R.}~\bibnamefont {{Perna}}}, \bibinfo {author} {\bibfnamefont {H.}~\bibnamefont {{Tanaka}}},\ and\ \bibinfo {author} {\bibfnamefont {I.}~\bibnamefont {{Bartos}}},\ }\href {https://doi.org/10.3847/1538-4357/ac45f8} {\bibfield  {journal} {\bibinfo  {journal} {Astrophys. J.}\ }\textbf {\bibinfo {volume} {927}},\ \bibinfo {eid} {41} (\bibinfo {year} {2022})},\ \Eprint {https://arxiv.org/abs/2112.01544} {arXiv:2112.01544 [astro-ph.HE]} \BibitemShut {NoStop}%
\bibitem [{\citenamefont {{DeLaurentiis}}\ \emph {et~al.}(2023)\citenamefont {{DeLaurentiis}}, \citenamefont {{Epstein-Martin}},\ and\ \citenamefont {{Haiman}}}]{DeLaurentiis2023}%
  \BibitemOpen
  \bibfield  {author} {\bibinfo {author} {\bibfnamefont {S.}~\bibnamefont {{DeLaurentiis}}}, \bibinfo {author} {\bibfnamefont {M.}~\bibnamefont {{Epstein-Martin}}},\ and\ \bibinfo {author} {\bibfnamefont {Z.}~\bibnamefont {{Haiman}}},\ }\href {https://doi.org/10.1093/mnras/stad1412} {\bibfield  {journal} {\bibinfo  {journal} {Mon. Not. R. Astron. Soc.}\ }\textbf {\bibinfo {volume} {523}},\ \bibinfo {pages} {1126} (\bibinfo {year} {2023})},\ \Eprint {https://arxiv.org/abs/2212.02650} {arXiv:2212.02650 [astro-ph.HE]} \BibitemShut {NoStop}%
\bibitem [{\citenamefont {{Li}}\ \emph {et~al.}(2023)\citenamefont {{Li}}, \citenamefont {{Dempsey}}, \citenamefont {{Li}}, \citenamefont {{Lai}},\ and\ \citenamefont {{Li}}}]{LiJiaru2023}%
  \BibitemOpen
  \bibfield  {author} {\bibinfo {author} {\bibfnamefont {J.}~\bibnamefont {{Li}}}, \bibinfo {author} {\bibfnamefont {A.~M.}\ \bibnamefont {{Dempsey}}}, \bibinfo {author} {\bibfnamefont {H.}~\bibnamefont {{Li}}}, \bibinfo {author} {\bibfnamefont {D.}~\bibnamefont {{Lai}}},\ and\ \bibinfo {author} {\bibfnamefont {S.}~\bibnamefont {{Li}}},\ }\href {https://doi.org/10.3847/2041-8213/acb934} {\bibfield  {journal} {\bibinfo  {journal} {Astrophys. J. Lett.}\ }\textbf {\bibinfo {volume} {944}},\ \bibinfo {eid} {L42} (\bibinfo {year} {2023})},\ \Eprint {https://arxiv.org/abs/2211.10357} {arXiv:2211.10357 [astro-ph.HE]} \BibitemShut {NoStop}%
\bibitem [{\citenamefont {{Whitehead}}\ \emph {et~al.}(2024)\citenamefont {{Whitehead}}, \citenamefont {{Rowan}}, \citenamefont {{Boekholt}},\ and\ \citenamefont {{Kocsis}}}]{Whitehead2024}%
  \BibitemOpen
  \bibfield  {author} {\bibinfo {author} {\bibfnamefont {H.}~\bibnamefont {{Whitehead}}}, \bibinfo {author} {\bibfnamefont {C.}~\bibnamefont {{Rowan}}}, \bibinfo {author} {\bibfnamefont {T.}~\bibnamefont {{Boekholt}}},\ and\ \bibinfo {author} {\bibfnamefont {B.}~\bibnamefont {{Kocsis}}},\ }\href {https://doi.org/10.1093/mnras/stae1430} {\bibfield  {journal} {\bibinfo  {journal} {Mon. Not. R. Astron. Soc.}\ }\textbf {\bibinfo {volume} {531}},\ \bibinfo {pages} {4656} (\bibinfo {year} {2024})},\ \Eprint {https://arxiv.org/abs/2309.11561} {arXiv:2309.11561 [astro-ph.GA]} \BibitemShut {NoStop}%
\bibitem [{\citenamefont {{Roedig}}\ \emph {et~al.}(2011)\citenamefont {{Roedig}}, \citenamefont {{Dotti}}, \citenamefont {{Sesana}}, \citenamefont {{Cuadra}},\ and\ \citenamefont {{Colpi}}}]{Roedig2011}%
  \BibitemOpen
  \bibfield  {author} {\bibinfo {author} {\bibfnamefont {C.}~\bibnamefont {{Roedig}}}, \bibinfo {author} {\bibfnamefont {M.}~\bibnamefont {{Dotti}}}, \bibinfo {author} {\bibfnamefont {A.}~\bibnamefont {{Sesana}}}, \bibinfo {author} {\bibfnamefont {J.}~\bibnamefont {{Cuadra}}},\ and\ \bibinfo {author} {\bibfnamefont {M.}~\bibnamefont {{Colpi}}},\ }\href {https://doi.org/10.1111/j.1365-2966.2011.18927.x} {\bibfield  {journal} {\bibinfo  {journal} {Mon. Not. R. Astron. Soc.}\ }\textbf {\bibinfo {volume} {415}},\ \bibinfo {pages} {3033} (\bibinfo {year} {2011})},\ \Eprint {https://arxiv.org/abs/1104.3868} {arXiv:1104.3868 [astro-ph.CO]} \BibitemShut {NoStop}%
\bibitem [{\citenamefont {{Zrake}}\ \emph {et~al.}(2021)\citenamefont {{Zrake}}, \citenamefont {{Tiede}}, \citenamefont {{MacFadyen}},\ and\ \citenamefont {{Haiman}}}]{Zrake2021}%
  \BibitemOpen
  \bibfield  {author} {\bibinfo {author} {\bibfnamefont {J.}~\bibnamefont {{Zrake}}}, \bibinfo {author} {\bibfnamefont {C.}~\bibnamefont {{Tiede}}}, \bibinfo {author} {\bibfnamefont {A.}~\bibnamefont {{MacFadyen}}},\ and\ \bibinfo {author} {\bibfnamefont {Z.}~\bibnamefont {{Haiman}}},\ }\href {https://doi.org/10.3847/2041-8213/abdd1c} {\bibfield  {journal} {\bibinfo  {journal} {Astrophys. J. Lett.}\ }\textbf {\bibinfo {volume} {909}},\ \bibinfo {eid} {L13} (\bibinfo {year} {2021})},\ \Eprint {https://arxiv.org/abs/2010.09707} {arXiv:2010.09707 [astro-ph.HE]} \BibitemShut {NoStop}%
\bibitem [{\citenamefont {{D'Orazio}}\ and\ \citenamefont {{Duffell}}(2021)}]{DOrazio2021}%
  \BibitemOpen
  \bibfield  {author} {\bibinfo {author} {\bibfnamefont {D.~J.}\ \bibnamefont {{D'Orazio}}}\ and\ \bibinfo {author} {\bibfnamefont {P.~C.}\ \bibnamefont {{Duffell}}},\ }\href {https://doi.org/10.3847/2041-8213/ac0621} {\bibfield  {journal} {\bibinfo  {journal} {Astrophys. J. Lett.}\ }\textbf {\bibinfo {volume} {914}},\ \bibinfo {eid} {L21} (\bibinfo {year} {2021})},\ \Eprint {https://arxiv.org/abs/2103.09251} {arXiv:2103.09251 [astro-ph.HE]} \BibitemShut {NoStop}%
\bibitem [{\citenamefont {{Siwek}}\ \emph {et~al.}(2023)\citenamefont {{Siwek}}, \citenamefont {{Weinberger}},\ and\ \citenamefont {{Hernquist}}}]{Siwek:CBDs:2023}%
  \BibitemOpen
  \bibfield  {author} {\bibinfo {author} {\bibfnamefont {M.}~\bibnamefont {{Siwek}}}, \bibinfo {author} {\bibfnamefont {R.}~\bibnamefont {{Weinberger}}},\ and\ \bibinfo {author} {\bibfnamefont {L.}~\bibnamefont {{Hernquist}}},\ }\href {https://doi.org/10.1093/mnras/stad1131} {\bibfield  {journal} {\bibinfo  {journal} {Mon. Not. R. Astron. Soc.}\ }\textbf {\bibinfo {volume} {522}},\ \bibinfo {pages} {2707} (\bibinfo {year} {2023})},\ \Eprint {https://arxiv.org/abs/2302.01785} {arXiv:2302.01785 [astro-ph.HE]} \BibitemShut {NoStop}%
\bibitem [{\citenamefont {{Franchini}}\ \emph {et~al.}(2024)\citenamefont {{Franchini}}, \citenamefont {{Prato}}, \citenamefont {{Longarini}},\ and\ \citenamefont {{Sesana}}}]{Franchini2024}%
  \BibitemOpen
  \bibfield  {author} {\bibinfo {author} {\bibfnamefont {A.}~\bibnamefont {{Franchini}}}, \bibinfo {author} {\bibfnamefont {A.}~\bibnamefont {{Prato}}}, \bibinfo {author} {\bibfnamefont {C.}~\bibnamefont {{Longarini}}},\ and\ \bibinfo {author} {\bibfnamefont {A.}~\bibnamefont {{Sesana}}},\ }\href {https://doi.org/10.1051/0004-6361/202449402} {\bibfield  {journal} {\bibinfo  {journal} {Astron. Astrophys.}\ }\textbf {\bibinfo {volume} {688}},\ \bibinfo {eid} {A174} (\bibinfo {year} {2024})},\ \Eprint {https://arxiv.org/abs/2402.00938} {arXiv:2402.00938 [astro-ph.HE]} \BibitemShut {NoStop}%
\bibitem [{\citenamefont {{Calcino}}\ \emph {et~al.}(2024)\citenamefont {{Calcino}}, \citenamefont {{Dempsey}}, \citenamefont {{Dittmann}},\ and\ \citenamefont {{Li}}}]{Calcino2024}%
  \BibitemOpen
  \bibfield  {author} {\bibinfo {author} {\bibfnamefont {J.}~\bibnamefont {{Calcino}}}, \bibinfo {author} {\bibfnamefont {A.~M.}\ \bibnamefont {{Dempsey}}}, \bibinfo {author} {\bibfnamefont {A.~J.}\ \bibnamefont {{Dittmann}}},\ and\ \bibinfo {author} {\bibfnamefont {H.}~\bibnamefont {{Li}}},\ }\href {https://doi.org/10.3847/1538-4357/ad4a53} {\bibfield  {journal} {\bibinfo  {journal} {Astrophys. J.}\ }\textbf {\bibinfo {volume} {970}},\ \bibinfo {eid} {107} (\bibinfo {year} {2024})},\ \Eprint {https://arxiv.org/abs/2311.13727} {arXiv:2311.13727 [astro-ph.HE]} \BibitemShut {NoStop}%
\bibitem [{\citenamefont {{Dittmann}}\ \emph {et~al.}(2025)\citenamefont {{Dittmann}}, \citenamefont {{Dempsey}},\ and\ \citenamefont {{Li}}}]{Dittmann2025}%
  \BibitemOpen
  \bibfield  {author} {\bibinfo {author} {\bibfnamefont {A.~J.}\ \bibnamefont {{Dittmann}}}, \bibinfo {author} {\bibfnamefont {A.~M.}\ \bibnamefont {{Dempsey}}},\ and\ \bibinfo {author} {\bibfnamefont {H.}~\bibnamefont {{Li}}},\ }\href {https://doi.org/10.48550/arXiv.2505.05555} {\bibfield  {journal} {\bibinfo  {journal} {arXiv e-prints}\ ,\ \bibinfo {eid} {arXiv:2505.05555}} (\bibinfo {year} {2025})},\ \Eprint {https://arxiv.org/abs/2505.05555} {arXiv:2505.05555 [astro-ph.HE]} \BibitemShut {NoStop}%
\bibitem [{\citenamefont {{Rowan}}\ \emph {et~al.}(2023)\citenamefont {{Rowan}}, \citenamefont {{Boekholt}}, \citenamefont {{Kocsis}},\ and\ \citenamefont {{Haiman}}}]{Rowan2023}%
  \BibitemOpen
  \bibfield  {author} {\bibinfo {author} {\bibfnamefont {C.}~\bibnamefont {{Rowan}}}, \bibinfo {author} {\bibfnamefont {T.}~\bibnamefont {{Boekholt}}}, \bibinfo {author} {\bibfnamefont {B.}~\bibnamefont {{Kocsis}}},\ and\ \bibinfo {author} {\bibfnamefont {Z.}~\bibnamefont {{Haiman}}},\ }\href {https://doi.org/10.1093/mnras/stad1926} {\bibfield  {journal} {\bibinfo  {journal} {Mon. Not. R. Astron. Soc.}\ }\textbf {\bibinfo {volume} {524}},\ \bibinfo {pages} {2770} (\bibinfo {year} {2023})},\ \Eprint {https://arxiv.org/abs/2212.06133} {arXiv:2212.06133 [astro-ph.GA]} \BibitemShut {NoStop}%
\bibitem [{\citenamefont {{Xue}}\ \emph {et~al.}(2025)\citenamefont {{Xue}}, \citenamefont {{Tagawa}}, \citenamefont {{Haiman}},\ and\ \citenamefont {{Bartos}}}]{Xue2025}%
  \BibitemOpen
  \bibfield  {author} {\bibinfo {author} {\bibfnamefont {L.}~\bibnamefont {{Xue}}}, \bibinfo {author} {\bibfnamefont {H.}~\bibnamefont {{Tagawa}}}, \bibinfo {author} {\bibfnamefont {Z.}~\bibnamefont {{Haiman}}},\ and\ \bibinfo {author} {\bibfnamefont {I.}~\bibnamefont {{Bartos}}},\ }\href {https://doi.org/10.48550/arXiv.2504.19570} {\bibfield  {journal} {\bibinfo  {journal} {arXiv e-prints}\ ,\ \bibinfo {eid} {arXiv:2504.19570}} (\bibinfo {year} {2025})},\ \Eprint {https://arxiv.org/abs/2504.19570} {arXiv:2504.19570 [astro-ph.HE]} \BibitemShut {NoStop}%
\bibitem [{\citenamefont {{Tagawa}}\ \emph {et~al.}(2020{\natexlab{b}})\citenamefont {{Tagawa}}, \citenamefont {{Haiman}}, \citenamefont {{Bartos}},\ and\ \citenamefont {{Kocsis}}}]{Tagawa2020_spin}%
  \BibitemOpen
  \bibfield  {author} {\bibinfo {author} {\bibfnamefont {H.}~\bibnamefont {{Tagawa}}}, \bibinfo {author} {\bibfnamefont {Z.}~\bibnamefont {{Haiman}}}, \bibinfo {author} {\bibfnamefont {I.}~\bibnamefont {{Bartos}}},\ and\ \bibinfo {author} {\bibfnamefont {B.}~\bibnamefont {{Kocsis}}},\ }\href {https://doi.org/10.3847/1538-4357/aba2cc} {\bibfield  {journal} {\bibinfo  {journal} {Astrophys. J.}\ }\textbf {\bibinfo {volume} {899}},\ \bibinfo {eid} {26} (\bibinfo {year} {2020}{\natexlab{b}})},\ \Eprint {https://arxiv.org/abs/2004.11914} {arXiv:2004.11914 [astro-ph.HE]} \BibitemShut {NoStop}%
\bibitem [{\citenamefont {{Farr}}\ \emph {et~al.}(2017)\citenamefont {{Farr}}, \citenamefont {{Stevenson}}, \citenamefont {{Miller}}, \citenamefont {{Mandel}}, \citenamefont {{Farr}},\ and\ \citenamefont {{Vecchio}}}]{2017Natur.548..426F}%
  \BibitemOpen
  \bibfield  {author} {\bibinfo {author} {\bibfnamefont {W.~M.}\ \bibnamefont {{Farr}}}, \bibinfo {author} {\bibfnamefont {S.}~\bibnamefont {{Stevenson}}}, \bibinfo {author} {\bibfnamefont {M.~C.}\ \bibnamefont {{Miller}}}, \bibinfo {author} {\bibfnamefont {I.}~\bibnamefont {{Mandel}}}, \bibinfo {author} {\bibfnamefont {B.}~\bibnamefont {{Farr}}},\ and\ \bibinfo {author} {\bibfnamefont {A.}~\bibnamefont {{Vecchio}}},\ }\href {https://doi.org/10.1038/nature23453} {\bibfield  {journal} {\bibinfo  {journal} {Nature}\ }\textbf {\bibinfo {volume} {548}},\ \bibinfo {pages} {426} (\bibinfo {year} {2017})},\ \Eprint {https://arxiv.org/abs/1706.01385} {arXiv:1706.01385 [astro-ph.HE]} \BibitemShut {NoStop}%
\bibitem [{\citenamefont {{Samsing}}\ \emph {et~al.}(2022)\citenamefont {{Samsing}}, \citenamefont {{Bartos}}, \citenamefont {{D'Orazio}}, \citenamefont {{Haiman}}, \citenamefont {{Kocsis}}, \citenamefont {{Leigh}}, \citenamefont {{Liu}}, \citenamefont {{Pessah}},\ and\ \citenamefont {{Tagawa}}}]{Samsing2022}%
  \BibitemOpen
  \bibfield  {author} {\bibinfo {author} {\bibfnamefont {J.}~\bibnamefont {{Samsing}}}, \bibinfo {author} {\bibfnamefont {I.}~\bibnamefont {{Bartos}}}, \bibinfo {author} {\bibfnamefont {D.~J.}\ \bibnamefont {{D'Orazio}}}, \bibinfo {author} {\bibfnamefont {Z.}~\bibnamefont {{Haiman}}}, \bibinfo {author} {\bibfnamefont {B.}~\bibnamefont {{Kocsis}}}, \bibinfo {author} {\bibfnamefont {N.~W.~C.}\ \bibnamefont {{Leigh}}}, \bibinfo {author} {\bibfnamefont {B.}~\bibnamefont {{Liu}}}, \bibinfo {author} {\bibfnamefont {M.~E.}\ \bibnamefont {{Pessah}}},\ and\ \bibinfo {author} {\bibfnamefont {H.}~\bibnamefont {{Tagawa}}},\ }\href {https://doi.org/10.1038/s41586-021-04333-1} {\bibfield  {journal} {\bibinfo  {journal} {Nature}\ }\textbf {\bibinfo {volume} {603}},\ \bibinfo {pages} {237} (\bibinfo {year} {2022})},\ \Eprint {https://arxiv.org/abs/2010.09765} {arXiv:2010.09765 [astro-ph.HE]} \BibitemShut {NoStop}%
\bibitem [{\citenamefont {{Tagawa}}\ \emph {et~al.}(2021{\natexlab{a}})\citenamefont {{Tagawa}}, \citenamefont {{Kocsis}}, \citenamefont {{Haiman}}, \citenamefont {{Bartos}}, \citenamefont {{Omukai}},\ and\ \citenamefont {{Samsing}}}]{Tagawa2021_gap}%
  \BibitemOpen
  \bibfield  {author} {\bibinfo {author} {\bibfnamefont {H.}~\bibnamefont {{Tagawa}}}, \bibinfo {author} {\bibfnamefont {B.}~\bibnamefont {{Kocsis}}}, \bibinfo {author} {\bibfnamefont {Z.}~\bibnamefont {{Haiman}}}, \bibinfo {author} {\bibfnamefont {I.}~\bibnamefont {{Bartos}}}, \bibinfo {author} {\bibfnamefont {K.}~\bibnamefont {{Omukai}}},\ and\ \bibinfo {author} {\bibfnamefont {J.}~\bibnamefont {{Samsing}}},\ }\href {https://doi.org/10.3847/1538-4357/abd555} {\bibfield  {journal} {\bibinfo  {journal} {Astrophys. J.}\ }\textbf {\bibinfo {volume} {908}},\ \bibinfo {eid} {194} (\bibinfo {year} {2021}{\natexlab{a}})},\ \Eprint {https://arxiv.org/abs/2012.00011} {arXiv:2012.00011 [astro-ph.HE]} \BibitemShut {NoStop}%
\bibitem [{\citenamefont {{Gayathri}}\ \emph {et~al.}(2021)\citenamefont {{Gayathri}}, \citenamefont {{Yang}}, \citenamefont {{Tagawa}}, \citenamefont {{Haiman}},\ and\ \citenamefont {{Bartos}}}]{Gayathri2021}%
  \BibitemOpen
  \bibfield  {author} {\bibinfo {author} {\bibfnamefont {V.}~\bibnamefont {{Gayathri}}}, \bibinfo {author} {\bibfnamefont {Y.}~\bibnamefont {{Yang}}}, \bibinfo {author} {\bibfnamefont {H.}~\bibnamefont {{Tagawa}}}, \bibinfo {author} {\bibfnamefont {Z.}~\bibnamefont {{Haiman}}},\ and\ \bibinfo {author} {\bibfnamefont {I.}~\bibnamefont {{Bartos}}},\ }\href {https://doi.org/10.3847/2041-8213/ac2cc1} {\bibfield  {journal} {\bibinfo  {journal} {Astrophys. J.}\ }\textbf {\bibinfo {volume} {920}},\ \bibinfo {eid} {L42} (\bibinfo {year} {2021})},\ \Eprint {https://arxiv.org/abs/2104.10253} {arXiv:2104.10253 [gr-qc]} \BibitemShut {NoStop}%
\bibitem [{\citenamefont {{Tagawa}}\ \emph {et~al.}(2021{\natexlab{b}})\citenamefont {{Tagawa}}, \citenamefont {{Kocsis}}, \citenamefont {{Haiman}}, \citenamefont {{Bartos}}, \citenamefont {{Omukai}},\ and\ \citenamefont {{Samsing}}}]{Tagawa2021_ecc}%
  \BibitemOpen
  \bibfield  {author} {\bibinfo {author} {\bibfnamefont {H.}~\bibnamefont {{Tagawa}}}, \bibinfo {author} {\bibfnamefont {B.}~\bibnamefont {{Kocsis}}}, \bibinfo {author} {\bibfnamefont {Z.}~\bibnamefont {{Haiman}}}, \bibinfo {author} {\bibfnamefont {I.}~\bibnamefont {{Bartos}}}, \bibinfo {author} {\bibfnamefont {K.}~\bibnamefont {{Omukai}}},\ and\ \bibinfo {author} {\bibfnamefont {J.}~\bibnamefont {{Samsing}}},\ }\href {https://doi.org/10.3847/2041-8213/abd4d3} {\bibfield  {journal} {\bibinfo  {journal} {Astrophys. J. Lett.}\ }\textbf {\bibinfo {volume} {907}},\ \bibinfo {eid} {L20} (\bibinfo {year} {2021}{\natexlab{b}})},\ \Eprint {https://arxiv.org/abs/2010.10526} {arXiv:2010.10526 [astro-ph.HE]} \BibitemShut {NoStop}%
\bibitem [{\citenamefont {{Bhaumik}}\ \emph {et~al.}(2025)\citenamefont {{Bhaumik}}, \citenamefont {{Gayathri}}, \citenamefont {{Bartos}}, \citenamefont {{Anglin}}, \citenamefont {{Carullo}}, \citenamefont {{Healy}}, \citenamefont {{Klimenko}}, \citenamefont {{Lange}}, \citenamefont {{Lousto}}, \citenamefont {{Mishra}},\ and\ \citenamefont {{Szczepa{\'n}czyk}}}]{2025PhRvD.111l3032B}%
  \BibitemOpen
  \bibfield  {author} {\bibinfo {author} {\bibfnamefont {S.}~\bibnamefont {{Bhaumik}}}, \bibinfo {author} {\bibfnamefont {V.}~\bibnamefont {{Gayathri}}}, \bibinfo {author} {\bibfnamefont {I.}~\bibnamefont {{Bartos}}}, \bibinfo {author} {\bibfnamefont {J.}~\bibnamefont {{Anglin}}}, \bibinfo {author} {\bibfnamefont {G.}~\bibnamefont {{Carullo}}}, \bibinfo {author} {\bibfnamefont {J.}~\bibnamefont {{Healy}}}, \bibinfo {author} {\bibfnamefont {S.}~\bibnamefont {{Klimenko}}}, \bibinfo {author} {\bibfnamefont {J.}~\bibnamefont {{Lange}}}, \bibinfo {author} {\bibfnamefont {C.}~\bibnamefont {{Lousto}}}, \bibinfo {author} {\bibfnamefont {T.}~\bibnamefont {{Mishra}}},\ and\ \bibinfo {author} {\bibfnamefont {M.~J.}\ \bibnamefont {{Szczepa{\'n}czyk}}},\ }\href {https://doi.org/10.1103/hwr5-scp4} {\bibfield  {journal} {\bibinfo  {journal} {\prd}\ }\textbf {\bibinfo {volume} {111}},\ \bibinfo {eid} {123032} (\bibinfo {year} {2025})},\ \Eprint {https://arxiv.org/abs/2410.15192} {arXiv:2410.15192 [gr-qc]} \BibitemShut
  {NoStop}%
\bibitem [{\citenamefont {{Fabj}}\ \emph {et~al.}(2020)\citenamefont {{Fabj}}, \citenamefont {{Nasim}}, \citenamefont {{Caban}}, \citenamefont {{Ford}}, \citenamefont {{McKernan}},\ and\ \citenamefont {{Bellovary}}}]{2020MNRAS.499.2608F}%
  \BibitemOpen
  \bibfield  {author} {\bibinfo {author} {\bibfnamefont {G.}~\bibnamefont {{Fabj}}}, \bibinfo {author} {\bibfnamefont {S.~S.}\ \bibnamefont {{Nasim}}}, \bibinfo {author} {\bibfnamefont {F.}~\bibnamefont {{Caban}}}, \bibinfo {author} {\bibfnamefont {K.~E.~S.}\ \bibnamefont {{Ford}}}, \bibinfo {author} {\bibfnamefont {B.}~\bibnamefont {{McKernan}}},\ and\ \bibinfo {author} {\bibfnamefont {J.~M.}\ \bibnamefont {{Bellovary}}},\ }\href {https://doi.org/10.1093/mnras/staa3004} {\bibfield  {journal} {\bibinfo  {journal} {Mon. Not. R. Astron. Soc.}\ }\textbf {\bibinfo {volume} {499}},\ \bibinfo {pages} {2608} (\bibinfo {year} {2020})},\ \Eprint {https://arxiv.org/abs/2006.11229} {arXiv:2006.11229 [astro-ph.GA]} \BibitemShut {NoStop}%
\bibitem [{\citenamefont {{Nasim}}\ \emph {et~al.}(2023)\citenamefont {{Nasim}}, \citenamefont {{Fabj}}, \citenamefont {{Caban}}, \citenamefont {{Secunda}}, \citenamefont {{Ford}}, \citenamefont {{McKernan}}, \citenamefont {{Bellovary}}, \citenamefont {{Leigh}},\ and\ \citenamefont {{Lyra}}}]{2023MNRAS.522.5393N}%
  \BibitemOpen
  \bibfield  {author} {\bibinfo {author} {\bibfnamefont {S.~S.}\ \bibnamefont {{Nasim}}}, \bibinfo {author} {\bibfnamefont {G.}~\bibnamefont {{Fabj}}}, \bibinfo {author} {\bibfnamefont {F.}~\bibnamefont {{Caban}}}, \bibinfo {author} {\bibfnamefont {A.}~\bibnamefont {{Secunda}}}, \bibinfo {author} {\bibfnamefont {K.~E.~S.}\ \bibnamefont {{Ford}}}, \bibinfo {author} {\bibfnamefont {B.}~\bibnamefont {{McKernan}}}, \bibinfo {author} {\bibfnamefont {J.~M.}\ \bibnamefont {{Bellovary}}}, \bibinfo {author} {\bibfnamefont {N.~W.~C.}\ \bibnamefont {{Leigh}}},\ and\ \bibinfo {author} {\bibfnamefont {W.}~\bibnamefont {{Lyra}}},\ }\href {https://doi.org/10.1093/mnras/stad1295} {\bibfield  {journal} {\bibinfo  {journal} {Mon. Not. R. Astron. Soc.}\ }\textbf {\bibinfo {volume} {522}},\ \bibinfo {pages} {5393} (\bibinfo {year} {2023})},\ \Eprint {https://arxiv.org/abs/2207.09540} {arXiv:2207.09540 [astro-ph.GA]} \BibitemShut {NoStop}%
\bibitem [{\citenamefont {{Wang}}\ \emph {et~al.}(2024)\citenamefont {{Wang}}, \citenamefont {{Zhu}},\ and\ \citenamefont {{Lin}}}]{2024MNRAS.528.4958W}%
  \BibitemOpen
  \bibfield  {author} {\bibinfo {author} {\bibfnamefont {Y.}~\bibnamefont {{Wang}}}, \bibinfo {author} {\bibfnamefont {Z.}~\bibnamefont {{Zhu}}},\ and\ \bibinfo {author} {\bibfnamefont {D.~N.~C.}\ \bibnamefont {{Lin}}},\ }\href {https://doi.org/10.1093/mnras/stae321} {\bibfield  {journal} {\bibinfo  {journal} {Mon. Not. R. Astron. Soc.}\ }\textbf {\bibinfo {volume} {528}},\ \bibinfo {pages} {4958} (\bibinfo {year} {2024})},\ \Eprint {https://arxiv.org/abs/2308.09129} {arXiv:2308.09129 [astro-ph.GA]} \BibitemShut {NoStop}%
\bibitem [{\citenamefont {{Samsing}}\ \emph {et~al.}(2018)\citenamefont {{Samsing}}, \citenamefont {{Askar}},\ and\ \citenamefont {{Giersz}}}]{2018ApJ...855..124S}%
  \BibitemOpen
  \bibfield  {author} {\bibinfo {author} {\bibfnamefont {J.}~\bibnamefont {{Samsing}}}, \bibinfo {author} {\bibfnamefont {A.}~\bibnamefont {{Askar}}},\ and\ \bibinfo {author} {\bibfnamefont {M.}~\bibnamefont {{Giersz}}},\ }\href {https://doi.org/10.3847/1538-4357/aaab52} {\bibfield  {journal} {\bibinfo  {journal} {Astrophys. J.}\ }\textbf {\bibinfo {volume} {855}},\ \bibinfo {eid} {124} (\bibinfo {year} {2018})},\ \Eprint {https://arxiv.org/abs/1712.06186} {arXiv:1712.06186 [astro-ph.HE]} \BibitemShut {NoStop}%
\bibitem [{\citenamefont {{Shaikh}}\ \emph {et~al.}(2023)\citenamefont {{Shaikh}}, \citenamefont {{Varma}}, \citenamefont {{Pfeiffer}}, \citenamefont {{Ramos-Buades}},\ and\ \citenamefont {{van de Meent}}}]{2023:Shaikh:ecc}%
  \BibitemOpen
  \bibfield  {author} {\bibinfo {author} {\bibfnamefont {M.~A.}\ \bibnamefont {{Shaikh}}}, \bibinfo {author} {\bibfnamefont {V.}~\bibnamefont {{Varma}}}, \bibinfo {author} {\bibfnamefont {H.~P.}\ \bibnamefont {{Pfeiffer}}}, \bibinfo {author} {\bibfnamefont {A.}~\bibnamefont {{Ramos-Buades}}},\ and\ \bibinfo {author} {\bibfnamefont {M.}~\bibnamefont {{van de Meent}}},\ }\href {https://doi.org/10.1103/PhysRevD.108.104007} {\bibfield  {journal} {\bibinfo  {journal} {Phys. Rev. D}\ }\textbf {\bibinfo {volume} {108}},\ \bibinfo {eid} {104007} (\bibinfo {year} {2023})},\ \Eprint {https://arxiv.org/abs/2302.11257} {arXiv:2302.11257 [gr-qc]} \BibitemShut {NoStop}%
\bibitem [{\citenamefont {{McMillin}}\ \emph {et~al.}(2025)\citenamefont {{McMillin}}, \citenamefont {{Wagner}}, \citenamefont {{Ficarra}}, \citenamefont {{Lousto}},\ and\ \citenamefont {{O'Shaughnessy}}}]{2025arXiv250722862M}%
  \BibitemOpen
  \bibfield  {author} {\bibinfo {author} {\bibfnamefont {P.}~\bibnamefont {{McMillin}}}, \bibinfo {author} {\bibfnamefont {K.~J.}\ \bibnamefont {{Wagner}}}, \bibinfo {author} {\bibfnamefont {G.}~\bibnamefont {{Ficarra}}}, \bibinfo {author} {\bibfnamefont {C.~O.}\ \bibnamefont {{Lousto}}},\ and\ \bibinfo {author} {\bibfnamefont {R.}~\bibnamefont {{O'Shaughnessy}}},\ }\href@noop {} {\bibfield  {journal} {\bibinfo  {journal} {{}}\ } (\bibinfo {year} {2025})},\ \Eprint {https://arxiv.org/abs/2507.22862} {arXiv:2507.22862 [gr-qc]} \BibitemShut {NoStop}%
\bibitem [{\citenamefont {{Buonanno}}\ and\ \citenamefont {{Damour}}(1999)}]{1999PhRvD..59h4006B}%
  \BibitemOpen
  \bibfield  {author} {\bibinfo {author} {\bibfnamefont {A.}~\bibnamefont {{Buonanno}}}\ and\ \bibinfo {author} {\bibfnamefont {T.}~\bibnamefont {{Damour}}},\ }\href {https://doi.org/10.1103/PhysRevD.59.084006} {\bibfield  {journal} {\bibinfo  {journal} {Phys. Rev. D}\ }\textbf {\bibinfo {volume} {59}},\ \bibinfo {eid} {084006} (\bibinfo {year} {1999})},\ \Eprint {https://arxiv.org/abs/gr-qc/9811091} {arXiv:gr-qc/9811091 [gr-qc]} \BibitemShut {NoStop}%
\bibitem [{\citenamefont {{Damour}}\ and\ \citenamefont {{Nagar}}(2009)}]{DamourNagar2009}%
  \BibitemOpen
  \bibfield  {author} {\bibinfo {author} {\bibfnamefont {T.}~\bibnamefont {{Damour}}}\ and\ \bibinfo {author} {\bibfnamefont {A.}~\bibnamefont {{Nagar}}},\ }\href {https://doi.org/10.48550/arXiv.0906.1769} {\bibfield  {journal} {\bibinfo  {journal} {arXiv e-prints}\ ,\ \bibinfo {eid} {arXiv:0906.1769}} (\bibinfo {year} {2009})},\ \Eprint {https://arxiv.org/abs/0906.1769} {arXiv:0906.1769 [gr-qc]} \BibitemShut {NoStop}%
\bibitem [{\citenamefont {Cotesta}\ \emph {et~al.}(2018)\citenamefont {Cotesta}, \citenamefont {Buonanno}, \citenamefont {Boh\'e}, \citenamefont {Taracchini}, \citenamefont {Hinder},\ and\ \citenamefont {Ossokine}}]{Cotesta:2018fcv}%
  \BibitemOpen
  \bibfield  {author} {\bibinfo {author} {\bibfnamefont {R.}~\bibnamefont {Cotesta}}, \bibinfo {author} {\bibfnamefont {A.}~\bibnamefont {Buonanno}}, \bibinfo {author} {\bibfnamefont {A.}~\bibnamefont {Boh\'e}}, \bibinfo {author} {\bibfnamefont {A.}~\bibnamefont {Taracchini}}, \bibinfo {author} {\bibfnamefont {I.}~\bibnamefont {Hinder}},\ and\ \bibinfo {author} {\bibfnamefont {S.}~\bibnamefont {Ossokine}},\ }\href {https://doi.org/10.1103/PhysRevD.98.084028} {\bibfield  {journal} {\bibinfo  {journal} {Phys. Rev. D}\ }\textbf {\bibinfo {volume} {98}},\ \bibinfo {pages} {084028} (\bibinfo {year} {2018})},\ \Eprint {https://arxiv.org/abs/1803.10701} {arXiv:1803.10701 [gr-qc]} \BibitemShut {NoStop}%
\bibitem [{\citenamefont {Taracchini}\ \emph {et~al.}(2012)\citenamefont {Taracchini}, \citenamefont {Pan}, \citenamefont {Buonanno}, \citenamefont {Barausse}, \citenamefont {Boyle}, \citenamefont {Chu}, \citenamefont {Lovelace}, \citenamefont {Pfeiffer},\ and\ \citenamefont {Scheel}}]{Taracchini:2012ig}%
  \BibitemOpen
  \bibfield  {author} {\bibinfo {author} {\bibfnamefont {A.}~\bibnamefont {Taracchini}}, \bibinfo {author} {\bibfnamefont {Y.}~\bibnamefont {Pan}}, \bibinfo {author} {\bibfnamefont {A.}~\bibnamefont {Buonanno}}, \bibinfo {author} {\bibfnamefont {E.}~\bibnamefont {Barausse}}, \bibinfo {author} {\bibfnamefont {M.}~\bibnamefont {Boyle}}, \bibinfo {author} {\bibfnamefont {T.}~\bibnamefont {Chu}}, \bibinfo {author} {\bibfnamefont {G.}~\bibnamefont {Lovelace}}, \bibinfo {author} {\bibfnamefont {H.~P.}\ \bibnamefont {Pfeiffer}},\ and\ \bibinfo {author} {\bibfnamefont {M.~A.}\ \bibnamefont {Scheel}},\ }\href {https://doi.org/10.1103/PhysRevD.86.024011} {\bibfield  {journal} {\bibinfo  {journal} {Phys. Rev. D}\ }\textbf {\bibinfo {volume} {86}},\ \bibinfo {pages} {024011} (\bibinfo {year} {2012})},\ \Eprint {https://arxiv.org/abs/1202.0790} {arXiv:1202.0790 [gr-qc]} \BibitemShut {NoStop}%
\bibitem [{\citenamefont {Ramos-Buades}\ \emph {et~al.}(2023)\citenamefont {Ramos-Buades}, \citenamefont {Buonanno},\ and\ \citenamefont {Gair}}]{Ramos-Buades:2023yhy}%
  \BibitemOpen
  \bibfield  {author} {\bibinfo {author} {\bibfnamefont {A.}~\bibnamefont {Ramos-Buades}}, \bibinfo {author} {\bibfnamefont {A.}~\bibnamefont {Buonanno}},\ and\ \bibinfo {author} {\bibfnamefont {J.}~\bibnamefont {Gair}},\ }\href {https://doi.org/10.1103/PhysRevD.108.124063} {\bibfield  {journal} {\bibinfo  {journal} {Phys. Rev. D}\ }\textbf {\bibinfo {volume} {108}},\ \bibinfo {pages} {124063} (\bibinfo {year} {2023})},\ \Eprint {https://arxiv.org/abs/2309.15528} {arXiv:2309.15528 [gr-qc]} \BibitemShut {NoStop}%
\bibitem [{\citenamefont {Dax}\ \emph {et~al.}(2023)\citenamefont {Dax}, \citenamefont {Green}, \citenamefont {Gair}, \citenamefont {P\"urrer}, \citenamefont {Wildberger}, \citenamefont {Macke}, \citenamefont {Buonanno},\ and\ \citenamefont {Sch\"olkopf}}]{Dax:2022pxd}%
  \BibitemOpen
  \bibfield  {author} {\bibinfo {author} {\bibfnamefont {M.}~\bibnamefont {Dax}}, \bibinfo {author} {\bibfnamefont {S.~R.}\ \bibnamefont {Green}}, \bibinfo {author} {\bibfnamefont {J.}~\bibnamefont {Gair}}, \bibinfo {author} {\bibfnamefont {M.}~\bibnamefont {P\"urrer}}, \bibinfo {author} {\bibfnamefont {J.}~\bibnamefont {Wildberger}}, \bibinfo {author} {\bibfnamefont {J.~H.}\ \bibnamefont {Macke}}, \bibinfo {author} {\bibfnamefont {A.}~\bibnamefont {Buonanno}},\ and\ \bibinfo {author} {\bibfnamefont {B.}~\bibnamefont {Sch\"olkopf}},\ }\href {https://doi.org/10.1103/PhysRevLett.130.171403} {\bibfield  {journal} {\bibinfo  {journal} {Phys. Rev. Lett.}\ }\textbf {\bibinfo {volume} {130}},\ \bibinfo {pages} {171403} (\bibinfo {year} {2023})},\ \Eprint {https://arxiv.org/abs/2210.05686} {arXiv:2210.05686 [gr-qc]} \BibitemShut {NoStop}%
\bibitem [{\citenamefont {{Hannam}}\ \emph {et~al.}(2022)\citenamefont {{Hannam}}, \citenamefont {{Hoy}}, \citenamefont {{Thompson}}, \citenamefont {{Fairhurst}}, \citenamefont {{Raymond}}, \citenamefont {{Colleoni}}, \citenamefont {{Davis}}, \citenamefont {{Estell{\'e}s}}, \citenamefont {{Haster}}, \citenamefont {{Helmling-Cornell}}, \citenamefont {{Husa}}, \citenamefont {{Keitel}}, \citenamefont {{Massinger}}, \citenamefont {{Men{\'e}ndez-V{\'a}zquez}}, \citenamefont {{Mogushi}}, \citenamefont {{Ossokine}}, \citenamefont {{Payne}}, \citenamefont {{Pratten}}, \citenamefont {{Romero-Shaw}}, \citenamefont {{Sadiq}}, \citenamefont {{Schmidt}}, \citenamefont {{Tenorio}}, \citenamefont {{Udall}}, \citenamefont {{Veitch}}, \citenamefont {{Williams}}, \citenamefont {{Yelikar}},\ and\ \citenamefont {{Zimmerman}}}]{2022:Hannam:GW200129}%
  \BibitemOpen
  \bibfield  {author} {\bibinfo {author} {\bibfnamefont {M.}~\bibnamefont {{Hannam}}}, \bibinfo {author} {\bibfnamefont {C.}~\bibnamefont {{Hoy}}}, \bibinfo {author} {\bibfnamefont {J.~E.}\ \bibnamefont {{Thompson}}}, \bibinfo {author} {\bibfnamefont {S.}~\bibnamefont {{Fairhurst}}}, \bibinfo {author} {\bibfnamefont {V.}~\bibnamefont {{Raymond}}}, \bibinfo {author} {\bibfnamefont {M.}~\bibnamefont {{Colleoni}}}, \bibinfo {author} {\bibfnamefont {D.}~\bibnamefont {{Davis}}}, \bibinfo {author} {\bibfnamefont {H.}~\bibnamefont {{Estell{\'e}s}}}, \bibinfo {author} {\bibfnamefont {C.-J.}\ \bibnamefont {{Haster}}}, \bibinfo {author} {\bibfnamefont {A.}~\bibnamefont {{Helmling-Cornell}}}, \bibinfo {author} {\bibfnamefont {S.}~\bibnamefont {{Husa}}}, \bibinfo {author} {\bibfnamefont {D.}~\bibnamefont {{Keitel}}}, \bibinfo {author} {\bibfnamefont {T.~J.}\ \bibnamefont {{Massinger}}}, \bibinfo {author} {\bibfnamefont {A.}~\bibnamefont {{Men{\'e}ndez-V{\'a}zquez}}}, \bibinfo {author} {\bibfnamefont {K.}~\bibnamefont
  {{Mogushi}}}, \bibinfo {author} {\bibfnamefont {S.}~\bibnamefont {{Ossokine}}}, \bibinfo {author} {\bibfnamefont {E.}~\bibnamefont {{Payne}}}, \bibinfo {author} {\bibfnamefont {G.}~\bibnamefont {{Pratten}}}, \bibinfo {author} {\bibfnamefont {I.}~\bibnamefont {{Romero-Shaw}}}, \bibinfo {author} {\bibfnamefont {J.}~\bibnamefont {{Sadiq}}}, \bibinfo {author} {\bibfnamefont {P.}~\bibnamefont {{Schmidt}}}, \bibinfo {author} {\bibfnamefont {R.}~\bibnamefont {{Tenorio}}}, \bibinfo {author} {\bibfnamefont {R.}~\bibnamefont {{Udall}}}, \bibinfo {author} {\bibfnamefont {J.}~\bibnamefont {{Veitch}}}, \bibinfo {author} {\bibfnamefont {D.}~\bibnamefont {{Williams}}}, \bibinfo {author} {\bibfnamefont {A.~B.}\ \bibnamefont {{Yelikar}}},\ and\ \bibinfo {author} {\bibfnamefont {A.}~\bibnamefont {{Zimmerman}}},\ }\href {https://doi.org/10.1038/s41586-022-05212-z} {\bibfield  {journal} {\bibinfo  {journal} {Nature}\ }\textbf {\bibinfo {volume} {610}},\ \bibinfo {pages} {652} (\bibinfo {year} {2022})},\ \Eprint
  {https://arxiv.org/abs/2112.11300} {arXiv:2112.11300 [gr-qc]} \BibitemShut {NoStop}%
\bibitem [{\citenamefont {{Varma}}\ \emph {et~al.}(2022)\citenamefont {{Varma}}, \citenamefont {{Biscoveanu}}, \citenamefont {{Islam}}, \citenamefont {{Shaik}}, \citenamefont {{Haster}}, \citenamefont {{Isi}}, \citenamefont {{Farr}}, \citenamefont {{Field}},\ and\ \citenamefont {{Vitale}}}]{2022PhRvL.128s1102V}%
  \BibitemOpen
  \bibfield  {author} {\bibinfo {author} {\bibfnamefont {V.}~\bibnamefont {{Varma}}}, \bibinfo {author} {\bibfnamefont {S.}~\bibnamefont {{Biscoveanu}}}, \bibinfo {author} {\bibfnamefont {T.}~\bibnamefont {{Islam}}}, \bibinfo {author} {\bibfnamefont {F.~H.}\ \bibnamefont {{Shaik}}}, \bibinfo {author} {\bibfnamefont {C.-J.}\ \bibnamefont {{Haster}}}, \bibinfo {author} {\bibfnamefont {M.}~\bibnamefont {{Isi}}}, \bibinfo {author} {\bibfnamefont {W.~M.}\ \bibnamefont {{Farr}}}, \bibinfo {author} {\bibfnamefont {S.~E.}\ \bibnamefont {{Field}}},\ and\ \bibinfo {author} {\bibfnamefont {S.}~\bibnamefont {{Vitale}}},\ }\href {https://doi.org/10.1103/PhysRevLett.128.191102} {\bibfield  {journal} {\bibinfo  {journal} {Phys. Rev. Lett.}\ }\textbf {\bibinfo {volume} {128}},\ \bibinfo {eid} {191102} (\bibinfo {year} {2022})},\ \Eprint {https://arxiv.org/abs/2201.01302} {arXiv:2201.01302 [astro-ph.HE]} \BibitemShut {NoStop}%
\bibitem [{\citenamefont {{Payne}}\ \emph {et~al.}(2022)\citenamefont {{Payne}}, \citenamefont {{Hourihane}}, \citenamefont {{Golomb}}, \citenamefont {{Udall}}, \citenamefont {{Davis}},\ and\ \citenamefont {{Chatziioannou}}}]{2022:Payne:glitch}%
  \BibitemOpen
  \bibfield  {author} {\bibinfo {author} {\bibfnamefont {E.}~\bibnamefont {{Payne}}}, \bibinfo {author} {\bibfnamefont {S.}~\bibnamefont {{Hourihane}}}, \bibinfo {author} {\bibfnamefont {J.}~\bibnamefont {{Golomb}}}, \bibinfo {author} {\bibfnamefont {R.}~\bibnamefont {{Udall}}}, \bibinfo {author} {\bibfnamefont {D.}~\bibnamefont {{Davis}}},\ and\ \bibinfo {author} {\bibfnamefont {K.}~\bibnamefont {{Chatziioannou}}},\ }\href {https://doi.org/10.1103/PhysRevD.106.104017} {\bibfield  {journal} {\bibinfo  {journal} {Phys. Rev. D}\ }\textbf {\bibinfo {volume} {106}},\ \bibinfo {eid} {104017} (\bibinfo {year} {2022})},\ \Eprint {https://arxiv.org/abs/2206.11932} {arXiv:2206.11932 [gr-qc]} \BibitemShut {NoStop}%
\bibitem [{\citenamefont {{Macas}}\ \emph {et~al.}(2024)\citenamefont {{Macas}}, \citenamefont {{Lundgren}},\ and\ \citenamefont {{Ashton}}}]{2024:Macas:glitch}%
  \BibitemOpen
  \bibfield  {author} {\bibinfo {author} {\bibfnamefont {R.}~\bibnamefont {{Macas}}}, \bibinfo {author} {\bibfnamefont {A.}~\bibnamefont {{Lundgren}}},\ and\ \bibinfo {author} {\bibfnamefont {G.}~\bibnamefont {{Ashton}}},\ }\href {https://doi.org/10.1103/PhysRevD.109.062006} {\bibfield  {journal} {\bibinfo  {journal} {Phys. Rev. D}\ }\textbf {\bibinfo {volume} {109}},\ \bibinfo {eid} {062006} (\bibinfo {year} {2024})},\ \Eprint {https://arxiv.org/abs/2311.09921} {arXiv:2311.09921 [gr-qc]} \BibitemShut {NoStop}%
\end{thebibliography}%
\end{document}